\documentclass[article]{JHEP}
\usepackage{amsmath,epsfig}
\usepackage{amssymb,amsfonts}
\usepackage{epsfig}
\relax
\renewcommand{\theequation}{\arabic{section}.\arabic{subsection}.\arabic{equation}}
\def\be{\begin{equation}}
\def\ee{\end{equation}}
\def\bea{\begin{eqnarray}}
\def\eea{\end{eqnarray}}
\def\be{\begin{equation}}
\def\ee{\end{equation}}

\newcommand\fverb{\setbox\pippobox=\hbox\bgroup\verb}
\newcommand\fverbdo{\egroup\medskip\noindent%
                        \fbox{\unhbox\pippobox}\ }
\newcommand\fverbit{\egroup\item[\fbox{\unhbox\pippobox}]}
\newbox\pippobox
\def\F{\Phi}

\def\e{\epsilon}
\def\h{\eta}
\def\ka{\kappa}
\def\z{\zeta}
\def\m{\mu}
\def\n{\nu}
\def\r{\rho}
\def\s{\sigma}
\def\t{\theta}
\def\sp{\;\;\;,\;\;\;}

\def\p{\partial}
\def\bp{\bar\partial}
\def\pb{\bar\partial}

\def\zb{\bar z}

\def\f{\varphi}
\def\a{\alpha}
\def\b{\beta}
\def\d{\delta}
\def\l{\lambda}
\def\g{\gamma}
\def\G{\Gamma}
\def\S{\Sigma}
\def\w{\Omega}
\def\ba{\begin{eqnarray}}
\def\ea{\end{eqnarray}}
\def\nb{\nonumber}
\def\jh{{\hat{\jmath}}}
\newcommand{\Real}{\mathbb{R}}

\def\np#1#2#3{Nucl. Phys. {\bf{B#1}} (#2) #3}

\title{String interactions in gravitational wave backgrounds}

\author{Giuseppe D'Appollonio\\
LPTHE, , Universit\'es Paris VI et VII, 4 pl Jussieu, \\
75252 Paris cedex 05, FRANCE\\{\tt E-mail:
giuseppe@lpthe.jussieu.fr}}
\author{
Elias Kiritsis\\
CPHT, UMR du CNRS 7644, Ecole Polytechnique,\\
 91128, Palaiseau, FRANCE\\
and\\
Department of Physics, University of Crete\\
71003 Heraklion, GREECE\\
{\tt E-mail: kiritsis@cpht.polytechnique.fr}
}

\preprint{\hepth{0305081} \\ CPHT-RR019.0503 \\ LPTHE-03-14}

\abstract{The non-compact CFT of a class of NS-supported pp-wave backgrounds is solved exactly.
The associated tree-level covariant string scattering amplitudes are calculated.
The S-matrix elements are well-defined, dual but not analytic as a function of $p^+$.
They have poles corresponding to physical intermediate states with
$p^+\not =0$ and logarithmic branch cuts due to on-shell exchange of spectral-flow images of
$p^+=0$ states.
When $\mu\to 0$ a smooth flat space limit is
obtained. The $\mu\to\infty$ limit, unlike the case of
RR-supported pp-waves, gives again a flat space theory.

}

\begin{document}

\maketitle 

\section{Introduction}

Plane gravitational wave backgrounds have been studied for a while
in the context of string theory. The original interest in these
backgrounds was motivated by the observation that, although the
underlying CFT is not known in general, the $\s$-model can be
shown to be conformally invariant to all orders in the
$\alpha'$-expansion \cite{orig}.

Further interest was sparked with the discovery \cite{nw} that
there are WZW models that describe such pp-wave backgrounds, where
the Killing symmetries give rise to current algebra on the
two-dimensional world-sheet.
This approach produced  a list of WZW and coset conformal theories with
pp-wave character \cite{kk,kkl,wzw,ors}.

The existence of the current algebra was used in \cite{kk} to organize
the operator algebra  of the Nappi-Witten (NW)  WZW model, compute the
exact spectrum, find a quasi-free-field resolution, and compute
the partition function (really the vacuum energy)  \cite{kkl}
which turns out to be equal to that of flat space\footnote{For recent work on this see \cite{sm}.}.
Moreover, pp-waves with a partially compact light-cone were analyzed and the
partition functions computed \cite{rt}.
A characteristic feature of the NW background was observed, namely
that there is an upper limit in $p^+$ due to stringy effects.
This was visible directly in the quasi-free-field resolution
\cite{kk} and emerged as a unitarity bound in the associated
string theory \cite{forgacs}.
The world-sheet and space-time supersymmetry of the NW background
was analyzed and shown to be half of the maximal supersymmetry
(compared to that of flat space) \cite{kkl} and it was also
pointed out that  such pp-wave CFTs are
T-dual (semiclassically) to flat space \cite{kk}.

Scattering amplitudes remained for a while out of reach.
An exception was  the $p^+=0$ sector of these theories that was shown to be
similar to that of flat space-time \cite{kk}.

Renewed interest in pp-wave backgrounds emerged after the
observation that, in the context of gravity/gauge-theory
correspondence, they are dual to alternative large-N limits of the
gauge theory.
In the prototype example of N=4 super Yang-Mills theory, this
corresponds to taking the limit $N\to \infty$ together with
$J\to
\infty$ (where $J$ is a U(1) charge of the SO(6) R-symmetry) keeping
the ratio $N/J^2$ fixed.
The associated limit of the dual gravitational background is the
Penrose limit of the $AdS_5\times S^5$ background studied recently
\cite{pen} where geodesics spinning  around $S^5$ were blown-up.

This observation provided with a new laboratory for the study of
AdS/CFT correspondence and holography in a context where the
dual string theory may be exactly solvable.
Indeed, the exact light-cone spectrum of the associated
Green-Schwarz $\s$-model could be computed \cite{metsaev}
and it was argued that it matches the one obtained from super-YM
theory\cite{bmn}.
This conjecture was further sharpened \cite{pp}  and today it stands
on a firm footing despite complications associated with higher order
mixing of the leading set of string-like states in the gauge
theory.

The gauge-theory/gravity correspondence involves backgrounds
containing non-trivial RR fields, and the only known string
description today is the light-cone Green Schwarz formalism.
Despite the simplicity of the light-cone sigma model for the
Penrose limit of $AdS_5\times S^5$, (a free theory of massive
two-dimensional fields), it turns out that the calculation
of non-trivial light-cone scattering amplitudes seems formidable,
and there is no consensus yet on  the details of the three-string vertex\cite{lc}.

There are other cases of similar holographic correspondences. The most
interesting involves the correspondence of  the Little String Theory (LST) to string
theory on the near horizon region of NS5-branes \cite{seiberg},
involving the $SU(2)_k$ CFT together with a Liouville direction as
well as six flat longitudinal directions.
The Penrose limit of this background, corresponding to blowing-up
a geodesic spinning around $S^3$ was shown to be the NW background
tensored with six flat non-compact coordinates \cite{gomis,kp}.
Such limits were considered earlier in the context of CFT in order
to produce pp-wave backgrounds \cite{ors,ks1,ks2}.
Other NS-supported pp-wave backgrounds corresponding to WZW models
are Penrose limits of several intersecting NS5 and F1
configurations \cite{kp,sj}.

The detailed formulation of holography in pp-wave backgrounds has
proven not to be straightforward.
There are several proposals in the literature  \cite{hol,kp}.
that share common  points but also have differences.
These reflect the position of the holographic screen (the analogue of the space-time
boundary in AdS/CFT) as well as the nature of the ``boundary"
observables.

In \cite{kp} it was argued that the S-matrix elements (defined in
a holographic way) are the observables of the bulk theory.
This is the only natural set of on-shell bulk observables in any
theory, and, as it was argued semiclassically in \cite{kp} and
will be shown rigorously in this paper, they exist and are calculable
on the string theory side.
S-matrix elements in a pp-wave metric can be defined in terms of
the evolution in light-cone time \cite{kp}.
There are several bases for writing the S-matrix elements.
In the oscillator basis, they are functions of $p^+$ and they are
infinite dimensional matrices indicating that the ``boundary
theory", is an infinite-N matrix model with $p^+$ as a parameter.
One could Fourier transform the $p^+$ dependence and introduce a
 null coordinate in the matrix theory.
 In a basis that diagonalizes the raising operators of the
 Heisenberg symmetry, they involve transverse continuous
 coordinates $x^i$, \cite{kp}.

An alternative organisation of the S-matrix amplitudes (that will
be used in this paper and that matches the one mostly used in \cite{kp})
is to introduce auxiliary charge variables.
These will transmute into coordinates of the holographic dual  as
in the case of the AdS$_3$/CFT correspondence \cite{ads3}.
Beyond the spectrum of string excitations in pp-wave backgrounds the
nature of ``boundary observables" is still open.

In this paper, we deal with the calculation of tree-level
scattering amplitudes in the pp-wave background obtained from the
Penrose limit of LST theory.
This involves the NW WZW model and six flat extra coordinates.
The are several reasons for considering this particular background:

{\bf (i)} It is supported by NS-flux and thus has a conventional CFT
description as a WZW model for a non-compact and non
semisimple group, the Heisenberg group $H_4$.
Consequently, we can quantize the system covariantly
and use the full machinery of CFT. A bonus is that
the $p^+=0$ sector, invisible in light-cone
quantization, will be present here.
There are also direct generalizations
of this model with several non-trivial transverse planes
and our techniques will be valid in the general case.

{\bf (ii)} It seems to contain all the important ingredients of similar
RR-supported backgrounds and more. In fact, in NS-supported
pp-wave backgrounds, we always have spectral flow and a related
extension of the spectrum of fundamental strings
that does not exist in the analogous
RR-supported backgrounds.

{\bf (iii)} The most important part of the parent theory (LST) surviving the Penrose limit,
namely the
$SU(2)_k$ CFT, is exactly solvable. The Penrose limit sends the level $k\to \infty$
with an appropriate  scaling of the spectrum.
The troublesome part of the parent theory, the linear dilaton,
disappears in the Penrose limit.
Thus, we are able to compute directly the Penrose limit of several LST
data and this will be a non-trivial independent check on our
calculations.
Another important factor here is that we can follow, if we wish, the
approach to the Penrose limit at large $k$, and establish a well
defined perturbation theory in $1/k$.

We should mention, that the two different classes of $SU(2)_k$ states
that will be relevant for our analysis,
have been noted before, \cite{bk} in a seemingly unrelated
construction of unitary representations of the $W_{\infty}$
algebra from large-k limits of the $SU(2)_k$ (really the parafermion) theory.
The two classes of states correspond
to states with $l\sim {\cal O}(k)$ (that here correspond to the
$V^{\pm}$ sector, i.e. states with non-zero $p^+$) and states with
$l\sim {\cal O}(\sqrt{k})$ (corresponding to the $V^0$ sector,
i.e. states with $p^+=0$).
It was also observed that states belonging to the second class have a
free-field operator algebra, a result that we will confirm here.

{\bf (iv)} As shown in \cite{kk}, the Heisenberg current algebra of the NW
pp-wave has a quasi-free-field realization. This provides with an
alternative method to compute the scattering amplitudes which
will be extremely useful. Moreover, according to the free-field realization,
primary fields
of the $H_4$ current algebra can be represented as orbifold twist fields
and therefore
the correlators we will compute in this paper can be considered as generating functions
for correlators of arbitrary-excited twist fields.

We should also  stress that the NW CFT is an excellent
laboratory to study the ramifications of non-compact curved CFTs.
Substantial progress has been made already in the SL(2,R) case
\cite{maloog1,maloog2,maloog3}, but several questions there, remain
unanswered. The NW CFT shares all nontrivial features of the
SL(2,R) CFT and as we show in this paper, is completely solvable, and all
S-matrix elements regular and computable. Moreover, it is one in a larger class of
non-compact CFTs that can be handled with the present techniques.

We will have three different techniques at our disposal in
order to calculate the NW CFT correlators and eventually the
string theory S-matrix elements:

{\bf (A)} Abstract current algebra techniques, namely solving
Knizhnik-Zamolodchikov (KZ)
equations and imposing monodromy and crossing symmetry on
correlators. This approach has the advantage, via the introduction
of the usual auxiliary charge coordinates, that it keeps track of the
structure of the whole Heisenberg algebra (infinite dimensional)
representations.
However, the equations to be solved for the four-point correlators
are partial differential equations in two variables, and additional
boundary conditions are needed.

{\bf (B)} The quasi-free-field realization. Here the computations
of the correlators are straightforward by a slight generalization
of the techniques of \cite{dfms}.
However, obtaining the full infinite-dimensional set of four-point correlators
is cumbersome in this approach.

It turns out that (A) and (B), used in conjunction, provides an
unambiguous and complete set of rules that will enable us to
calculate exactly all three-point and four-point correlators
of the NW WZW model. Actually we will solve the KZ equation
and present explicit expressions for the conformal blocks.
Then, we can use the third method as an independent check on
our results.

{\bf (C)} The direct Penrose (large-k) limit of the $SU(2)_k$ correlators,
first computed in \cite{zf}.
The advantage of this approach is obvious. There are two
difficulties however: the limiting amplitudes have divergent normalizations
(that reflect the presence of continuous momentum-conserving
$\delta$-functions) and the four-point correlators which in the limit have
an infinite number of blocks are very hard to deal with.

Our findings can be described as follows:

There are two basic sets of primary operators, corresponding to
the standard Heisenberg current algebra representations \cite{kk}:

$\bullet$ The
``discrete-like" $V^{\pm}$ sector involving operators with
$0<|p^+|<1$. The superscript $\pm$ indicates positive or negative $p^+$.
$V^+$ is conjugate to $V^-$.
$V^{\pm}$ are organized into semi-infinite lowest-weight
or highest-weight representations of the Heisenberg algebra.
The transverse plane spectrum of such representations is discrete
in agreement with the fact that in this sector there is an
effective harmonic oscillator potential confining them around the
center of the plane.

The cutoff $p^+=1$ is stringy in origin and descents from the
$l\leq {k\over 2}$ cutoff of the parent $SU(2)_k$ theory.
It is  also a unitarity cutoff: standard current algebra
representations with $p^+>1$ will contain physical negative-norm states in
the associated string theory \cite{forgacs}.

$\bullet$ The ``continuous-like" $V^0$ sector involves operators
with $p^+=0$. The transverse-plane spectrum of such operators is
continuous in agreement
with the fact that there is no potential for such operators.
The interactions among states in this sector are
similar to the flat space theory.

We will recover arbitrary values of $p^+$ using the spectral flow that shifts
$p^+$ by integers \cite{kk,kp}.
The spectral-flowed current algebra representations have
$L_0$ eigenvalues not bounded from below.
Such representations correspond to states deep-down the $SU(2)_k$
current algebra representations, related to the top states by the
action of the affine Weyl group of $\widehat{SU(2)}$.
They are crucial for the closure of the full
operator algebra and the consistency of interactions, as they
are in the parent $SU(2)_k$ theory.
Their semiclassical picture involves long-strings \cite{kp}, much like the
AdS$_3$ case \cite{maloog1}. At $p^+=1$, the harmonic oscillator
potential fails to squeeze the string since it is compensated by
the NS-antisymmetric tensor background. Such long strings generate
a sequence of new ground-states that are related to the current
algebra ground states by the spectral flow \cite{kp}.

There are  three types of three-point functions: $\langle
V^{+}V^{+}V^{-}\rangle$, $\langle
V^{+}V^{-}V^{0}\rangle$, $\langle
V^{0}V^{0}V^{0}\rangle$ and their conjugates.
The first two have quantum structure constants that are
non-trivial functions of the $p^+_i$ of $V^{\pm}$ and $\vec p$ of
$V^{0}$.

The third is the same as in flat space (with $p^+=0$) since the
$V^0$ operators are standard free vertex operators.

There are several types of four-point functions among the $V^{\pm,0}$
operators:

\bigskip
$\bullet$ $\langle
V^+(p^+_1)V^+(p^+_2)V^+(p^+_3)V^-(-p^+_1-p^+_2-p^+_3)\rangle$ and their conjugates.
Such correlators factorize on $V^{\pm}$ representations and
have a finite number of conformal blocks. They are given by powers of
$_2F_1$-hypergeometric functions.

\bigskip
$\bullet$ $\langle
V^+(p^+_1)V^+(p^+_2)V^-(-p^+_3)V^-(p^+_3-p^+_1-p^+_2)\rangle$.
These
correlators factorize on $V^{\pm}$ representations and their spectral-flowed images
for generic values of $p^+_i$.
They have an infinite number of conformal blocks and they are given by
exponentials of hypergeometric functions.
For special values of the momenta, for example $p^+_1=p^+_3$, they
factorize onto the $V^0$ representations and their spectral-flowed
images.
In this case, they develop logarithmic behavior in the cross-ratio which signals the presence
of the continuum of intermediate operators.
It should be noted that the logarithmic behavior here does not
indicate that we are dealing with a logarithmic CFT (argued to
describe cosets of pp-wave nature \cite{lcft}) but rather signals
the presence of a continuum of intermediate states.

\bigskip
$\bullet$ $\langle
V^+(p^+_1)V^+(p^+_2)V^-(-p^+_1-p^+_2)V^0(\vec p)\rangle$.
Such correlators have an infinite number of conformal blocks. They factorize
on $V^{\pm}$ representations, and they are given in terms of the
$_3F_2$-hypergeometric functions.

\bigskip
$\bullet$ $\langle
V^+(p^+)V^-(-p^+)V^0(\vec p_1)V^0(\vec p_2)\rangle$. These
correlators have an infinite number of conformal blocks.
They factorize on $V^{\pm}$ or $V^0$ representations according to the  channel.

\bigskip
$\bullet$ $\langle
V^0(\vec p_1)V^0(\vec p_2)V^0(\vec p_3)V^0(-\vec p_1-\vec p_2-\vec p_3)\rangle$.
These correlators have a single conformal block and are the same
as in flat space.

Starting from the solution of the NW CFT, we can calculate the
three-point and the four-point S-matrix elements of the associated
(super)string theory on this background.
They have the following salient features:

\bigskip
$\bullet$ The S-matrix elements exist, i.e. they are free of
spurious singularities.

\bigskip
$\bullet$ They are {\tt non-analytic} as functions of the external $p^+$
momenta.

\bigskip
$\bullet$ They are ``dual", i.e. crossing symmetric as in string
theory in flat space.

\bigskip
$\bullet$ The four-point S-matrix elements have two types of
singularities: Poles associated to the propagation of intermediate
physical states with $p^+\not =0$, as well as logarithmic branch
cuts signaling the propagation of a continuum of intermediate
physical states which are spectral flow images of  $p^+=0$ states.
Such branch cuts are very much
similar to those appearing in field theory S-matrix elements with
massless on-shell intermediate states.
However, because of momentum conservation such branch cuts can
appear only in loops in field theory, while here they already
appear at tree-level.

\bigskip
$\bullet$ The states corresponding to $p^+=0$ do not give rise to physical
states except when there is a tachyon in the spectrum.
This would be the case in a bosonic NW$\times R^{22}$ string
theory, but not in the superstring NW$\times R^{6}$ corresponding
to the Penrose limit of LST.
However, spectral-flowed
$V^0$ states with $p^+ \in \mathbb{Z}$ integer, can be physical both in the
massless (i.e. level-zero) sector as well as in massive sectors.

\bigskip
$\bullet$ The S-matrix elements we compute are naturally functions of the
auxiliary charges variables $x,\bar x$, used to keep track of the
Heisenberg symmetry of the pp-wave, as well as of $p^+$.
In analogy with $AdS_3$, we can identify these variables as
auxiliary coordinates of the holographic dual. Introducing
also $x^-$, the Fourier conjugate of $p^+$, we obtain ``boundary" coordinates
that match the holographic setup of \cite{kp}.

\bigskip
$\bullet$ The corresponding light-cone quantization \cite{forgacs} of this theory can
be compared with the covariant quantization employed in this
paper.
In the light-cone gauge one obtains the same spectrum for $p^+\not
=0$. The $p^+=0$ states are not accessible in the light-cone
gauge.
The presence and structure of the spectral flow is also visible
in the light-cone gauge, but again, {\tt not} the spectral flow
images of the $p^+=0$ sector.

The structure of this paper is as follows. In section 2 we discuss
the Penrose limit of NS5-branes both at the $\sigma$-model level
and at the CFT (current algebra) level. We also present
semiclassical expressions for the string vertex operators in this background
and discuss marginal
deformations of the theory generated by  current-current
perturbations, which result in backgrounds that may or may not be in
the pp-wave family.
In section 3 we describe the representation theory of the $H_4$
current algebra, as well as its free-field resolution.
In section 4 we present the three-point correlation functions and
the associated operator product expansions.
In section 5 we solve the $KZ$ equation for the various classes of correlators
and give explicit expressions for the conformal blocks and
the four-point correlators.
In section 6 we describe the transformation properties of the
four-point conformal blocks , relevant for the monodromy invariance
of the four-point amplitudes.
In section 7 we describe the non-trivial affine null vectors of the $H_4$ current algebra and
the associated null-vector equations
and we use them to study three-point couplings involving spectral flowed states.
In section 8 we describe the computation of correlators of
spectral-flowed operators of the theory.
In section 9 we analyze the structure of string S-matrix elements obtained upon
integration of the CFT correlators.
In section 10, we describe the flat space limit of the theory.
In section 11, the light-cone quantization is described and the two
approaches compared.
In section 12, we describe generalized backgrounds that can be
solved by our methods, with qualitatively similar physics.
In section 13 we comment on the relevance of our results for the
putative holographic description of these backgrounds.
Finally section 14 contains our conclusions and an outlook.

In appendix \ref{b} we present in detail the Penrose limit of $SU(2)_k$
correlators and other data, and confirm where such a limit is
possible, the results presented in the main body of the paper,
obtained by other methods.
In appendix \ref{c} we review  the calculation of arbitrary
four-point correlators of twist fields.
In appendix \ref{d} we present the details of the calculation of
CFT four-point amplitudes.
In appendix \ref{e} we derive in more details the marginal
deformations of the NW pp-wave.

\section{The Penrose Limit}
\setcounter{equation}{0}

The Nappi-Witten model is a WZW model based on a non-semisimple group, namely the
Heisenberg group $H_4$ defined by the following commutation relations
\be
[P^+,P^-] = - 2i K \ , \hspace{1cm}
[J,P^{\pm}] = \mp i P^{\pm} \ .
\label{lab01}
\ee
The background metric and antisymmetric tensor field
entering the world-sheet $\s$-model
\be
ds^2 = -2 du dv - \frac{\m^2 \r^2}{4} du^2 + d\r^2 + \r^2 d\f^2 \ , \hspace{1cm}
B_{\f u} = \frac{\m \r^2}{2} \ ,
\label{lab1}
\ee
describe a gravitational wave. In this section
we study this model from two different points of view.
From the $\s$-model perspective, the gravitational wave in $(\ref{lab1})$
can be considered as the Penrose limit of some curved space-time. There are various
possible choices but we will concentrate on a very simple one, a WZW model
of the form $U(1)_{{\rm time}} \times SU(2)_k$.
Similarly, from an algebraic point of view, the affine algebra of the $H_4$ WZW model
can be considered as a contraction of the  $U(1)_{{\rm time}} \times SU(2)_k$
WZW model. In the contraction, the level $k$ is sent to infinity and
the quantum number of the states scaled in such a way that
the representations of the original algebra organize themselves in representations
of the contracted one.
The two points of view are closely related.

We will first consider the Penrose limit of the near horizon geometry
of $k=N-2$ NS5-branes, blowing-up a null geodesic spinning along the sphere $S^3$.
The metric, antisymmetric tensor and dilaton profiles are
\cite{ns5}
\bea
ds^2 &=& -dt^2 + dr^2+ k(d\psi^2\cos^2\theta+d\theta^2+\sin^2\theta
d\phi^2)+ (dx^i)^2\\
H_{\psi\theta\phi}&=&\cos\theta\sin \theta\\
g_s^2&=&e^{-2 r/\sqrt{k}}
\label{la1}
\eea
and are described by the exact CFT $R^6\times SU(2)_k\times R_Q$ where $R_Q$ represents the
Liouville 1-d CFT corresponding to the radial coordinate.

In order to perform the Penrose limit we define
\be
\frac{t}{\sqrt{k}}+\psi = u \ , \hspace{1cm}
\frac{t}{\sqrt{k}}-\psi = 2 \l^2 v \ ,  \hspace{1cm}
\theta=\l \rho\ .
\label{la2}
\ee
In the limit $\l\to 0$ keeping $\l^2 k = 1$ we obtain
\be\hspace*{-1cm}
ds^2
= - 2du dv - \frac{\r^2}{4}du^2 +
d\r^2+\r^2d\phi^2 + (dy^i)^2 \ .
\label{la3}
\ee
The dilaton gradient disappears in this limit. We have obtained the Nappi-Witten geometry times a
six-dimensional  flat Euclidean space \cite{gomis,kp}.

\subsection{Current algebra}
\setcounter{equation}{0}

Here, we discuss the Penrose limit from the point of view of the world-sheet CFT.
As shown in \cite{ors}, WZW models based on non-compact and non-semi-simple groups can be
obtained by the contraction of a current algebra of the form $G \times H$ where $G$ is a simple
group and $H$ is a compact subgroup of $G$. The contraction can be formulated as a large-level
limit, whose semiclassical nature reflects itself in the integral value of the central
charge which coincides with the number of space-time dimensions.
As remarked in \cite{lcft}, if one performs a similar contraction starting with
a current algebra given by the product of a coset model and a free boson, the final result
is a logarithmic CFT.

We now discuss in some detail how the Nappi-Witten wave arises as a limit of an
$U(1)_{\rm time}\times SU(2)_k$ WZW model, paying particular attention to
the relation between the quantum numbers labeling the representations of the two
CFTs, since we shall use them in the study of the limit of the three-point couplings
of $SU(2)_k$, presented in detail in Appendix A.
The original current algebra is given by
\be
J^a(z)J^b(0)={k\over 2}{\delta^{ab}\over
z^2}+i\epsilon^{abc}{J^c(0)\over z}+{\rm RT} \ ,
\label{la4}
\ee
\be
J^{0}(z)J^{0}(0)=-{k\over 2}{1\over z^2}+{\rm RT} \ ,
\label{la5}
\ee
where here and in the following ${\rm RT}$ stands for additional terms that are regular in the limit
$z \rightarrow 0$.
If we define the raising and lowering operators as $ J^{\pm}=(J^1\pm i J^2)$ then
\be
J^+(z)J^-(0)={k\over z^2}+{2J^3\over z}+{\rm RT} \ , \hspace{0.4cm}
J^3(z)J^{\pm}(0)= \pm {J^{\pm}\over z}+{\rm RT} \ .
\label{la6}
\ee
On the other hand,  the $H_4$ current algebra of the Nappi-Witten model reads
\ba
P^+(z)P^-(w) &\sim& \frac{2}{(z-w)^2} - \frac{2iK(w)}{z-w}+{\rm RT} \ , \nb \\
J(z)P^{\pm}(w) &\sim& \mp i\frac{P^\pm(w)}{z-w}+{\rm RT} \ , \nb \\
J(z)K(w) &\sim& \frac{1}{(z-w)^2}+{\rm RT} \ .
\label{c-ope}
\ea
The most general contraction of  $U(1)\times SU(2)_k$ $\to H_4$ performed by taking $k\to \infty$
involves, up to changing the signs of the charges,
a real parameter $b$. The relations between the currents of the two models are
\be
K(z)={i\over k}\left[(2-b)J^0(z)+bJ^3(z)\right] \ , \hspace{0.4cm} J(z)=i(J^0(z)-J^3(z))
\ , \hspace{0.4cm} P^{\pm}=\sqrt{2\over k}J^{\pm} \ .
\label{cur-con}
\ee
The limit $b\to \infty$ is singular as $K$ and $J$ coincide.
Another special choice for the parameter $b$ is $b=0$ where the structure of the representations
of $SU(2)$ remains intact.

Before investigating how the limit connects the highest-weight
representations (hwr)
of the two CFTs, we recall the structure of the $H_4$ affine algebra.
The commutation relations, following from the OPE in (\ref{c-ope}) are
\be
[P^+_n,P^-_m] = 2n~
\d_{n+m,0} - 2i K_{n+m}
\ , \hspace{0.6cm}
[J_n, P^{\pm}_m] = \mp i P^{\pm}_{n+m}
\ , \hspace{0.6cm}
[J_n,K_m] = n~ \d_{n+m,0} \ .
\label{la7}
\ee
The horizontal subalgebra generated by the zero modes is
\be
[P^+,P^-] = - 2i K \ , \hspace{1cm}
[J,P^{\pm}] = \mp i P^{\pm} \ .
\label{la8}
\ee
There are two independent Casimir operators. One is simply
the central element of the algebra, $K$, while
the other is given by the combination
\be
C= \frac{1}{2} ( P^+P^- + P^-P^+ ) + 2JK \ .
\label{la9}
\ee
Eigenstates of $J$ and $K$ are defined according to
\be
K |p,j> = i p |p,j> \ , \hspace{1cm} J |p,j> = i j |p,j> \ ,
\label{la10}
\ee
(the $J$ and $K$ generators are anti-hermitian, while $(P^+)^{\dagger} = P^-$).
Since
\be
JP^{\pm}|p,j> = i (j \mp 1) P^{\pm}|p,j> \ ,
\label{la11}
\ee
$P^+$ lowers and $P^-$ raises the $J$ eigenvalue of a state.

The $H_4$ group has three types of unitary representations:
\begin{itemize}

\item $V^+_{p,\jh}$ representations
\footnote{In reference \cite{kk} $V^0$,
$V^+$, $V^-$ representations were called type I, II and III respectively}.
These are
lowest-weight representations, generated by acting with $P^-$
on a state $|p,\jh>_+$  satisfying $P^+ |p,\jh>_+ = 0$.
For these representations, unitarity implies $p>0$,
the spectrum of $J$ is given by $\{\jh + n \}_{n \in
\mathbb{N}}$ and $C= -2p ( \jh - \frac{1}{2})$.
Moreover, we have
\be
P^+ |p,\jh+n> = \sqrt{2p n} |p,\jh+n-1 > \sp
P^- |p,\jh+n> = \sqrt{2p (n+1)} |p,\jh+n+1 >
\label{la12}
\ee

\item  $V^-_{|p|,\jh}$ representations. These are
highest-weight representations, generated acting with $P^+$
on a state $|p,\jh>_-$  which satisfies $P^- |p,\jh>_- = 0$.
For such representations $p<0$, the spectrum of $J$ is given by $\{\jh - n \}_{n \in
\mathbb{N}}$ and $C= -2p ( \jh + \frac{1}{2})$.
As before, we have
\be
P^+ |p,\jh-n> = \sqrt{-2p (n+1)} |p,\jh-n-1 > \sp
P^- |p,\jh-n> = \sqrt{-2p n} |p,\jh-n+1> \ .
\label{la13}
\ee
Note that the representation
$V^-_{p,-\jh}$ is the representation conjugate to $V^+_{p,\jh}$.

\item $V^0_{s,\jh}$ representations. Such
representations have $p=0$ and do not contain neither a lowest nor a highest-weight state.
The spectrum of $J$ is given by $\{\jh + n \}_{n \in
\mathbb{Z}}$ with $\jh \in [-1/2,1/2)$ and $C= s^2$, $s \in \mathbb{R}^+$ .
Finally,
\be
P^{\pm}|s,\jh+n> = s |s,\jh+n \mp 1 > \ .
\label{la14}
\ee
\end{itemize}

The irreducible representations of the current algebra are, as usual,
built on the infinite dimensional unitary representations of the zero-mode algebra.
The associated ``level-zero" states satisfy
\be
J^a_{n>0}|R_i\rangle=0 \ .
\label{hwr}
\ee
They are generated from the vacuum by the affine-primary fields
\be
J^a(z)R_{i}(w)=T^a_{ij}{R_j(w)\over z-w} +{\rm RT}
\label{hw} \ .
\ee
Since the group is not semi-simple, the standard quadratic form is degenerate
and we can not use it to express the stress energy tensor in the usual Sugawara form.
It is however still possible to introduce a non-degenerate metric \cite{nw}
and to represent the stress energy tensor as a bilinear combination of the currents
\be
T={1\over2}\left[{1\over 2}\left(P^+P^-+P^-P^+\right)+2JK+K^2\right] \ .
\label{th4}
\ee
As is common when dealing with non-compact groups, an indefinite metric
appears in the operator algebra and the Hilbert space created by acting with the
modes of the currents contains negative-norm states.
If we want that the $H_4$ WZW model, considered as part of a string theory background, leads
to a spectrum of physical string states free of ghosts, we have to consider
representations of the $H_4$ current algebra whose
base is a unitary representations of the global $H_4$ algebra, described
before. We start then with three sets of affine representations: those
associated with $V^{\pm}_{p,\jh}$ representations, whose conformal
dimension is
\be
h_{p,\jh} = \frac{|p|}{2}(1-|p|) - p\jh \ , \hspace{2cm} p \ne 0 \ ,
\label{la15}
\ee
and those associated with $V^0_{p,\jh}$ representations, whose conformal dimension is
\be
h_{s,\jh} = \frac{s^2}{2} \ .
\label{la16}
\ee

The absence of negative-norm states imposes a further restriction on  $V^{\pm}_{p,\jh}$
representations, namely $p<1$ \cite{forgacs}. As we will discuss in more detail later,
states with $p=1$ are null states. Arbitrary real values for the momentum $p$
are recovered including other representations of the current algebra, that are not
highest-weight, as spectral-flowed images of the usual
representations. We will discuss them in more detail in the next section.

We can now follow how the  $H_4$ affine representations arise as limits
of the $U(1)_{\rm time}\times SU(2)_k$ representations.
First, the limit of the Sugawara stress tensor of the
original CFT gives:
\be
T=-{1\over k}:J^0J^0:+{:J^aJ^a:\over k+2}\to {1\over
2}\left[{1\over 2}\left(P^+P^-+P^-P^+\right)+2JK+K^2\right] \ ,
\label{la17}
\ee
which coincides with (\ref{th4}).
It is interesting to note that the last term comes from the one-loop correction
in the $SU(2)_k$ $\sigma$-model.
This is an explanation of an earlier observation \cite{nw}
that the last term indeed seemed  as a one-loop term.

We introduce in the standard way the $SU(2)\times U(1)$ quantum numbers
\be
J^0_0|q,l,m\rangle =q |q,l,m\rangle \ , \hspace{1cm}
J^3_0|q,l,m\rangle =m|q,l,m\rangle \ ,
\label{la18}
\ee
and use the relations between the currents displayed in (\ref{cur-con})
to derive the following relations between the charges
(in the limit $k\to \infty$, $p$ and $j$, as well as the corresponding
charges for the anti-holomorphic sector, are kept fixed)
\be
q={1\over 2}(kp+b j)\ , \hspace{1cm}
l={1\over
2}\left ( k|p|+ \frac{p}{|p|}(b-2) \jh \right ) \ , \hspace{1cm}
m={1\over
2}(kp+(b-2) j)\label{mscale} \ .
\ee

For generic $b$ the eigenvalues of $L_0$ on the primary states are
\be
h = -{q^2\over k}+{l(l+1)\over k+2}\to -{b\over
2}p (j-\jh) - |p|\jh+{1\over 2}|p|(1-|p|) \ ,
\label{la21}
\ee
and it is evident that the $H_4$  and  the SU(2) representations do not coincide.
Only the $b=0$ contraction preserves the structure of SU(2) representations and from
now on we will focus on this case.

In the limit $k \rightarrow \infty$, the states that are sitting at the top of
an $SU(2)$ representation give rise to a lowest-weight representation of
$H_4$,
while the states  sitting at the bottom of
an $SU(2)$ representation give rise to a highest-weight representation of $H_4$.
More precisely, $V^+_{p,\jh}$ representations result from states characterized by
\be
l = \frac{1}{2}(kp-2\jh) \ , \hspace{1cm}
m = \frac{1}{2}(kp-2(\jh+n)) \ , \hspace{1cm} p > 0  \ ,
\label{la24}
\ee
while $V^-_{|p|,\jh}$ representations result from states characterized by
\be
l = \frac{1}{2}(k|p| + 2\jh) \ , \hspace{1cm}
m = \frac{1}{2}(kp -2(\jh-n)) \ , \hspace{1cm} p < 0  \ .
\label{la26}
\ee
Note that each $SU(2)$ representation, being self-conjugate, splits in
two conjugate $H_4$ representations.
Finally, the states in a $V_{s,\jh}^0$ representation correspond to states in
the middle of an SU(2) representation with $q,m,\bar q, \bar m\sim O(1)$ as
$l=\sqrt{\frac{k}{2}}s+O(1)\to \infty$.

\subsection{The $\sigma$-model view point}
\setcounter{equation}{0}

It is instructive to perform the same limit not at the algebraic level (as we
did in the previous subsection)  but on the fields that appear in the Lagrangian of the
world-sheet $\s$-model.
Here, we will ignore the radial as well as the other five flat directions
which are present in the $NS5$ background since they do not play an important
role in the Penrose limit,
and we will keep only the time direction as well as the $SU(2)_k$ WZW model.
Let us parameterize the $SU(2)$ group manifold using the Euler angles
\be
g=\exp\left[i\chi{\sigma^3\over 2}\right]
\exp\left[i\theta{\sigma^1\over
2}\right]\exp\left[i\phi{\sigma^3\over 2}\right] \ ,
\label{la30}
\ee
with $0 \le \chi \le 4 \pi$,  $0 \le \phi \le 2 \pi$
and $0 \le \theta  \le  \pi$. From the action
\be
S= {k\over 8 \pi}\int d^2z\left[\p\theta\pb\theta+\p\chi\pb\chi+\p\phi\pb\phi
+2\cos \theta~\p \chi\pb \phi-4\p t\pb t\right] \ ,
\label{la29}
\ee
where we added the time direction, we read the background metric and the antisymmetric
tensor field
\be
ds^2=\frac{k \a^{'}}{4}\left[d\chi^2+d\phi^2+d\theta^2+2\cos\theta d\chi
d\phi-4 dt^2\right]\sp B_{\chi\phi}= \frac{k \a^{'}}{4}\cos{\theta} \ ,
\label{la31}
\ee
where $k  \a^{'} =R^2$ and $R$ is the radius of $S^3$.
The holomorphic and anti-holomorphic currents are
\be
J^{\pm}=\frac{k}{2}e^{\mp i\chi}\left[\partial\theta\pm
i\sin\theta\partial\phi\right]\sp J^3=\frac{k}{2}[\partial\chi+\cos\theta
\partial\phi]\sp J^0=-k\p t \ ,
\label{la32}
\ee
\be
\bar J^{\pm}= \frac{k}{2}e^{\pm i\phi}\left[\pb\theta\mp
i\sin\theta\pb\chi\right]\sp \bar J^3=\frac{k}{2}[\pb\phi+\cos\theta
\pb\chi]\sp \bar J^0=-k\pb t \ .
\label{la33}
\ee
The limit we are interested in is better described by first changing variables to
\be
\chi=\a+\f\sp \phi=\a-\f\sp \theta =2r \ ,
\label{la34}
\ee
where $0 \le \a \ , \f \ , r \le 2 \pi$.
The metric becomes
\be
ds^2={k \a^{'} }[dr^2+\sin^2 rd\f^2+\cos^2
rd\a^2-dt^2] \ .
\label{la35}
\ee
We then set
\be
\a= \frac{\l^2 v}{\m} - \frac{\m u}{2}   \sp
t= \frac{\l^2 v}{\m} + \frac{\m u}{2}   \sp
r = \l \r \ ,
\label{la36}
\ee
and take the limit $ \l \rightarrow 0$, $k \rightarrow \infty$ keeping
$k \l^2 = 1 $.
The resulting background is the Nappi-Witten gravitational wave
\footnote{From now on we will set $\alpha'=1$.}
\be
ds^2 = -2 du dv - \frac{\m^2 \r^2}{4} du^2 + d\r^2 + \r^2 d\f^2 \ , \hspace{1cm}
B_{\f u} = \frac{\m \r^2}{2} \ .
\label{la37}
\ee
The same form of the metric results from the following parameterization of the
$H_4$ group manifold
\be
g =e^{\frac{u}{2}J} e^{ i \frac{\bar{\z}}{\sqrt{2}}P^- + i \frac{\z}{\sqrt{2}}P^+}
e^{\frac{u}{2}J- 2v K} \ ,
\label{la39}
\ee
where  $\z = \r e^{i \f}$ and $\tilde{\z} = \r e^{-i \f}$. The currents
\ba
K &=&  -i \p u \ , \hspace{0.2cm}
J = -2 i \p v -i \m^2 \r^2 \p u -2 i \m \r^2 \p \f  \ , \hspace{0.2cm}
P^{\pm} =
\sqrt{2} e^{\pm i\m u} \p \left [ \r  e^{\mp i(\f+\frac{\m}{2}u)} \right ] \ , \nb \\
&&\label{la40}\\
\bar{K} &=&  -i \bar{\p} u \ , \hspace{0.2cm}
\bar{J} = - 2 i \bar{\p} v -i \m^2 \r^2 \bar{\p} u +2 i \m \r^2 \bar{\p} \f  \ , \hspace{0.2cm}
\bar{P}^{\pm} = \sqrt{2} e^{\mp i\m u} \bar{\p} \left [ \r  e^{\mp i(\f-\frac{\m}{2}u)} \right ] \
,\nb
\ea
precisely match the Penrose limit of the $SU(2)_k \times U(1)$ currents
in (\ref{la32}-\ref{la33}).
The dependence of the currents on the $\s$-model fields suggests the possibility of realizing
the $H_4$ algebra in terms of free fields. Indeed if we change variables setting $\f =
\phi - \frac{\m}{2}u$ for the left currents and
$\f = \phi + \frac{\m}{2}u$ for the right currents and we define $y = \r e^{i \phi}$ we can write
\ba
K &=&  - i \p u \ ,  \hspace{0.4cm}
J = - 2 i \p v  - i \m \r^2 \p \phi \ , \hspace{0.4cm}
P^{+} = \sqrt{2} e^{ i\m u} \p y \ , \hspace{0.4cm}
P^{-} = \sqrt{2} e^{- i\m u} \p \tilde{y} \ ,
\nb \\
&&\label{la401}\\
\bar{K} &=&  - i \bar{\p} u \ , \hspace{0.4cm}
\bar{J} = - 2 i \bar{\p} v  + i \m \r^2 \bar{\p} \phi \ , \hspace{0.4cm}
\bar{P}^{+} = \sqrt{2} e^{- i\m u} \bar{\p} y  \ , \hspace{0.4cm}
\bar{P}^{-} = \sqrt{2} e^{i\m u} \bar{\p} \tilde{y}   \ .\nb
\ea
In the next section, the free-field realization of the $H_4$ algebra will be explained in detail.

We conclude this section discussing the semiclassical form of the vertex
operators corresponding to the states in the various $H_4$ representations.
Vertex operators for the primary fields can be represented
in terms of the $\s$-model fields as local expressions not containing derivatives.
A natural choice \cite{dvv} is to take as
a complete basis for the space of functions on the group manifold
the matrix elements of group operators between states in irreducible representations.
For $V^{\pm}_{p,\jh}$ representations the generating function for such matrix
elements is
\be
\F^{\pm}_{p,\jh} = e^{ \mp ipv + i\jh u - \frac{\m p}{2} \z \tilde{\z}
+ i \m p \z x e^{\pm\frac{i \m u}{2}}+ i \m p \tilde{\z}\bar{x} e^{\pm \frac{i \m
u}{2}}+\m p x \bar{x}e^{ \pm i \m u}} \ ,
\label{la41}
\ee
where we have introduced the formal charge variables $x,\bar x$ to
keep track of different states inside the representation. They
will be described in more detail in the next section where we will use
them to write in a compact way the operator algebra of the $H_4$ model.

Expanding the previous expression in powers of $x$ and $\bar{x}$, we can see that
the semiclassical vertex operators
for the states that form a $V^{\pm}_{p,\jh}$ representation are given by the product
of a gaussian and a generalized
Laguerre polynomial, and thus coincide with the wave-functions for a two-dimensional
harmonic oscillator in radial
coordinates, which describes the semiclassical
periodic motion in the transverse plane.
The expansion can be performed using the following generating function
\be
e^{xy + r(x-y)} = \sum_{n,m=0}^\infty \frac{r^{m-n}}{m!} L_n^{m-n}(r^2)x^m y^n \ ,
\label{laa1}
\ee
and the result is
\be
\F^{+}_{p,\jh} =  e^{ - ipv + i\jh u - \frac{\m p}{2} \z \tilde{\z}}
\sum_{n,\bar{n}=0}^\infty ~(i x)^n (-i\bar{x})^{\bar{n}}~
e^{\frac{i \m u}{2}(n+\bar{n})} ~ \frac{(\m p)^n}{n!}
~\z^{n-\bar{n}}~
L_{\bar{n}}^{n-\bar{n}}(\m p ~\z ~\tilde{\z}) \ .
\label{laa2}
\ee
Comparing this expression with the expansion in $(\ref{lb9})$, we
can read the semiclassical form of the primary fields in an $H_4$
representation. Moreover, if we take into account the relation between
$H_4$ primary fields and orbifold twist fields implied by the free-field
realization of the $H_4$ algebra \cite{kk}, we obtain a nice
representation of the latter in terms of a gaussian (for the unexcited
twist field) and Laguerre polynomials (for the excited twist fields).
One can reconstruct the expansion in $(\ref{laa2})$ acting with the
generators of the Killing symmetries
on the wave-function of the ground state of a $V^{\pm}_{p\jh}$ representation.
The generators are $K = \bar{K} = - \frac{\p}{\p v}$ and
\ba
P^+ &=& -\sqrt{2}e^{-i \frac{\m u}{2}}\left
[ i \frac{\p}{\p \z} - \frac{\m \tilde{\z}}{2}\frac{\p}{\p v} \right] \ ,
\hspace{.8cm}
P^- = -\sqrt{2}e^{i \frac{\m u}{2}}\left [ i \frac{\p}{\p \tilde{\z}}
+ \frac{\m \z}{2} \frac{\p}{\p v} \right] \ , \nb \\
\bar{P}^+ &=& -\sqrt{2}e^{-i \frac{\m u}{2}}\left [ i \frac{\p}{\p \tilde{\z}}
- \frac{\m \z}{2} \frac{\p}{\p v}
 \right] \ ,
\hspace{.8cm}
\bar{P}^- = -\sqrt{2}e^{i \frac{\m u}{2}}\left [ i \frac{\p}{\p \z}
+ \frac{\m \tilde{\z}}{2}\frac{\p}{\p_v} \right] \ , \\
J &=& \frac{\p}{\p u} + i \frac{\m}{2} \left [ \z  \frac{\p}{\p \z}
- \tilde{\z} \frac{\p}{\p \tilde{\z}} \right ] \ , \hspace{1.5cm}
\bar{J} = \frac{\p}{\p u} - i \frac{\m}{2} \left [ \z  \frac{\p}{\p \z}
- \tilde{\z} \frac{\p}{\p \tilde{\z}} \right ] \ . \nb
\label{laa3}
\ea

The semiclassical expression for the $V^0_{s,\jh}$ vertex operators is
\be
\F^{0}_{s,\jh} = e^{ iju + \frac{i s}{\sqrt{2}}
\left [ \z \sqrt{\frac{x}{\bar{x}}} + \tilde{\z} \sqrt{\frac{\bar{x}}{x}} \right ]}
\sum_{n \in \mathbb{Z}}~ (x \bar{x})^n ~e^{-i n \m u}   \ ,
\label{la42}
\ee
where $x = e^{i \a}$ and the sum over $n$ imposes the constraint $x \bar{x} e^{i \m u} = 1$.
The coefficients of the expansion of this vertex operators in $x$ and $\bar{x}$ are Bessel functions
describing a free motion in the transverse plane, in accord with the fact that the $p=0$ sector does
not feel the quadratic potential. Comparing $(\ref{la42})$
with the usual expression for a tachyonic vertex operators
we see that we can identify $s$ with the modulus of the momentum in the two-plane and
$\frac{1}{2}(\bar{\a}-\a)$
with its phase.

Vertex operators for the graviton, for the dilaton and for the
antisymmetric tensor are given by descendant fields and their
correlation functions can be derived from those of the
primary fields using the standard Ward identities of the current algebra.

The addition of four free fermions is all we need to make
the model superconformal. Actually \cite{kkl}, the resulting
$\s$-model realizes an $N=4$ superconformal algebra.

\subsection{(Current)$^2$ deformations of NW pp-waves.}
\setcounter{equation}{0}

Here we describe two families of CFTs that are related  to the NW model
by  marginal deformations generated by its currents.
Since the Cartan of $H_4$ is two-dimensional the moduli space of deformations is
four-dimensional.
There are, however, two inequivalent choices for the commuting currents, not
related by the action of the group.
The first choice consists of the generators $K,J$ and the other of
the generators $K, P_1$.

The former deformation is most easily described in
coordinates exhibiting the polar angle in the transverse plane:
\be
S_{NW}={1\over 2\pi}\int d^2z\left[2\p v\pb u+2\p u\pb
v-\r^2\p u\pb u+\p \r\pb \r+\right.
\label{la412}
\ee
$$\left.+
\r^2\p \f\pb \f+\r^2(\p u\pb
\f-\p\f\pb u)\right] \ .
$$
The non-trivial deformation is associated with the current $J$.

The deformed geometry is described by the metric
\be
ds^2={4R^2(dv)^2+4dvdu+\r^2[-(du)^2+(d\f)^2]\over R^2
\r^2+1}+d\r^2
\label{origi} \ , \ee
the antisymmetric tensor
\be
B_{v\f}={2R^2\r^2\over R^2\r^2+1}\sp B_{u\f}={\r^2\over R^2\r^2+1} \ ,
\label{la424}
\ee
and the dilaton
\be
\Phi=-{1\over 2}\log[R^2 \r^2+1] \ .
\label{la43}
\ee

The undeformed model corresponds to $R=0$.
The background is smooth for $R^2$ positive and is not of the pp-wave type.
For $R^2<0$ it has a naked singularity.
Upon T-duality, it is mapped to flat Minkowski space or other dual
versions of it. This is a generalization of the fact that the NW model is mapped by T-duality
to flat Minkowski space \cite{kk}.

The other inequivalent set of deformations are generated by the
$K$
and $P_1$ currents. It is convenient here to change coordinates
and write the NW $\sigma$-model as \cite{kk}:

\be
S={1\over 2\pi}\int d^2z\left[\p y^i\bp y^i+2\cos u\pb y^1\p
y^2+\p u\pb v+\p v\bp u\right] \ ,
\label{la44}
\ee
which bears a strong similarity to the $SU(2)$ WZW model.
The line of marginal deformations, associated to the current
$P_1$,
parameterized by a positive real variable R,
is given
by \cite{gk}

\be
S(R)={1\over 2\pi}\int d^2z\left[2\p v\pb u+2\p u\pb v +
\p x^1\pb x^1+\p x^2\pb x^2+\right.
\label{la45}
\ee
$$
+{[1-2R^2+(1-R^2)\cos(2u)](x^1)^2+R^2[R^2-2+(1-R^2)\cos(2u)](x^2)^2\over
(\cos^2(u)+R^2\sin^2(u))^2}\p u\pb u-
$$
$$\left.
-2\arctan[R\tan(u)](\p x^1\pb x^2-\pb x^1\p x^2)\right] \ ,
$$
while  the dilaton is
\be
\Phi=-{1\over 2}\log\left[{1\over R}\cos^2(u)+R~\sin^2(u)\right]+{\rm constant} \ ,
\label{la46}
\ee
after changing to Brinkman coordinates.
The undeformed model corresponds in this case to $R=1$.
Note that there is an $R\to 1/R$ duality of the background fields
accompanied by the reparametrization $x^1\leftrightarrow x^2$ and
$u\to u+\pi/2$.
At the ends of the line, the theory becomes a direct product of a
real line CFT and the coset of the NW theory obtained by gauging $P_1$
\cite{kk}.
This class of space-times are of the pp-wave type.
More details, and fully deformed $\s$ models can be found in
appendix \ref{e}.

\section{Current algebra representation theory}
\setcounter{equation}{0}

Before moving to the actual computation of the three and four-point correlators,
we present in this section a more detailed discussion of some important features
that the $H_4$ WZW model shares with other non-compact WZW model, in particular
with $AdS_3$.
We will first introduce an auxiliary coordinate $x$ in order to
organize the infinite number of fields that form an $H_4$ representation in a
single field. We will then discuss the spectral-flowed representations of
the affine algebra, representations that must be included
in the spectrum of the model in order to obtain a closed operator algebra
even though they are not highest-weight representations.
Finally we will describe a free-field realization of the $H_4$
algebra and show that in this setting, its primary fields can be expressed
in terms of orbifold twist-fields and the usual flat-space
tachyonic vertex operators.

Since we are  dealing with infinite dimensional representations,
it is useful to realize the global
$H_4$ algebra as an algebra of operators acting on a suitable space of functions.
For $V^{\pm}$ representations we consider
power series of the form
\be
\f_{\pm}(x) = \sum_{n=0}^{\infty} \f_n \frac{(x\sqrt{|p|})^n}{\sqrt{n!}} \ ,
\label{lb1}
\ee
where $x$ is a formal complex variable.
When $p>0$, the operators realizing the $H_4$ algebra are
\footnote{We rescaled $x$ to $\sqrt{p}~x$ with respect to the more
standard harmonic oscillator choice $\sqrt{2p} ~x$ and $\sqrt{2p}~ \p_x$
in order to avoid factor of $\sqrt{p}$ in the following expressions.}
\ba
P_0^+ &=&  \sqrt{2} p ~x \ , \hspace{1cm}
P_0^- = \sqrt{2} ~\p_x \ , \nb \\
J_0 &=& i(\jh + x \p_x ) \ , \hspace{1cm}
K_0 = i p \ ,
\label{h1}
\ea
while for $p < 0$ they are
\ba
P_0^+ &=&  \sqrt{2} ~\p_x  \ , \hspace{1cm}
P_0^- = \sqrt{2}|p| ~x \ , \nb \\
J_0 &=& i(\jh - x \p_x ) \ , \hspace{1cm}
K_0 = i p \ .
\label{h2}
\ea
In this way, all the states in a given representations are collected
in a single function of $x$.

The monomials $b_n = \frac{(x\sqrt{p})^n}{\sqrt{n!}}$ form an
orthonormal basis if we define the scalar product using the measure
\be
\int d \n \equiv \frac{p}{\pi} \int d^2 x \ e^{-p x^*x} \ ,
\label{lb2}
\ee
where $x^*$ is the complex conjugate of $x$.
The right-moving algebra is similarly realized as an algebra of operators acting on
an independent variable $\bar{x}$. The operators are defined exactly as before and
the measure is
\be
\int d \bar{\n} \equiv \frac{p}{\pi} \int d^2 \bar{x}
\ e^{-p \bar{x}^*\bar{x}} \ .
\label{lb3}
\ee

For $V^0_{s,\jh}$ representations we consider power series of the form
\be
\f_0(x) = \sum_{n = - \infty }^{\infty} \f_n x^n \ ,
\label{lb4}
\ee
where $x = e^{i\a}$ is a phase and the zero-mode operators are
given by
\ba
P_0^+ &=& s x  \ , \hspace{1cm}
P_0^- = \frac{s}{x} \ , \nb \\
J_0 &=& i(\jh + x \p_x ) \ , \hspace{1cm}
K_0 = 0 \ .
\label{h3}
\ea
In this case the measure is simply
\be
\frac{1}{2 \pi} \int_0^{2\pi} d \a \ .
\label{lb5}
\ee
The relation between $V^{\pm}_{p,\jh}$ and $V^0_{s,\jh}$ representations can be considered as a further
contraction of the $H_4$ group to the isometry group of the two-dimensional plane.

We proceed in precisely the same way for the representations of the current algebra:
we introduce a complex variable $x$ and
regroup together the infinite number of fields that appear in a given representation defining
\ba
\F^+_{p,\jh}(z,x) &=& \sum_{n=0}^{\infty} R^+_{p,\jh;n}(z) \frac{(x\sqrt{p})^{n}}{\sqrt{n!}} \ ,
\ p>0 \ , \nb \\
\F^-_{p,\jh}(z,x) &=& \sum_{n=0}^{\infty} R^-_{p,\jh;n}(z) \frac{(x\sqrt{|p|})^{n}}{\sqrt{n!}} \ ,
\ p<0 \ , \nb \\
\F^0_{s,\jh}(z,x) &=& \sum_{n= - \infty }^{\infty}
R^0_{s,\jh;n}(z) x^{n} \ ,
\ s \ge 0 \ ,
\label{lb9}
\ea
which are respectively the primary fields for $V^+$, $V^-$ and $V^0$ representations.
The OPEs in (\ref{hw}) translate to
\ba
P^+(z)\F^+_{p,\jh}(w,x) &=& \sqrt{2p}~x~~ {  \F^+_{p,\jh}(w,x)\over z-w}+{\rm RT} \ , \nb \\
P^-(z)\F^+_{p,\jh}(w,x) &=& \sqrt{2}~ \p_x ~ ~ {  \F^+_{p,\jh}(w,x)\over z-w}+ {\rm RT}\ , \nb \\
J(z) \F^+_{p,\jh}(w,x) &=& i(\jh + x \p_x ) {\F^+_{p,\jh}(w,x)\over z-w}+{\rm RT} \ , \nb \\
K(z) \F^+_{p,\jh}(w,x) &=& i p  {\F^+_{p,\jh}(w,x)\over z-w}+ {\rm RT}\ ,
\label{v-ope}
\ea
and similar expressions for the other vertex operators.
We will refer to the variables $x$, $\bar{x}$ as to the {\it dual} variables, in analogy with
the $AdS_3$ case \cite{ads3}.
The OPEs in (\ref{v-ope}) are the central elements for deriving Ward
identities and the Knizhnik-Zamolodchikov equations.
We can introduce an $x$-dependence also in the currents, using the
generator of translations in $x$. For $p>0$ we have
\be
J^a(z,x) = e^{\frac{x}{\sqrt{2}}P_0^-}J^a(z,0)e^{-\frac{x}{\sqrt{2}}P_0^-} \ ,
\label{lb10}
\ee
and for $p<0$
\be
J^a(z,x) = e^{\frac{x}{\sqrt{2}}P_0^+}J^a(z,0)e^{-\frac{x}{\sqrt{2}}P_0^+} \ ,
\label{lb11}
\ee
where the index $a$ labels the four $H_4$ currents.
Therefore the $x$ dependent currents are
\ba
P^-(z,x) &=& P^-(z) \ , \hspace{0.8cm}
P^+(z,x) = P^+(z) + i \sqrt{2} x K(z)  \ , \nb \\
J(z,x) &=& J(z) - \frac{ix}{\sqrt{2}}P^-(z) \ , \hspace{0.8cm}
K(z,x) = K(z) \ ,
\label{xcp}
\ea
when $p>0$ and
\ba
P^+(z,x) &=& P^+(z) \ , \hspace{0.8cm}
P^-(z,x) = P^-(z) - i \sqrt{2} x K(z)  \ , \nb \\
J(z,x) &=& J(z) + \frac{ix}{\sqrt{2}}P^+(z) \ , \hspace{0.8cm}
K(z,x) = K(z) \ ,
\label{xcm}
\ea
when $p<0$.

We also need to know how the tensor product of $H_4$ irreducible representations
decomposes in a direct sum  of irreducible representations,
since such a decomposition corresponds to the (semi-classical) fusion rules between
highest-weight representations of the current algebra.
The tensor products between  $V^{\pm}$ representations decompose according to
(a detailed analysis can be found in \cite{cg})
\ba
V^+_{p_1,\jh_1} \otimes V^+_{p_2,\jh_2}
&=& \sum_{n=0}^{\infty} V^+_{p_1+p_2,\jh_1+\jh_2+n} \ , \nb \\
V^+_{p_1,\jh_1} \otimes V^-_{p_2,\jh_2}
&=& \sum_{n=0}^{\infty} V^+_{p_1-p_2,\jh_1+\jh_2-n} \ , \hspace{1.2cm}
\ p_1 > p_2 \ , \nb \\
V^+_{p_1,\jh_1} \otimes V^-_{p_2,\jh_2} &=& \sum_{n=0}^{\infty}
V^-_{p_2-p_1,\jh_1+\jh_2+n} \ , \hspace{1cm}
 \ p_1 < p_2 \ , \nb \\
V^-_{p_1,\jh_1} \otimes V^-_{p_2,\jh_2} &=& \sum_{n=0}^{\infty} V^-_{p_1+p_2,\jh_1+\jh_2-n} \ .
\label{tensor}
\ea
Moreover, we have
\be
V^+_{p,\jh_1} \otimes V^-_{p,\jh_2} = \int_{0}^\infty s~ds~ V^0_{s,\jh_1+\jh_2} \ ,
\label{lb7}
\ee
and
\be
V^+_{p, \jh_1} \otimes V^0_{s,\jh_2} =  \sum_{n=- \infty}^{\infty} V^+_{p,\jh_1+\jh_2+n} \ .
\label{lb8}
\ee

The action of a generic $H_4$ group element (modulo the trivial
action of $K$) on functions of $x$ is as follows
\be
f_{\jh}(x)\to b^{\jh}~e^{ax}~f(bx+c) \ ,
\ee
where $a,b,c$ arbitrary.
This should be compared with
\be
f_l(x)\to (cx+d)^{2l}~f\left({ax+b\over cx+d}\right)\sp ad-bc=1
\ee
in the SU(2)/SL(2) case.

\subsection{Spectral flow}
\setcounter{equation}{0}

The operator content of the $H_4$ WZW model is not exhausted
by the highest-weight representations of the affine algebra.
These representations are generated acting with the negative modes of the currents
$J^a_{-n}$ on an highest-weight state $|\psi>$ which is annihilated by all positive modes
$J^a_{n}|\psi> = 0$, $n>0$. However, as it was first observed for $AdS_3$
\cite{maloog1,maloog3} and for $SU(2)_k$ models at fractional level \cite{gaberdiel},
they do not form a closed operator algebra since, in the fusion of two
of them, new representations can appear that are not highest-weight.
These new representations are called spectral-flowed representations, since they
can be defined as highest-weight representations of an isomorphic algebra related
to the original one by the operation of spectral flow.
Spectral flow acts as an integer shift in the mode numbers of the lowering and rising
generators as well as an integer shift in the value of the Cartan generators of the
finite Lie algebra.
In our case the action of the spectral flow is defined as follows
\cite{kk}
\be
\tilde{P}^{\pm}_n = P^{\pm}_{n\mp w} \ , \hspace{0.8cm}
\tilde{K}_n = K_n -i w \d_{n,0} \ ,  \hspace{0.8cm}
\tilde{J}_n = J_n  \ ,  \hspace{0.8cm} \tilde{L}_n = L_n -i w J_n \ ,
\label{sfmodes}
\ee
where $w \in \mathbb{Z}$.
In appendix \ref{b6} we show that there is a two parameter spectral flow
of the $H_4$ algebra. In appendix \ref{b7} by studying the Penrose
contraction of characters we show that the spectral flow relevant
for the NW theory is the one discussed here.

The new modes in (\ref{sfmodes}) generate
an algebra $\tilde{H}_{4,w}$ isomorphic to the original one. Let us
denote a highest-weight representation of this algebra as $\Omega_w(\F^{\pm}_{p,\jh})$.
The eigenvalue of $K_0$ on the states in this representation is $p+w$.
However, a highest-weight representation of $\tilde{H}_{4,w}$
is not a highest-weight representation of the original algebra.
This is evident because the conditions obeyed by a highest-weight state of
the $\tilde{H}_{4,w}$ become, in terms of the original modes
\be
P^+_{n}|\psi> = 0 \ , \hspace{0.4cm} n > -w \hspace{1cm}
P^-_{n}|\psi> = 0 \ , \hspace{0.4cm} n > w \ ,
\label{lb12}
\ee
and therefore, the state is annihilated only by the modes with $n$ bigger
than a fixed positive integer. Moreover, the spectrum of $L_0$ is generically unbounded from
below.
In only two cases does the action of the spectral flow give
back a highest-weight representation. These two cases are
\be
\w_{-1}(\F^+_{p,\jh}) = \F^-_{1-p,\jh} \ , \hspace{1cm}
\w_{1}(\F^-_{p,\jh}) = \F^+_{1-p,\jh} \ .
\label{sf1}
\ee

It is important to notice that when $w > 0$, all the states in an affine
representation $\w_w(\F^+)$ or $\w_w(\F^0)$
belong to some lowest-weight representations of the horizontal algebra
and that when $w < 0$, all the states in an affine
representation $\w_w(\F^-)$ or $\w_w(\F^0)$
belong to some highest-weight representations of the horizontal algebra.
Therefore representations that are neither highest nor lowest-weight
representations of the horizontal algebra appear only in $\F^0$.

Fusion rules between spectral-flowed representations can be derived using
\cite{gaberdiel}
\be
\w_{w_1}(\F_1) \otimes \w_{w_2}(\F_2) = \w_{w_1+w_2}(\F_1 \otimes \F_2) \ .
\label{sf2}
\ee

As we will show explicitly when studying the factorization properties of the four-point correlators,
these representations appear in the fusion of the highest-weight representations and
therefore we have to include them in the spectrum of the CFT.
With the addition of the spectral-flowed representations we can cover
all possible values of $p$. One can understand the necessity of including them also
with a semiclassical reasoning, following \cite{maloog1}.
From the semiclassical point of view, spectral flow generates new solutions of the
equations of motion by conjugating a given one with an element of the loop group not
continuously connected to the identity.
A similar analysis
was performed in \cite{kp} where it was shown that, at the semiclassical level, highest-weight
representations correspond to geodesics on $H_4$. When $p \ne 0$ the particle moves linearly
in $u$ and performs a periodic motion in the plane while for $p=0$ the particle has $u=const$
and moves along a straight line in the plane. The spectral flow images of these trajectories
also have a $\s$ dependence: the particle becomes a string with a winding in the plane.
Such a winding number is related to the integer index of the spectral flow and  is not
conserved. In the case we consider its non-conservation is evident since only the total $p$
is conserved.
Here, unlike the SL(2,R) case, the necessity of spectral-flowed representations is also
visible in the parent $U(1)\times SU(2)_k$ theory.

We will eventually  impose the mass-shell condition $(L_0 - 1)|\psi> = 0$  on a state in a
spectral-flowed representation $|\psi> = |\tilde{p},j,\tilde{N},h>$, where
$h$ is the conformal dimension of the state in the internal CFT, which is needed to construct
a critical string background,
$\tilde{p}$ and $j$ are the eigenvalues of
$\tilde{K}_0$ and $J_0$ respectively and $\tilde{N}$ the
level with respect to the $\tilde{H}_{4,w}$ algebra.
Moreover we can write $j = \jh + n$ with $n \in \mathbb{Z}$, $n \ge - \tilde{N}$.
The result is
\be
j = \frac{1}{\tilde{p}+w} \left [ \frac{\tilde{p}(1-\tilde{p})}{2}+h+\tilde{N}+n\tilde{p}-1 \right ] \ ,
\label{lb13}
\ee
when $\tilde{p}>0$. Similarly for $|\psi> = |s,j,\tilde{N},h>$,
\be
j = \frac{1}{w} \left [ \frac{s^2}{2}+h+\tilde{N}-1 \right ] \ .
\label{lb14}
\ee
The restriction $0<\tilde{p}<1$ for $\F^\pm$ representations leads to
\ba
&wj& <  h+\tilde{N}-1  < wj + \jh  \ , \hspace{0.4cm} {\rm when} \ \  \jh \ge 0 \ , \nb \\
&wj& + \jh <  h+\tilde{N}-1  < wj  \ , \hspace{0.4cm} {\rm when} \ \ \jh \le 0 \ ,
\label{lb15}
\ea
and it is clear from the previous expressions that whenever $h$ saturates
the upper or the lower bound, there is a
continuous representation with the same light-cone energy.
Each time $p \in \mathbb{Z}$, there is a point of contact between discrete
and continuous representations. This is similar to the $AdS_3$ WZW model where,
when $j=1/2$, there is a point of contact between the discrete and the continuous
representations.
The structure of the spectrum is indeed very similar to the one of $AdS_3$,
the only difference being the absence of the mass gap $j > 1/2$.
The presence of a continuum of states for $p=1$, signal the fact that starting
from $p \ge 1$ we have to construct the Hilbert space of string states from a different
vacuum state.

The spectrum of our model is then given by $\F^+_{p,\jh}$ representations, with $p<1$ and
$\jh \in \mathbb{R}$ together with their
spectral-flowed images  $\w_w(\F^+_{p,\jh})$ with $w \in \mathbb{N}$,
by $\F^-_{p,\jh}$ representations, with $p<1$ and
$\jh \in \mathbb{R}$ together with their
spectral-flowed images  $\w_{-w}(\F^+_{p,\jh})$ with $w \in \mathbb{N}$, and by
$\F^0_{s,\jh}$ representations, with $\jh \in [-1/2,1/2)$ together with their
spectral-flowed images  $\w_{w}(\F^0_{S,\jh})$ with $w \in \mathbb{Z}$.
Since the $u$ and $v$ coordinates are non-compact, the spectrum of $\jh$ is continuous and
we consider left-right symmetric combination of the representations. This also implies
that the amounts of spectral flow in the left and right sector have to coincide.

In Appendix A we show how the
spectral-flowed representations arise in the contraction of the
$SU(2) \times U(1)$ CFT.

\subsection{The quasi-free-field resolution}
\setcounter{equation}{0}

Before we start discussing the correlation functions between primary fields
we recall the  quasi-free-field resolution of the $H_4$ algebra \cite{kk}, since
it
provides
an interesting relation between $H_4$ correlators and correlators between
twist fields in orbifold models.
Moreover, in these variables the action of the spectral flow is extremely simple.
We introduce a pair of free bosons  $u,v$ with $<v(z)u(w)> = \log{(z-w)}$ and
a complex boson $y=y_1+i y_2$, $\tilde{y}=y_1-i y_2$
with $<y(z)\tilde{y}(w)> = -2 \log{(z-w)}$. One can then verify
that the following currents
\ba
J &=& \p v \ , \hspace{2cm} K =  \p u \ , \nb \\
P^+ &=& i e^{-iu} \p y \ , \hspace{1cm} P^- = i e^{iu} \p \tilde{y} \ ,
\label{lb16}
\ea
satisfy the $H_4$ operator algebra. The description of the
primary fields for the $V^{\pm}$ representations requires, in this quasi-free-field formalism,
the introduction of twist fields $ H^{\mp}_{p}(z)$ characterized by the
following OPEs
\ba
\p y(z) H^-_{p}(w) &\sim& (z-w)^{-p}T_p^-(w) \ , \hspace{1cm}
\p \tilde{y}(z) H^-_{p}(w) \sim (z-w)^{-1+p}H^{-(1)}_p(w) \ , \nb \\
\p y(z) H^+_{p}(w) &\sim& (z-w)^{-1+p} H^{+(1)}_p(w)\ , \hspace{.3cm}
\p \tilde{y}(z) H^+_{p}(w) \sim (z-w)^{-p} T_p^+(w)\ .
\label{lb17}
\ea
where $T_p^{\pm}(w)$,   $H^{\pm(1)}_p(w)$ are excited twist
fields.

The ground state of a $V^{\pm}$ representation is then given by
\be
R^{\pm}_{p,\jh;0}(z) = e^{i (\jh u(z) \pm p v(z))} H^{\mp}_{p}(z) \ ,
\label{t1}
\ee
and the other states are obtained through the action of $P_0^{\mp}$.
We recall that in the presence of a twist field $H_p^-$, the mode expansion
of a complex free boson reads
\be
\p y = \sum_{n \in Z} \a_{n+p} z^{-n-1-p} \ , \hspace{1cm}
\p \tilde{y} = \sum_{n \in Z} \tilde{\a}_{n-p} z^{-n-1+p} \ , \hspace{1cm}
\label{lb18}
\ee
where the oscillators obey the following commutation relations
\be
[\tilde{\a}_{n-p},\a_{m+p}] = (n-p) \d_{n+m,0}  \ ,
\label{lb19}
\ee
and act on the state created by $H_p^-$ according to
\be
\a_{n+p} |H_p^-> = 0 \ , \hspace{0.4cm} n \ge 0 \hspace{1cm}
\tilde{\a}_{n-p} |H_p^-> = 0 \ , \hspace{0.4cm} n \ge 1 \ .
\label{lb20}
\ee
In the presence of a twist field $H_p^+$, one has exactly the same expressions
up to exchanging $\a_n$ with $\tilde{\a}_n$.
Consider now a representation with $p>0$; starting from the vertex
operator in (\ref{t1}) we can generate all the other ground-states
acting with $P^-$ and the result is
\be
(P_0^-)^n R^+_{p,\jh;0}(z) = e^{i((\jh+n)u+pv)} H^{-(n)}_p(z) \ ,
\label{t2}
\ee
where we introduced the excited twist field $ H^{-(n)}_p$ defined as
\be
| H^{-(n)}_p> \ = \ \tilde{\a}^n_{-p} | H^{-}_p> \ .
\label{lb21}
\ee
The conformal dimension of the vertex operator in
(\ref{t2}) is
\be
\frac{p}{2}(1-p) + np -p(\jh+n) = \frac{p}{2}(1-p) -p\jh \ ,
\label{lb22}
\ee
that is, the additional contribution to the conformal dimension due to the
excited twist field is exactly compensated by the shift in $\jh$.

Proceeding in the same way for $V^-$ representations
one can write the other states in the multiplet as
\be
(P_0^+)^n R^-_{p,\jh;0}(z) = e^{i((\jh-n)u-pv)} H^{+(n)}_{p}(z) \ ,
\label{t3}
\ee
where the excited twist field $ H^{+(n)}_{p}$
is defined as
\be
| H^{+(n)}_{p}> \ = \ \a^n_{-p} | H^{+}_{p}> \ .
\label{lb23}
\ee

The vertex operators for $V^0$ representations are
\be
\F^0_{s,\jh}(z,x) = e^{i \jh u + \frac{i s }{2} \left ( y  \sqrt{\frac{x}{\bar{x}}}
+ \tilde{y} \sqrt{\frac{\bar{x}}{x}}  \right )} \sum_{n \in \mathbb{Z}} (x \bar{x})^n e^{i n \m u}  \ .
\label{lb24}
\ee
If we set $ \frac{\bar{x}}{x} = e^{2 i\f}$,
we see that we are dealing with a collection of standard tachyonic vertex
operators carrying a momentum $p = p_1 + i p_2 = s e^{i\f}$.

Spectral flow by $w$ units corresponds in this setting to the multiplication by
the operator $e^{iwv}$. Indeed defining
\be
\w_{w}(R^+_{p,\jh;0}(z)) =  e^{i(\jh~ u+(p+w)v)} H^{-}_p(z) \ , \hspace{1cm}
\w_{-w}(R^-_{p,\jh;0}(z)) =  e^{i(\jh~u-(p+w)v)} H^{+}_p(z) \ ,
\label{lb26}
\ee
it is easy to verify from the OPE that the corresponding states satisfy
\ba
P^+_n|p+w> &=& 0 \ , \hspace{0.4cm} n \ge -w \hspace{1cm}
P^-_n|p+w> = 0 \ , \hspace{0.4cm} n \ge w+1  \ , \nb \\
P^+_n|p+w> &=& 0 \ , \hspace{0.4cm} n \ge w+1 \hspace{1cm}
P^-_n|p+w> = 0 \ , \hspace{0.4cm} n \ge -w  \ .
\label{lb27}
\ea
We will make use of this quasi-free-field representation to compute
correlators between spectral-flowed states in section \ref{sfll}.

As we have seen, the primary vertex operators
are actually unchanged with respect to the corresponding ones in flat-space when
$p \in \mathbb{Z}$, while in the other cases
the operator $e^{i(p_1y_1+p_2y_2)}$, which describes free propagation in the
transverse plane is replaced by a twist field $H^{\pm}_{p}$, which describes
the bounded motion.

\section{Three-point couplings and the operator product expansion}
\setcounter{equation}{0}

In this section, we discuss the two- and three-point amplitudes between primary
vertex operators of the $H_4$ CFT. Following \cite{zf}, we will present
the OPE in a way that is well suited for the study of the factorization
of the amplitudes. As we shall see, the structure of the three-point couplings is completely fixed
by conformal invariance and invariance under global $H_4$ transformations, up to constants
which are the quantum structure constants of the operator algebra. We present the structure constants
in this section even though we
will derive them  from the study of the factorization of the four-point
correlators in a later section.

As we already explained,
the relevant local conformal primary fields depend, apart from the
two-dimensional coordinates
$z,\bar z$, on two further variables $x$ and $\bar{x}$, that encode the states of the
left and right infinite-dimensional representations of the left and right $H_4$ current algebras of the
conformal field theory.
Putting left and right together we consider the local fields
\be
\F^{a}_{q}(z,\bar z;x,\bar{x}) \ ,
\label{lc1}
\ee
where $a$ labels the different representations $(a \in \{+,-,0 \} )$ and
$q$ stands for the set of charges needed to completely specify the given
representation, that is $q = (p,\jh)$ when $a = \pm$ and $q= (s,\jh)$
when $a=0$.
Typical correlators are of the form
\be
A_n = \left \langle \prod_{i=1}^n~\F^{a_i}_{q_i}(z_i,\bar z_i;x_i,\bar x_i) \right \rangle \ .
\label{lc2}
\ee

We start with the two-point functions, fixing in this way the normalization of our operators.
The two-point function between a $\F^{+}$ operator and its conjugate $\F^{-}$ is
\be
 <\F^+_{p_1,\jh_1}(z_1,\bar{z}_1,x_1,\bar{x}_1)\F^-_{p_2,\jh_2}(z_2,\bar{z}_2,x_2,\bar{x}_2)>
= \d (p_1-p_2) \d (\jh_1+\jh_2) \frac{e^{-p_1(x_1x_2+\bar{x}_1\bar{x}_2)}}{|z_{12}|^{4h}} \ ,
\label{lc3}
\ee
where $h$ is the scaling dimension given in (\ref{la15}).
The measure on the $p$ and $\jh$ quantum numbers
is
\be
\int d \s_{\pm} \equiv \int_0^1 dp  \int_{-\infty}^{\infty} d \jh .
\label{lc4}
\ee
In terms of the component fields,
\be
<R^+_{p_1,\jh_1;n,\bar{n}}(z_1,\bar{z}_1)R^-_{p_2,\jh_2;m,\bar{m}}(z_2,\bar{z}_2)>
= \d (p_1-p_2) \d (\jh_1+\jh_2) \d_{n,m}  \d_{\bar{n},\bar{m}}
\frac{(-1)^{n+\bar{n}}}{|z_{12}|^{4h}} \ .
\label{lc5}
\ee
For $\F^0$ operators, the two-point function is
\ba
& &
<\F^0_{s_1,\jh_1}(z_1,\bar{z}_1,\a_1,\bar{\a}_1)\F^0_{s_2,\jh_2}(z_2,\bar{z}_2,\a_2,\bar{\a}_2)>=
\nb \\
&=&
\frac{(2 \pi)^2}{|z_{12}|^{4h}}
\frac{\d (s_1-s_2)}{s_1}~ \d(\jh_1+\jh_2) ~\d(\a_1-\a_2)~\d(\bar{\a}_1-\bar{\a}_2)
 \ ,
\label{lc6}
\ea
where we used the conventional normalization for the two-point
function of tachyon vertex operators in flat space, namely
$\d^{(2)}(\vec{p}_1+\vec{p}_2)$.
The measure over the Casimir $s$ and the $\jh$ quantum number is
\be
\int d \s_0 \equiv \int_0^\infty ds \ s  \int_{-1/2}^{1/2} d \jh .
\label{lc7}
\ee
From $(\ref{lc6})$ it follows for the component fields
\ba
& & <R^0_{s_1,\jh_1;n,\bar{n}}(z_1,\bar{z}_1)R^0_{s_2,\jh_2;m,\bar{m}}(z_2,\bar{z}_2)>=  \nb \\
&=&
\frac{\d (s_1-s_2)}{s_1}
\d(\jh_1+\jh_2)
\d_{n+m,0}~\d_{\bar{n}+\bar{m},0} ~\frac{(-1)^{n+\bar{n}}}{|z_{12}|^{4h}}  \ .
\label{lc8}
\ea

We now turn to the three-point couplings. Their general form is
\ba
&&
<\F^a_{q_1}(z_1,\bar{z}_1,x_1,\bar{x}_1)\F^b_{q_2}(z_2,\bar{z}_2,x_2,\bar{x}_2)
\F^c_{q_3}(z_3,\bar{z}_3,x_3,\bar{x}_3)>
\nb \\
&=& \frac{{\cal C}_{abc}(q_1,q_2,q_3)
D_{abc}(x_1,x_2,x_3;\bar{x}_1,\bar{x}_2,\bar{x}_3)}
{|z_{12}|^{2(h_1+h_2-h_3)}|z_{13}|^{2(h_1+h_3-h_2)}|z_{23}|^{2(h_2+h_3-h_1)}} \ .
\label{lc9}
\ea
In the previous expression, the ${\cal C}_{abc}(q_1,q_2,q_3)$ are the quantum structure constants of the
CFT and, as before, the labels $a$, $b$ and $c$ distinguish between the different types of representations
($a=+,-,0$) while $q_1$, $q_2$ and $q_3$ stand for the set of charges describing the
corresponding state, $(p,\jh)$ or $(s,\jh)$.
As usual, the $z$ dependence is completely fixed by conformal invariance
and the functions $D$, which encode the $x$ dependence of the three-point couplings,
are completely fixed as well by the Ward identities of the $H_4$ algebra.
They are the classical Clebsch-Gordan coefficients of the left and
right-moving $H_4$ algebras.

Using these quantities, the OPE can be written as
\be
\F^a_{q_1}(z_1,\bar z_1,x_1,\bar x_1) \F^b_{q_2}(z_2,\bar z_2,x_2,\bar x_2) =
\label{ope-gen}\ee
$$
=\int d \s_c
\int d \n_3 \int d \bar{\n}_3 \
D_{ab}{}{}^c(x_1,x_2,x_3^*;\bar{x}_1,\bar{x}_2,\bar{x}_3^* )~
{\cal C}_{ab}{}{}^c~~
\frac{[ \F^c_3 (z_2,\bar z_2,x_3,\bar x_3) ]}{ |z_{12}|^{2(h_1+h_2-h_3)}} \ ,
$$
where $[ \F^c_3 ]$ denotes the affine family comprising the primary $\F^c_3$ and all of
its descendants
and the measures $d \n$, $d \bar{\n}$ are as defined in (\ref{lb2}), (\ref{lb3})
and (\ref{lb5}).
Indexes are lowered and raised using the two-point functions (\ref{lc3}) and
(\ref{lc6}).
When we raise or lower an index in $D_{abc}$ or in  $C_{abc}$,
we have to replace a $V^+$ with a $V^-$
representation and to change the sign of $x$ and $\jh$.

The OPE between  the components of the various fields $\F^a$ can be obtained
by expanding both sides of (\ref{ope-gen}) in powers of $x$, $\bar{x}$
and performing the integral in $x_3$, $\bar{x}_3$.

As an explicit example consider the
fusion between two $\F^+$ representations
\be
[\F^+_{p_1,\jh_1}] \otimes [\F^+_{p_2,\jh_2}] = \sum_{n=0}^\infty [\F^+_{p_1+p_2, \jh_1+\jh_2+n}] \ .
\ee
The function $D_{++-}$ which appears in the three-point coupling is
\be
D_{++-}(x_1,x_2,x_3,\bar{x}_1 ,\bar{x}_2 ,\bar{x}_3)
=  \left | e^{-x_3(p_1x_1+p_2x_2)}(x_2-x_1)^{-L} \right |^2
\d(p_3-p_1-p_2) \d_{\mathbb{N}}(-L) \ ,
\label{lc10}
\ee
where  $L = \jh_1+\jh_2+\jh_3$ and $\d_{\mathbb{N}}(a) \equiv \sum_{n=0}^\infty \d(a-n)$.
Thus,
the coupling is non-vanishing only when $L$ is a non-positive integer.
The two $\d$-functions keep track of the structure of the tensor products
displayed in (\ref{tensor}).
In this and in the following equations we use the shorthand $|f(x,z)|^2$ for $f(x,z) f(\bar{x},\bar{z})$.
Similarly, the D-function appearing in the OPE is
\be
D_{++}{}{}^+(x_1,x_2,x^*_3,\bar{x}_1 ,\bar{x}_2 ,\bar{x}^*_3)
=  \left | e^{x^*_3(p_1x_1+p_2x_2)}(x_2-x_1)^{-L} \right |^2
\d(p_3-p_1-p_2) \d_{\mathbb{N}}(-L) \ ,
\label{lc11}
\ee
where now  $L = \jh_1+\jh_2-\jh_3$.
Finally, the OPE between the component fields reads
\ba
& & R^+_{p_1,\jh_1;n_1,\bar{n}_1}(z_1,\bar{z}_1)R^+_{p_2,\jh_2;n_2,\bar{n}_2}(z_2,\bar{z}_2)
= \sum_{n=0}^{\infty}  \frac{
{\cal C}_{++}{}{}^+ (p_1,\jh_1;p_2,\jh_2;p_1+p_2,\jh_1+\jh_2+n)}{|z_{12}|^{2(h_1+h_2-h_n)}}  \nb \\
& &\sum_{m,\bar{m}} D_{++}{}^+[n_1,n_2;n,m] D_{++}{}^+[\bar{n}_1,\bar{n}_2;n,\bar{m}]
[R^+_{p_1+p_2,\jh_1+\jh_2+n;m,\bar{m}}(z_2,\bar{z}_2)] \ ,
\label{lc12}
\ea
with $n+m=n_1+n_2$ and  $\bar{n}+\bar{m}=\bar{n}_1+\bar{n}_2$.
The coefficients
\ba
D_{++}{}^+[n_1,n_2;n,m] &=& p_1^{-\frac{n_1}{2}}p_2^{\frac{m+n_1-n}{2}}p_3^{-\frac{m}{2}} n!
\sqrt{(n_1)!(n_2)!m!} \nb \\
& & \sum_{k} \frac{ (-1)^{n_1-k}}{k!(m-k)!(n_1-k)!(n_2-k)!} \left ( \frac{p_1}{p_2} \right )^k \ ,
\label{lc13}
\ea
coincide, up to a factor we absorbed in the structure constant  ${\cal C}_{abc}$,
with the Clebsch-Gordan coefficients for the $H_4$ group \cite{cg}.
In the previous expression, the index $k$ takes all values
such that the summand is well defined and non-zero.

Similar expressions can be derived in the other cases and we
describe them in turn.
The OPE between $\F^+$ and $\F^-$ vertex operators produces
$\F^+$ vertex operators when $p_1>p_2$ and $\F^0$ vertex operators
when $p_1=p_2$.
Indeed the fusion rules are in the first case
\be
[\F^+_{p_1,\jh_1}] \otimes [\F^-_{p_2,\jh_2}] = \sum_{n=0}^\infty [\F^+_{p_1-p_2,\jh_1+\jh_2-n}] \ .
\label{ope-pmp}
\ee
The relevant $D$ function for this OPE is
\be
D_{+--}(x_1,x_2,x_3,\bar{x}_1,\bar{x}_2,\bar{x}_3)
=  \left | e^{-x_1(p_2x_2+p_3x_3)}
\left ( x_2 - x_3 \right )^{L} \right |^2 \d(p_3-p_1+p_2) \d_{\mathbb{N}}(L) \ ,
\label{lc14}
\ee
with $L = \jh_1+\jh_2+\jh_3$
and the coefficients appearing in the component  expansion are
\ba
D_{+-}{}^+[n_1,n_2;n,m] = &=& p_1^{-\frac{n_1}{2}}p_2^{\frac{n_2}{2}-n}p_3^{\frac{m}{2}} n!
\sqrt{(n_1)!(n_2)!m!} \nb \\
& & \sum_{k} \frac{ (-1)^{n_1+m-k}}{k!(m-k)!(n-k)!(n_1-m+k)!} \left ( \frac{p_2}{p_3} \right )^k \ ,
\label{lc15}
\ea
non vanishing only when $m-n=n_1-n_2$.
On the other hand, when $p_1=p_2$ the fusion rules are
\be
[\F^+_{p,\jh_1}]  \otimes [\F^-_{p,\jh_2}] = \int_0^\infty ds \ s  \ [\F^0_{s,\jh_3}] \ ,
\label{ope-pm0}
\ee
where $\jh_3 \in [-1/2,1/2)$ is given by  $ \jh_3 + a = \jh_1+\jh_2 $, $a \in \mathbb{Z}$.
Moreover
\be
D_{+-0}(x_1,x_2,x_3,\bar{x}_1,\bar{x}_2,\bar{x}_3) = \left |
e^{- p_1x_1x_2-\frac{s}{\sqrt{2}}\left (x_2 x_3 +
\frac{x_1}{x_3}\right)} x_3^{-L} \right |^2
\d(p_1-p_2) \ ,
\label{lc16}
\ee
where $L = \jh_1+\jh_2+\jh_3 = -a$ and $x_3 = e^{i \a_3}$.
The  coefficients appearing in the component  expansion are
\be
D_{+-}{}^0[n_1,n_2;a,m] =  \sqrt{n_1!n_2!} \left ( \frac{s}{\sqrt{2p}} \right )^{n_2-n_1} \sum_l
\frac{(-1)^{n_2+l}}{l!(n_1-l)!(a-m+l)!} \left (\frac{s^2}{2p} \right )^l \ ,
\label{lc17}
\ee
non zero only for  $n_1-n_2 = m - a$.
The fusion product between $\F^+$ and $\F^0$ operators is
\be
[\F^+_{p,\jh_1}]  \otimes [\F^0_{s,\jh_2}] = \sum_{n=-\infty}^\infty [\F^+_{p,\jh_1+\jh_2+n}] \ .
\label{ope-pm01}
\ee
The relevant coefficients for the OPE between components are
\be
D_{+0}{}^+[n_1,n_2;n,m] =  \sqrt{n_1!m!} \left ( \frac{s}{\sqrt{2p}} \right )^{n_1-m} \sum_l
\frac{(-1)^{n_1-m+l}}{l!(m-l)!(n_1-m+l)!} \left (\frac{s^2}{2p} \right )^l \ ,
\label{lc18}
\ee
non-vanishing only when $n+m=n_1+n_2$.
Note that
\be
D_{+-}{}^0[n_1,n_2;a,m] = (-1)^{n_1}D_{+0}{}^+[n_2,m;a,n_1] \ .
\label{lc19}
\ee
Finally we consider the three-point couplings between $\F^0$ representations.
The $P^+$ and $P^-$ constraints amounts to momentum conservation in
polar coordinates
\be
s_3^2 = s_1^2+s_2^2+2s_1s_2 \cos{\g} \ , \hspace{1cm}
s_3e^{i\h} = -s_1-s_2e^{i \g} \ ,
\label{conserv}
\ee
where $\g=\a_2-\a_1$ and $\h=\a_3-\a_1$. Similar equations can be written for $\bar{P}^{\pm}$ and
combining them with the $J$ and the $\bar{J}$ constraints we obtain
\be
D_{000}(\a_1,\a_2,\a_3,\bar{\a}_1,\bar{\a}_2,\bar{\a}_3)
=  e^{iL(\a_1+\bar{a}_1)}
\frac{ 8 \pi^2 \d(\g+\bar{\g})\d(\h + \bar{\h})}{\sqrt{4s_1^2s_2^2-(s_3^2-s_1^2-s_2^2)^2}}
\d_{\mathbb{Z}}(L) \ ,
\label{lc20}
\ee
where  $L = \jh_1+\jh_2+\jh_3$ and the angles
$\g$ and  $\h$ are fixed by Eqs. (\ref{conserv}).

We now turn to the quantum structure constants of the operator algebra.
The three-point couplings will be derived in the next section by studying the factorization
of the four-point amplitudes. They can also be
derived taking the Penrose limit of the $SU(2)_k \times U(1)_k$ model, as we will show
in Appendix A, and the two results agree (up to normalization). Moreover, they can be compared with
the three-point couplings for twist fields computed in \cite{dfms}.
The three-point coupling that appears in the OPE of two $\F^+$ is
\be
{\cal C}_{++}{}{}^+ (q_1,q_2,q_3) = \frac{1}
{\G(1+\jh_3-\jh_1-\jh_2)}
\left [ \frac{\g(p_3)}{\g(p_1)\g(p_2)} \right ]^{\frac{1}{2}+\jh_3-\jh_1-\jh_2} \  ,
\label{cppp}
\ee
where we defined
\be
\g(x) = \frac{\G(x)}{\G(1-x)} \ .
\label{lc21}
\ee
We also recall that in this case, $p_3=p_1+p_2$ and $\jh_3 = \jh_1+\jh_2+n$, $n \in \mathbb{N}$.
When we have one $\F^+$ and one $\F^-$ operator with $p_1>p_2$
the coupling is
\be
{\cal C}_{+-}{}^+(q_1,q_2,q_3) =  \frac{1}
{\G(1-\jh_3+\jh_1+\jh_2)}
\left [ \frac{\g(p_1)}{\g(p_2)\g(p_3)} \right ]^{\frac{1}{2}-\jh_3+\jh_1+\jh_2} \  ,
\label{cpmp}
\ee
where  $p_3=p_1-p_2$ and $\jh_3 = \jh_1+\jh_2-n$, $n \in \mathbb{N}$.
If on the other hand $p_1<p_2$ we obtain
\be
{\cal C}_{+-}{}^-(q_1,q_2,q_3) =  \frac{1}
{\G(1+\jh_3-\jh_1-\jh_2)}
\left [ \frac{\g(p_2)}{\g(p_1)\g(p_3)} \right ]^{\frac{1}{2}+\jh_3-\jh_1-\jh_2} \  ,
\label{cpmm}
\ee
where  $p_3=p_2-p_1$ and $\jh_3 = \jh_1+\jh_2+n$, $n \in \mathbb{N}$.
Moreover ${\cal C}_{++-}={\cal C}_{--+}$ up to changing the sign of all the
$\jh_i$ and similarly for the other couplings.
We will often use a short-hand notation for the
three-point couplings writing for instance \be
{\cal C}_{++}{}^+(q_1,q_2,n) \ ,
\label{lc22}
\ee
to denote the coupling (\ref{cppp}) with $\jh_3 = \jh_1 + \jh_2 + n$.

A coupling of particular interest is the coupling between two discrete and one continuous
representations. It is given by
\be
{\cal C}_{+-0}(p,\jh_1;p,\jh_2;s,\jh_3) = e^{\frac{s^2}{2}[\psi(p)+\psi(1-p)-2\psi(1)]} \ ,
\label{lc23}
\ee
where $\psi(x) = \frac{d \ln{\G(x)}}{dx}$ is the digamma function.
We introduce a short-hand notation also for this coupling setting
\be
{\cal C}_{+-0}(p,s) \equiv {\cal C}_{+-0}(p,\jh_1;p,\jh_2;s,\jh_3) \ .
\label{lc24}
\ee

The couplings ${\cal C}_{abc}$ are symmetric in the indexes.
The normalization of the coupling ${\cal C}_{+-0}$ is fixed by the requirement
\be
<\F^+_{p_1,\jh_1}\F^-_{p_2,\jh_2}\F^0_{0,0}> \ \ = \ \ <\F^+_{p_1,\jh_1}\F^-_{p_2,\jh_2}> \ ,
\label{lc25}
\ee
since the identity operator is contained in $\F^0_{0,0}$.
The normalization of the other three-point couplings has been fixed factorizing
the four-point amplitude first on a channel where the intermediate states belong
to $\F^0$ representations
so that we can fix its overall normalization using $(\ref{lc23})$ and then factorizing
it on a channel where the intermediate states belong to $\F^{\pm}$ representations
in order to read the other ${\cal C}_{abc}$ couplings.

\section{Four-point correlators}
\setcounter{equation}{0}

A very powerful method for the computation of correlation functions in a CFT
is to resort to a free field realization of the theory, as has been
done for the Virasoro minimal models \cite{df} or for $SU(2)_k$ \cite{dot}.
As we have already explained, the $H_4$ algebra has a quasi-free-field realization
and the primary states are collections of twist fields. A natural
way of computing correlators would therefore be to compute correlation functions
for twist fields, following \cite{dfms}. We will discuss this method in Appendix \ref{c}.

In this section we choose a different approach and
compute the correlators solving directly the Knizhnik-Zamolodchikov
(KZ) equation \cite{kz}, since in this way the factorization properties of the amplitudes and the
constraints imposed by the $H_4$ symmetry are more transparent.
We will compute the conformal blocks that can appear in the intermediate channels for
a given collection of external states and then we will reconstruct the four-point
correlator summing these conformal blocks in such a way as to construct a monodromy
invariant combination.
We will see in a very clear context two generic features of WZW models related to
$\s$-models with a non-compact target-space:
the number of conformal blocks is infinite and fusion does not close on
highest-weight representations of the affine algebra but it is necessary to
include the spectral-flowed representations.
Moreover, since we are using a covariant formalism,
we can study in detail states with $p=0$, which belong to continuous representations,
and their couplings with $p \ne 0$ states, which belong to discrete representations.

The KZ equation is a consequence of the existence of the null vector
\be
\left [ L_{-1} - \frac{1}{2} (P^-_{-1}P_0^+
+P^+_{-1}P_0^-) - J_{-1}K_0 -  K_{-1}J_0 - K_{-1}K_0 \right ] \left | V \right \rangle  \ ,
\label{ld1}
\ee
in an arbitrary highest-weight representation $V$ of the affine algebra.
When we insert this null vector in a four-point amplitude we obtain a
partial differential equation of the form
\be \p_{z_i} A = \sum_{j=1,j \ne i}^4
\frac{1}{z_{ij}} \left [ \frac{1}{2}(D^+_iD^-_j+D^-_iD^+_j)
+D^J_iD^K_j+D^K_iD^J_j+D^K_iD^K_j \right ] A \ .
\label{ld2bis}
\ee
In the previous equation, the four-point function
$A =  A(z_i, \bar{z}_i, x_i, \bar{x}_i)$ is a function of
the world-sheet variables $z_i$, $\bar{z}_i$, the insertion points
of the four vertex operators, and of the charge variables $x_i$, $\bar{x}_i$.
The $D^{J^a}_i$ are the operators displayed in Eq. (\ref{h1}), (\ref{h2}) and (\ref{h3}),
which act on the $x$ variables and realize the $H_4$ algebra, the
index $i$ labeling the vertex operator they are acting on.
The precise operators that appear in the KZ equation depend
on the choice of the external states.

As usual, we can reduce the dependence of $A(z_i,\bar{z}_i, x_i, \bar{x}_i)$
on the $z_i$ and $\bar{z}_i$ to the dependence on the two cross-ratios $z$ and $\bar{z}$
using the global $SL_2(\mathbb{C})$ symmetry. Using the global $H_4$ symmetry
we can similarly reduce the dependence on the group  variables $x_i$ and $\bar{x}_i$
to the dependence on two invariant variables  $x$ and $\bar{x}$.
We then write
\be
A(z_i, \bar{z}_i, x_i, \bar{x}_i)
= \prod_{i<j}^4 |z_{ij}|^{2 \left (\frac{h}{3}-h_i-h_j \right) }
K(x_i,\bar{x}_i){\cal A}(z,\bar{z}, x, \bar{x}) \ ,
\label{gs-ci}
\ee
where $h=\sum_{i=1}^4h_i$ and the cross-ratios $z$, $\bar{z}$ are defined according to
\be
z = \frac{z_{12}z_{34}}{z_{13}z_{24}} \ , \hspace{1cm}z = \frac{\bar{z}_{12}\bar{z}_{34}}
{\bar{z}_{13}\bar{z}_{24}} \ .
\label{ld2}
\ee
The form of the function $K$ and the expression of $x$ and $\bar{x}$ in terms
of the $x_i$ and $\bar{x}_i$ are fixed by the $H_4$ symmetry but are different
for different types of correlators and we will show them explicitly in the next
sub-sections.
We start our discussion with correlators involving only $\F^\pm$ vertex operators
which can be divided in two types: correlators containing three states with
$\F^+$ and one $\F^-$ ($<+++->$ correlators)
and correlators containing two $\F^+$ and two $\F^-$ ($<+-+->$ correlators).
We then consider correlators containing also $\F^0$ operators which can
also be divided in two types: $<++- \ 0>$ and $<+- \ 0 \ 0>$ correlators.
Moreover we note that states in $V^0$ representations can also
appear as intermediate states in correlators between $\F^\pm$ states
whenever their propagation is allowed by the kinematics.

Using the operator algebra in $(\ref{ope-gen})$, we can rewrite each four-point function
as a sum over intermediate representations of the affine algebra.
The functions appearing in this decomposition are called conformal blocks and are
in principle fixed by the algebra, while the three-point couplings represent the dynamical
input of the theory. We can choose to decompose the four-point functions in different ways
and all of them must agree due to the associativity of the operator algebra.
In terms of the cross-ratios in (\ref{ld2}), the $s$-channel
limit $z_1 \sim z_2$ corresponds to $z \sim 0$ while
the $t$-channel limit $z_1 \sim z_3$ and the $u$-channel limit $z_1 \sim z_4$
correspond respectively to $z \sim \infty$ and $z \sim 1$.

In the following, we will derive the KZ equation for the various classes of correlators
and we will solve it presenting explicit expressions for the conformal blocks.
The four-point function is then given by a monodromy invariant combination of the conformal
blocks. The coefficients multiplying the conformal blocks coincide with the product
of the three-point couplings once the four-point amplitude has been normalized in
a way consistent with the two point-functions.
Further details about the solutions of the KZ equations can be found in Appendix \ref{d}.

\subsection{$<+++->$ correlators}
\setcounter{equation}{0}

Consider a correlator of the form
\be
<\F^+_{p_1,\jh_1}(z_1,\bar{z}_1,x_1,\bar{x}_1)
\F^+_{p_2,\jh_2}(z_2,\bar{z}_2,x_2,\bar{x}_2)\F^+_{p_3,\jh_3}(z_3,\bar{z}_3,x_3,\bar{x}_3)
\F^-_{p_4,\jh_4}(z_4,\bar{z}_4,x_4,\bar{x}_4)> \ ,
\label{ld3}
\ee
with
\be
p_1+p_2+p_3 = p_4 \ ,
\label{ld4}
\ee
as required by momentum conservation.
From the global $H_4$ symmetry constraints we can derive
\be
K(x_i, \bar{x}_i)
= \left | e^{-x_4(p_1x_1+p_2x_2+p_3x_3)}(x_3-x_1)^{-L} \right |^2 \ ,
\label{gs-pppm}
\ee
where $L=\jh_1+\jh_2+\jh_3+\jh_4$ and
\be
x = \frac{x_2-x_1}{x_3-x_1} \ , \hspace{1cm}
\bar{x} = \frac{\bar{x}_2-\bar{x}_1}{\bar{x}_3-\bar{x}_1} \ .
\label{ld5}
\ee
From the decomposition of the tensor products of $H_4$ representations
displayed in Eq. (\ref{tensor}) it follows that the
correlator vanishes for $L > 0$ while for $L \le 0$ it decomposes
in the sum of a finite number of conformal blocks $N= |L| +1$
which reflects the propagation in the $s$-channel of the
representations $\F^+_{p_1+p_2,\jh_1+\jh_2+n}$ with $n = 0,
...,|L|$.
We also note that only states with $p \ne 0$ can appear in the intermediate channels.
When passing to the conformal blocks, we can write
\be
{\cal A}(z,\bar{z}, x,\bar{x}) \sim \sum_{n=0}^{|L|} {\cal F}_n(z,x) \bar{{\cal F}}_n
(\bar{z},\bar{x}) \ ,
\label{ld6}
\ee
and setting ${\cal F}_n = z^{\ka_{12}}(1-z)^{\ka_{14}}F_n$ where
\ba
\ka_{12} &=& h_1+h_2-\frac{h}{3}-\jh_2p_1-\jh_1p_2-p_1p_2 \ , \nb \\
\ka_{14} &=&  h_1+h_4-\frac{h}{3}-\jh_4p_1+\jh_1p_4+p_1p_4 -p_1-L(p_2+p_3) \ ,
\label{ka-pp}
\ea
the KZ equation becomes
\ba
\p_z F_n(x,z) &=& \frac{1}{z} \left [ -(p_1x+p_2x(1-x))\p_x +Lp_2x \right ]
F_n(x,z) \nb \\
&-& \frac{1}{1-z} \left [ (1-x)(p_2x+p_3)\p_x+Lp_2(1-x) \right ] F_n(x,z) \ .
\label{ld7}
\ea
The conformal blocks are
\be
F_n(z,x) = f^n(z,x) (g(z,x))^{|L|-n} \ , \hspace{1cm} n = 0, ..., |L| \ ,
\label{ld8}
\ee
where
\be
f(z,x) = \frac{z^{1-p_1-p_2} p_3}{1-p_1-p_2}\f_0 - x z^{-p_1-p_2}\f_1 \ , \hspace{1cm}
g(z,x) = \g_0 -\frac{xp_2}{p_1+p_2} \g_1 \ ,
\label{ld9}
\ee
and
\ba
\f_0 &=& F(1-p_1,1+p_3,2-p_1-p_2,z) \ , \hspace{0.6cm}
\g_0 = F(p_2,p_4,p_1+p_2,z) \ ,  \nb \\
\f_1 &=& F(1-p_1,p_3,1-p_1-p_2,z) \ , \hspace{0.6cm}
\g_1 = F(1+p_2,p_4,1+p_1+p_2,z) \ .
\label{ld10}
\ea
where $F(a,b,c,z)$ is the standard $_1F_2$-hypergeometric
function.

The four-point function, given by a monodromy invariant combination of the
conformal blocks, is
\be
{\cal A}(z,\bar{z},x,\bar{x})
= |z|^{2\ka_{12}}|1-z|^{2\ka_{14}}
\frac{ \sqrt{C_{12}C_{34}} }{|L|!} \left ( C_{12}|f(z,x)|^2+C_{34}|g(z,x)|^2 \right )^{|L|} \ ,
\label{ld12}
\ee
where
\be
C_{12} =  \frac{\g(p_1+p_2)}{\g(p_1)\g(p_2)} \ , \hspace{1cm}
C_{34} = \frac{\g(p_4)}{\g(p_3)\g(p_4-p_3)} \ .
\label{ld13}
\ee
The monodromy invariance of this correlators is manifest  around $z=0$ and  can be easily verified
around $z=1$ and $z=\infty$.
Moreover it can be expressed as a sum over all conformal blocks
with the appropriate three-point couplings (we use
the short-hand notation introduced in the previous section)
\be
{\cal A}(z,\bar{z},x,\bar{x}) = \sum_{n=0}^{|L|} {\cal C}_{++-}(q_1,q_2,n)
{\cal C}_{+-+}(q_3,q_4,|L|-n)|{\cal F}_n(z,x)|^2 \ .
\label{ld11}
\ee
Moreover, in the $t$ and in the $u$ channels the correlator
factorizes as expected from (\ref{tensor}).
In Appendix \ref{b} we show how to get exactly
the same correlator taking a suitable limit of the
$SU(2)$ correlators computed in \cite{zf}.

\subsection{$<+-+->$ correlators}
\setcounter{equation}{0}

Correlators of the form
\be
<\F^+_{p_1,\jh_1}(z_1,\bar{z}_1,x_1,\bar{x}_1)
\F^-_{p_2,\jh_2}(z_2,\bar{z}_2,x_2,\bar{x}_2)
\F^+_{p_3,\jh_3}(z_3,\bar{z}_3,x_3,\bar{x}_3)
\F^-_{p_4,\jh_4}(z_4,\bar{z}_4,x_4,\bar{x}_4)> \ ,
\label{ld14}
\ee
are more interesting.
First of all, according to (\ref{tensor})
they involve an infinite number of conformal blocks. Moreover, for particular
values of the momenta of the external states (for instance
$p_1=p_2$ and $p_3=p_4$), the correlator can be factorized
on states with $p=0$. Finally, we can see explicitly that
spectral-flowed representations appear
in the fusion of highest-weight representations of the affine algebra.
The $H_4$ Ward identities require
\be
p_1+p_3=p_2+p_4 \ ,
\label{ld15}
\ee
and give the function $K$
\be
K(x_i, \bar{x}_i) = \left |
e^{-p_2x_1x_2-p_3x_3x_4-(p_1-p_2)x_1x_4}
(x_1-x_3)^{-L} e^{-\frac{x}{4}(p_1-2p_2-p_3)} \right |^2 \ ,
\label{gs-pmpm2}
\ee
where  $L=\jh_1+\jh_2+\jh_3+\jh_4$ as well as
\be
x = (x_1-x_3)(x_2-x_4) \ , \hspace{1cm} \bar{x} = (\bar{x}_1-\bar{x}_3)(\bar{x}_2-\bar{x}_4) \ .
\label{ld16}
\ee
Proceeding as before we pass to the conformal blocks and we set
${\cal F}_n = z^{\ka_{12}}(1-z)^{\ka_{14}}F_n(x,z)$ where
\ba
\ka_{12} &=& h_1+h_2-\frac{h}{3}+p_1p_2-\jh_2p_1+\jh_1p_2-p_2 \ ,  \nb \\
\ka_{14} &=&  h_1+h_4-\frac{h}{3}+p_1p_4-\jh_4p_1+\jh_1p_4-p_4 \ .
\label{ka-pm}
\ea
We then arrive at the following form for the KZ equation
\ba
z(1-z) \p_z F_n(x,z) &=& \left [ x\p^2_x +\left ( ax+1-L \right )\p_x
+\frac{x}{4}(a^2-b^2) + \r_{12} \right ] F_n(x,z) \nb \\
&+& z \left [ -2ax \p_x +\frac{x}{4}(b^2-c^2)  - \r_{12} - \r_{14} \right ]   F_n(x,z) \ ,
\label{kz-pmpm-nf}
\ea
where
\be
2a = p_1+p_3 \ , \hspace{0.5cm} b = p_1-p_2 \ , \hspace{0.5cm}
c= p_2-p_3 \ ,
\label{ld17}
\ee
and
\be
\r_{12} = \frac{(1-L)}{2}(a-b) \ ,  \hspace{1cm}
\r_{14} = \frac{(1-L)}{2}(a-c)\ .
\label{ld18}
\ee
According to (\ref{tensor}) when $p_1>p_2$ and $L \le 0$
in the $s$-channel flow the representations $\F^+_{p_1-p_2,\jh_1+\jh_2+n}$
with $n \in \mathbb{N}$.
The conformal blocks are
\be
F_n(z,x) = \n_n \frac{e^{xg_1(z)}}{(f_1(z))^{1-L}}
L^{|L|}_n(x  \g_{\psi}(z)) \psi(z)^n \ , \hspace{1cm}  n \in \mathbb{N} \ ,
\label{ld19}
\ee
where $L^{|L|}_n$ is the n-th generalized Laguerre polynomial,
\ba
\psi(z) &=& \frac{f_2(z)}{f_1(z)}  \ , \hspace{1cm}  \g_{\psi}(z) =
-z(1-z)\p \ln{\psi} \ ,  \hspace{1cm} \n_n = \frac{n!}{(p_1-p_2)^n} \ , \nb \\
q(z) &=&  \frac{p_1-2p_2-p_3}{4} + z p_3 \ , \hspace{1cm}   g_1(z) =
q(z)-z(1-z)\p \ln{f_1} \ ,
\label{ld20}
\ea
and
\ba
f_1(z) &=& F(p_3,1-p_1,1-p_1+p_2,z) \ , \nb \\
f_2(z) &=& z^{p_1-p_2}F(p_4,1-p_2,1-p_2+p_1,z) \ .
\label{ld21}
\ea
The full correlator is given by
\be
{\cal A}(z,\bar{z}, x, \bar{x})
= \frac{\mu_L  |z|^{2\ka_{12}}|1-z|^{2\ka_{14}}}{S^{1+|L|}}
\left |e^{xq(z)-xz(1-z)\p \ln{S}} \right |^2
\left ( \frac{u}{2} \right )^{-|L|}
I_{|L|}(u) \ ,
\label{pmk-corr}
\ee
where  $I_{\a}$ is the standard  Bessel function of imaginary argument,
 we have introduced the following quantities
\be
r = \frac{C_{12}C_{34}}{(p_1-p_2)^2} \ , \hspace{1cm}
C_{12} = \frac{\g(p_1)}{\g(p_2)\g(p_1-p_2)} \ , \hspace{1cm}
C_{34} = \frac{\g(p_4)}{\g(p_3)\g(p_4-p_3)} \ ,
\label{ld24}
\ee
and defined
\be
S = |f_1|^2 - r |f_2|^2 \ , \hspace{1cm}
\mu_{L} = C_{12}^{\frac{1}{2}} C_{34}^{\frac{1}{2}+|L|} \ ,
\label{ld25}
\ee
as well as
\be
u = \frac{2\sqrt{r}|xz(1-z)\p \psi|}{1-r|\psi|^2}  = \frac{2\sqrt{r}|xz(1-z)W(f_1,f_2)|}{S} \ ,
\label{ld26}
\ee
with $W(f_1,f_2)$ the Wronskian of the two solutions of the hyper-geometric equation,
\be
W(f_1,f_2) = (p_1-p_2)z^{p_1-p_2-1}(1-z)^{p_2-p_3-1} \ .
\label{ld27}
\ee

The correlator in $(\ref{pmk-corr})$ factorizes as
\be
{\cal A}(z,\bar{z}, x, \bar{x})
= \sum_{n=0}^\infty {\cal C}_{+--}(q_1,q_2,n){\cal C}_{+-+}(q_3,q_4,n-L)|{\cal F}_{n}(z,x)|^2 \ ,
\label{ld22}
\ee
where we used again the short-hand notation for the three-point couplings
introduced in the previous section. Similar expressions can be written in the $t$ and in the $u$
channel.
We have derived (\ref{pmk-corr}) for $p_1>p_2$. The correlator for $p_1 < p_2$ is given
by the same expression up to exchanging $p_1$ with $p_2$, $p_3$ with $p_4$ and changing the
sign of the $\jh_i$ and there is a similar
decomposition in conformal blocks.
Moreover when $L>0$ one can show that
\be
{\cal A}(L) = |x|^{2L}|z|^{2Lb}|1-z|^{-2Lc} {\cal A}(-L) \ ,
\label{ld28}
\ee
where ${\cal A}(-L)$ is the amplitude we computed before.

We have been dealing so far only with conformal blocks pertaining to the propagation of
states belonging to the discrete series. For particular values of the
external momenta, also states with $p=0$ can flow in the intermediate
channel and as a consequence it has to be possible in these cases to
represent the correlator as an integral over the
continuous representations. A correlator of the form
$<p,-p,l,-l>$ factorizes on $\F^0$ representations in the $s$ channel, while a correlator
of the form $<p,-p,p,-p>$ both in the $s$ and in the $u$ channel.
Let us consider the former. The conformal blocks are
\be
F_{s}(z,x) = \frac{e^{xg_1(z)}}{(c_1(z))^{1-L}}
e^{\frac{s^2}{2} \r(z)} (xz(1-z)\p \r)^{\frac{L}{2}} J_{|L|}(v) \ ,
\label{ld29}
\ee
where
\be
\r(z) = \frac{c_2(z)}{c_1(z)}  \ , \hspace{1cm}
v = s\sqrt{-2xz(1-z)\p \r(z)}  \ ,
\label{ld30}
\ee
and
\ba
c_1(z) &=& F(l,1-p,1,z) \ , \nb \\
c_2(z) &=& [ \ln{z}+2\psi(1)-\psi(l)-\psi(1-p)]  c_1(z) \nb \\
&+& \sum_{n=0}^{\infty} \frac{(l)_n(1-p)_n}{n!^2}
[\psi(l+n)+\psi(1-p+n) - 2 \psi(n+1)] z^n \ ,
\label{ld31}
\ea
where
\be
(a)_n \equiv \frac{\G(a+n)}{\G(a)} \ .
\label{ld32}
\ee
Moreover
\be
q(z) = lz - \frac{p+l}{4} \ , \hspace{1cm} g_1(z) =q(z)  -z(1-z) \p_z \ln{c_1} \ .
\label{ld33}
\ee
The four-point amplitude is
\be
{\cal A}(z,\bar{z},x,\bar{x}) =
\frac{ |z|^{2\ka_{12}}|1-z|^{2\ka_{14}}}{S^{1+|L|}} \left |e^{xq(z)-xz(1-z)\p \ln{S}} \right |^2
\left( \frac{u}{2} \right )^{-|L|} I_{|L|}(u) \ ,
\label{cont-k-corr}
\ee
where
\be
S = -|c_1|^2 \{ 2[\psi(p)+\psi(1-p)-2\psi(1)] + \r + \bar{\r} \} \ , \hspace{0.5cm}
u = \frac{2|xz(1-z)W(c_1,c_2)|}{S} \ ,
\label{ld35}
\ee
and the Wronskian now is
\be
W(c_1,c_2) = \frac{(1-z)}{z}^{p-l-1} \ .
\label{ld36}
\ee
Using the coupling between states in discrete and states in continuous
representations we can represent it as
\be
{\cal A}(z,\bar{z},x,\bar{x})
= \int_0^\infty ds s  {\cal C}^2_{+-0}(p,s) |{\cal F}_{s}(z,x)|^2 \ .
\label{ld34}
\ee
The four-point amplitude in (\ref{cont-k-corr}) can also be obtained taking
the limit $p_1 \rightarrow p_2$ in the four-point amplitude in (\ref{pmk-corr})
as discussed in Appendix C.

\subsection{$<++- \ 0>$ correlators}
\setcounter{equation}{0}

Consider a correlator of the form
\be
<\F^+_{p_1,\jh_1}(z_1,\bar{z}_1,x_1,\bar{x}_1)
\F^+_{p_2,\jh_2}(z_2,\bar{z}_2,x_2,\bar{x}_2)\F^-_{p_3,\jh_3}(z_3,\bar{z}_3,x_3,\bar{x}_3)
\F^0_{s,\jh_4}(z_4,\bar{z}_4,x_4,\bar{x}_4)> \ ,
\label{ld37}
\ee
with
\be
p_1+p_2 = p_3 \ ,
\label{ld38}
\ee
as required by momentum conservation.
From the global $H_4$ symmetry constraints we can derive
\be
K(x_i, \bar{x}_i)
= \left |
e^{-x_3(p_1x_1+p_2x_2)- \frac{s}{\sqrt{2}} \frac{x_3}{x_4} - \frac{s}{2\sqrt{2}}(x_1+x_2)x_4}
(x_1-x_2)^{-L} \right |^2  \ ,
\label{ld39}
\ee
where $L=\jh_1+\jh_2+\jh_3+\jh_4$ and $x= (x_1-x_2)x_4$.
When passing to the conformal blocks we can write
\be
{\cal A}(z,\bar{z}, x,\bar{x}) \sim \sum_{n = 0 }^{\infty} {\cal F}_n(z,x) \bar{{\cal F}}_n
(\bar{z},\bar{x}) \ ,
\label{ld40}
\ee
and setting ${\cal F}_n = z^{\ka_{12}}(1-z)^{\ka_{14}} x^L F_n$ where
\be
\ka_{12} = h_1+h_2-\frac{h}{3} -p_1\jh_2-p_2\jh_1-p_1p_2 \ ,
\hspace{1cm}
\ka_{14} = h_1+h_4-\frac{h}{3}  -p_1\jh_4 + L p_1 - \frac{s^2}{4} \ ,
\label{ld41}
\ee
the KZ equation reads
\be
z(1-z) \p_z F_n = -\left[ p_3x\p_x+\frac{s}{2\sqrt{2}}(p_1-p_2)x \right ]
F_n + z \left [ \left ( p_2x-\frac{s}{\sqrt{2}} \right ) \p_x
- \frac{sp_2}{2\sqrt{2}}x \right ] F_n \ .
\label{kz-ppmo}
\ee

The conformal blocks are
\be
F_n(z,x) = (s \f(z)+x\g(z))^ne^{s^2 \h(z) +sx \psi(z)} \ ,
\label{ld42}
\ee
with $n \ge 0$ and
\ba
\f(z) &=& \frac{z^{1-p_3}}{\sqrt{2}(1-p_3)}F(1-p_1,1-p_3,2-p_3,z) \ , \nb \\
\g(z) &=& -z^{-p_3}(1-z)^{p_1} \ , \nb \\
\psi(z) &=& -\frac{1}{2\sqrt{2}} +\frac{p_2}{\sqrt{2}p_3}(1-z)F(1+p_2,1,1+p_3,z) \ , \nb \\
\h(z) &=& -\frac{z p_2}{2p_3} \ {}_3F_2(1+p_2,1,1;1+p_3,2;z)  - \frac{1}{4} \ln{(1-z)} \ .
\label{ld43}
\ea
The four-point function is then given by
\ba
{\cal A}(z,\bar{z},x,\bar{x})
&=&  \sqrt{C_{12}}C_{+-0}(p_3,s) |x|^{2 L}|z|^{2\ka_{12}}|1-z|^{2\ka_{14}} \nb \\
& & e^{ C_{12}|s\f+x\g|^2+s^2(\h+\bar{\h})+s(x\psi+\bar{x}\bar{\psi})} \ ,
\label{ld44}
\ea
where
\be
C_{12} = \frac{\g(p_1+p_2)}{\g(p_1)\g(p_2)} \ .
\label{ld45}
\ee
This correlator can be expressed as a sum over the conformal blocks
with the appropriate three-point couplings (we use
the short-hand notation introduced in the previous section)
\be
{\cal A}(z,\bar{z},x,\bar{x}) = \sum_{n=0}^{\infty} {\cal C}_{++-}(q_1,q_2,n)
{\cal C}_{-0+}(p_3,s)|{\cal F}_n(z,x)|^2 \ .
\label{ld46}
\ee

\subsection{$<+- \ 0 \ 0>$ correlators}
\setcounter{equation}{0}

Consider finally a correlator of the form
\be
<\F^+_{p,\jh_1}(z_1,\bar{z}_1,x_1,\bar{x}_1)
\F^-_{p,\jh_2}(z_2,\bar{z}_2,x_2,\bar{x}_2)\F^0_{s_3,\jh_3}(z_3,\bar{z}_3,x_3,\bar{x}_3)
\F^0_{s_4,\jh_4}(z_4,\bar{z}_4,x_4,\bar{x}_4)> \ .
\label{ld47}
\ee
The Ward identities give
\be
{\cal K}(x_i,\bar{x}_i) =  \left | e^{-px_1x_2 - \frac{x_1}{\sqrt{2}} \left ( \frac{s_3}{x_3}
+\frac{s_4}{x_4} \right )- \frac{x_2}{\sqrt{2}} ( s_3 x_3 +s_4 x_4)} x_3^{-L} \right |^2   \ ,
\label{ld48}
\ee
where $L = \sum_{i=1}^4 \jh_i$ and $x = \frac{x_4}{x_3}$.

The structure of this correlator is simpler if we decompose it in conformal blocks
around $z=1$ writing
\be
{\cal A}(u,\bar{u}, x,\bar{x}) \sim \sum_{n \in \mathbb{Z}} {\cal F}_n(u,x) \bar{{\cal F}}_n
(\bar{u},\bar{x}) \ ,
\label{ld49}
\ee
where $u=1-z$. Setting ${\cal F}_n = u^{\ka_{14}}(1-u)^{\ka_{12}} F_n$ where
\be
\ka_{14} = h_1+h_4-\frac{h}{3} -p \jh_4 - \frac{s_4^2}{2} \ ,
\hspace{1cm}
\ka_{12} =   \frac{s_3^2+s_4^2}{2} - \frac{h}{3} \ ,
\label{ld50}
\ee
we obtain the following KZ equation
\be
\p_u F_n = - \frac{1}{u} \left [ p x \p_x  + \frac{s_3s_4 x}{2} \right ] F_n
- \frac{1}{1-u} \frac{s_3s_4}{2} \left ( x + \frac{1}{x} \right ) F_n \ .
\label{ld51}
\ee
The conformal blocks are
\be
F_n(u,x) = (x \g(u))^n e^{x \omega(u) + \frac{\psi(u)}{x}} \ ,
\label{ld52}
\ee
with $n \in \mathbb{Z}$ and
\be
\g(u) = u^{-p} \ , \hspace{0.4cm}
\omega(u) = - \frac{s_3s_4}{2p} F_p(u) \ ,  \hspace{0.4cm}
\psi(u) =  - \frac{s_3s_4}{2(1-p)} u F_{1-p}(u) \ ,
\label{ld53}
\ee
where $F_p(u) = F(p,1,1+p,u)$ and $F_{1-p}(u) = F(1-p,1,2-p,u)$.
The four-point function is then given by
\be
{\cal A}(u,\bar{u},x,\bar{x})
=  {\cal C}_{+-0}(p,s_3)  {\cal C}_{+-0}(p,s_4) |u|^{2\ka_{12}}|1-u|^{2\ka_{14}}
\left | e^{x \omega(u) + \frac{\psi(u)}{x}} \right |^2 \sum_{n \in \mathbb{Z}}
\left | x u^{-p} \right |^{2n} \ .
\label{ld54}
\ee
The monodromy invariance of this correlator is more evident if we reorganize the series in $(\ref{ld54})$ as
\be
\left | e^{x \omega(u) + \frac{\psi(u)}{x}} \right |^2 \sum_{n \in \mathbb{Z}} \left | x u^{-p} \right |^{2n}
= \sum_{l,m \in \mathbb{Z}} x^l \bar{x}^m |u|^{-p(l+m)} \chi^{\frac{m-l}{2}} I_{|m-l|}(R) \ ,
\label{ld55}
\ee
where
\ba
\chi &=& \frac{u p F_{1-p} + (1-p) |u|^{2p} \bar{F}_p}{\bar{u} p \bar{F}_{1-p} + (1-p) |u|^{2p} F_p} \ ,
\nb \\
R^2 &=& s^2_3 s^2_4 |u|^{-2p} \left ( u \frac{F_{1-p}}{1-p} + |u|^{2p} \frac{\bar{F}_p}{p} \right )
\left ( \bar{u} \frac{\bar{F}_{1-p}}{1-p} + |u|^{2p} \frac{F_p}{p} \right ) \ .
\label{ld56}
\ea
In Appendix C we explain how to compute this correlation function using the free-field realization
of the $H_4$ algebra.

\section{Transformations of the conformal blocks}
\setcounter{equation}{0}

In the previous section we computed the four-point correlation functions
between highest-weight representations of the $H_4$ WZW model and derived
explicit expressions for the corresponding conformal blocks.
Due to the fact that each four-point correlator can be factorized around
$z = 0$, $1$ and $\infty$, we have at our disposal three different sets
of conformal blocks, related between themselves by linear transformations
known as fusion and braiding transformations.
Actually, fusion and braiding matrices are usually discussed in the context of
rational CFT. In that case, since
by definition the Hilbert space of the model decomposes in a finite number of
representations of the affine algebra, they are finite dimensional matrices.
Duality transformations for non-compact CFTs have been investigated
mostly for the Liouville model and the $H_3^+$ WZW model \cite{feb}.
The $H_4$ WZW model is another case where it is possible to derive
explicit expressions for these matrices.

Here we discuss a particular example, computing the matrix that implements the change
of basis from the $z=0$ to the $z=1$ conformal blocks for a $<+-+->$ correlator with
$L \le 0$.
When the intermediate states in both channels belong to the discrete
series,
we have an infinite matrix $c^L_{n,m}$, $n,m \in \mathbb{N}$. When
the intermediate states in one channel belong to the continuous series
and in the other channel
to the discrete series we will have a matrix $c^L_n(s)$ with a discrete index
$n \in \mathbb{N}$ and a continuous index $s \in \mathbb{R}^+$.

Let us start from the first case.
The conformal blocks around $z=0$ are
\be
F_n(z,x) = \n_n \frac{e^{x g_1(z)}}{(f_1(z))^{1-L}}
L_n^{|L|}(x \g_\psi(z))\psi(z)^n \ ,
\label{lt1}
\ee
where the quantities that appear in this expression were defined in
$(\ref{ld20})$ and $(\ref{ld21})$.
Around $z=1$ we have (setting $u=1-z$)
\be
G_n(u,x) = \r_n  \frac{e^{x\omega_1(u)}}{(\f_1(u))^{1+|L|}} L_n^{|L|}(x \g_\xi(u))\xi(u)^n \ ,
\label{lt7}
\ee
where
\be
\xi(u) = \frac{\f_2(u)}{\f_1(u)}  \ , \hspace{0.68cm}
\g_{\xi}(u) = u(1-u)\p_u \ln{\xi(u)} \ , \hspace{0.68cm}
q(u) = \frac{p_1-2p_2-p_3}{4} + u p_3 \ ,
\label{lt9}
\ee
and
\ba
\f_1(u) &=& F(p_3,1-p_1,1-p_2+p_3,u) \ , \hspace{1cm}
\f_2(u) = u^{p_2-p_3}F(1-p_4,p_2,1+p_2-p_3,u) \nb \\
\omega_i(u) &=& q(u)+u(1-u)\p_u \ln{\f_i(u)} \ , \ i=1,2 \ ,
\hspace{1cm} \r_n = \frac{n!}{(p_2-p_3)^n} \ .
\label{lt10}
\ea

The $f_i$ and $\f_i$ are related by the transformation formulas for the
hypergeometric function
\be
f_1(z) = r_1\f_1(1-z)+s_1\f_2(1-z) \ , \hspace{1cm}
f_2(z) = r_2\f_1(1-z)+s_2\f_2(1-z) \ ,
\label{lt11}
\ee
with
\ba
r_1 &=& \frac{\G(\g)\G(\g-\a-\b)}{\G(\g-\a)\G(\g-\b)} \ , \hspace{1.7cm}
s_1 = \frac{\G(\g)\G(\a+\b-\g)}{\G(\a)\G(\b)} \ , \nb \\
r_2 &=& \frac{\G(2-\g)\G(\g-\a-\b)}{\G(1-\a)\G(1-\b)} \ , \hspace{1cm}
s_2 = \frac{\G(2-\g)\G(\a+\b-\g)}{\G(1-\g+\a)\G(1-\g+\b)} \ ,
\label{lt12}
\ea
and
\be
\a = p_3 \ , \hspace{0.5cm} \b = 1-p_1 \ , \hspace{0.5cm}
\g= 1-p_1+p_2 \ .
\label{lt13}
\ee

The duality transformation between the
two bases of conformal blocks can be written as
\be
F_n(z,x) = \sum_{m=0}^\infty c^L_{nm} G_m(u,x)  \ .
\label{lt14}
\ee
In order to compute the matrix $c^L_{nm}$ we use the orthogonality of
the Laguerre polynomials
\be
\int_0^\infty dx e^{-x} x^\a L_n^\a(x) L_m^\a(x) = \d_{n,m} \frac{\G(\a+n+1)}{n!} \ ,
\label{lt15}
\ee
and the following integral
\be
\int_0^\infty dx e^{- \s x} x^\a L_n^\a(\l x) L_m^\a(\m x) = \frac{\G(m\!+\!n\!+\!\a\!+\!1)}{m!\ n!}
\frac{(\s\!-\!\l)^n(\s\!-\!\m)^m}{\s^{n+m+\a+1}} F(\!-\!m,\!-\!n,\!-\!m\!-\!n\!-\!\a, \t) \ ,
\label{lt16}
\ee
where
\be
\t = \frac{\s(\s-\l-\m)}{(\s-\l)(\s-\m)} \ .
\label{lt17}
\ee
The result is
\be
c^L_{nm} = \frac{\G(m+n+|L|+1)}{m!\G(m+|L|+1)} \frac{1}{r_1^{n+m+|L|+1}}
\left ( \frac{r_2}{p_1\!\!-\!\!p_2} \right )^n [ (p_3\!\!-\!\!p_2)s_1 ]^m
F(-m,-n,-m-n-|L|; \t) \ ,
\label{lt18}
\ee
where
\be
\t = \frac{r_1s_2}{r_2s_1}
= \frac{\sin{\pi p_4}\sin{\pi p_2}}{\sin{\pi p_1}\sin{\pi p_3}} \ .
\label{lt19}
\ee

Let us now pass to the second case.
As before we consider a correlator of the form $<+-+->$ with $p_1=p_4=p$ and
$p_2=p_3=q$ in such a way that for $z \sim 0$ the correlator factorizes on
$V^\pm$ representations while for  $z \sim 1$ it factorizes on
$V^0$ representations and use
\be
c_1(u) = r_1 f_1(1-u) + s_1 f_2(1-u) \ , \hspace{1cm}
c_2(u) = r_2 f_1(1-u) + s_2 f_2(1-u) \ ,
\label{lt20}
\ee
with
\ba
r_1 &=& \frac{\G(p-q)}{\G(1-q)\G(p)} \ , \hspace{2cm} r_2 = \frac{1}{(p-q)s_1}
+ r_1 (2\psi(1)-\psi(q)-\psi(1-p)) \ ,  \nb \\
s_1
&=& \frac{\G(q-p)}{\G(1-p)\G(q)} \ , \hspace{2cm}
s_2 = s_1(2\psi(1)-\psi(q)-\psi(1-p))  \ .
\label{lt21}
\ea
In this case the fusion matrix is defined according to
\be
F_s(u,x) = \sum_{m=0}^\infty c^L_m(s) F_m(z,x) \ ,
\label{lt22}
\ee
and we need the following integral
\be
\int_0^\infty dt t^{\frac{\n}{2}}e^{-\b t}L_n^\n(\a t)J_\n(\sqrt{t}y) =
2^{-\n}\b^{-\n-n-1}(\b-\a)^ny^\n e^{-\frac{y^2}{4\b}}L_n^\n \left ( \frac{\a y^2}{4\b(\a-\b)}
\right ) \ .
\label{lt23}
\ee
The result is
\be
c_n^{L}(s) = (-1)^{n+|L|+1}\frac{s^{|L|}(p-q)^n}{(|L|+n)!}\frac{s_1^n}{r_1^{n+|L|+1}}
e^{-\frac{s^2}{2}(\psi(q)+\psi(1-p)-2\psi(1)+\t)} L_n^{|L|} \left ( -\frac{s^2}{2} \t \right )
\ ,
\label{lt24}
\ee
where
\be
\t = \frac{\pi \sin{\pi (p-q)}}{\sin{\pi p}\sin{\pi q}} \ .
\label{lt25}
\ee

Using similar techniques one can compute the fusion and braiding matrices
for the other classes of correlators.

\section{Null vectors}
\setcounter{equation}{0}

The structure of the representations of the affine $H_4$ algebra is very simple
since generically they do not contain affine null vectors, and we do not have to quotient
out the corresponding null submodule. The only exception is given by the
representations $\F^{\pm}_{1,\jh}$ which contain a null vector at level one.
Using the currents in (\ref{xcp}) and
(\ref{xcm}), the corresponding fields can be expressed as
\be
\psi_{-1}(z,x) = P^-_{-1}(x) \F^-_{1,\jh}(z,x) \ , \hspace{1cm}
\psi_1(z,x) = P^+_{-1}(x) \F^+_{1,\jh}(z,x) \ .
\label{le1}
\ee
It is easy to verify that they are annihilated
by all the modes $P_n^{\pm}(x), J_n(x), K_n(x)$ with $n>0$.
Due to the presence of the null vectors, we can derive a first order
differential equation for correlators involving vertex operators
with $p= \pm 1$. In this section we will compute correlators involving null vectors
and we will use them to study the three-point couplings between states in
highest weight representations with states in spectral flowed representations.
In the following we will concentrate on $ \F^-_{1,\jh}$.
Let us start with a three-point coupling of the form
\be
\langle \F^+_{p_1,\jh_1}(z_1,x_1)\F^+_{p_2,\jh_2}(z_2,x_2)\F^-_{1,\jh_3}(z_3,x_3) \rangle \ .
\label{le2}
\ee
{}From the null state condition it follows that
\be
(\jh_1+\jh_2+\jh_3) {\cal C}_{++-} = 0 \ ,
\label{le3}
\ee
and therefore ${\cal C}_{++-}=0$ unless $\jh_1+\jh_2+\jh_3 = 0$. As a consequence,
only one conformal block can appear in the decomposition of
a four-point function containing $ \F^-_{1,\jh}$.
Consider first a correlator of the form
$<+++->$ with $p_4 = 1$.
The equation implied by the null state is
\be
(x-z) \p_x {\cal A} + L {\cal A} = 0 \ ,
\label{le4}
\ee
where $L = \sum_{i=1}^4 \jh_i$ and we use the same notations as in section $5.1$.
A similar equation holds for $\bar{z}$ and $\bar{x}$.
The solution is
\be
{\cal A} = |z-x|^{-2L} G(z)H(\bar{z}) \ ,
\label{le5}
\ee
where $G$ and $H$ are arbitrary functions of their argument, fixed by the KZ equation.
According to the decomposition in (\ref{ld11}), the conformal block contributing to the correlator
is the one with $n=|L|$ and the correlator reads
\be
{\cal A}(z,\bar{z},x,\bar{x}) = \frac{ \sqrt{C_{12}C_{34}} }{|L|!} C_{12}^{|L|}
|z|^{2\ka_{12}}|1-z|^{2\ka_{14}}
|f(z,x)|^{2|L|} \ ,
\label{le6}
\ee
where
\be
f(z,x) = z^{-1+p_3}(1-z)^{p_1-1}(z-x) \ .
\label{le7}
\ee
Note however that when $p_4 \rightarrow 1$, the four-point correlator
vanishes due to the factor $C_{34}$ in the normalization.
The four-point function we have to use in this case is the spectral flow
of a correlator of the form $<++- \ 0>$, as explained in section 8.

The null vector equation for correlators of the form $<+ - + - >$ with $p_4=1$
reads
\be
\left [ x\p_x -p_2xz -L -\frac{p_1-2p_2-p_3}{4}x  \right ] {\cal A} = 0 \ ,
\label{le8}
\ee
where  $L = \sum_{i=1}^4 \jh_i$ and we now use the same notations as in section $5.2$.
The solution is
\be
{\cal A}(z,\bar{z},x,\bar{x}) = \left | x^L e^{\frac{p_1-2p_2-p_3}{4}x +p_2 z x} \right |^2
G(z)H(\bar{z}) \ .
\label{le9}
\ee
{}From the constraint on the three-point couplings it follows that this correlator
can be non-zero only for $L \ge 0$ and that only the conformal block with
$n=L$ contributes to it. The correlator in $(\ref{pmk-corr})$ (using also $(\ref{ld28})$)
becomes
\be
{\cal A}(z,\bar{z},x,\bar{x}) =  \frac{ \sqrt{C_{12}C_{34}} }{L!} C_{12}^{L}
\left | x^L e^{\frac{p_1-2p_2-p_3}{4}x +p_2 z x} \right |^2
|z|^{2\ka_{12}+2L(1-p_3)}|1-z|^{2\ka_{14}+2(1-p_1)} \ .
\label{le10}
\ee
Again, due to the normalization factor, the
correlator vanishes and we have to use the spectral flow of
a correlator of the form $<++- \ 0>$.

We can use the correlators in $(\ref{le6})$ and $(\ref{le10})$
to compute three-point couplings involving spectral flowed states.
Following \cite{maloog3}, we define the operator generating spectral flow by
one unit as follows
\be
\S^{\pm}(z,\bar{z}) = \lim_{p \rightarrow 1} \frac{1}{\sqrt{\g(p)}}
\F^{\pm}_{p,0}(z,\bar{z},0,0) \ ,
\label{le11}
\ee
that is as the $n=\bar{n}=0$ component of the field $\S^{\pm}(z,\bar{z},x,\bar{x})$.
Four-point functions with one insertion of the operator $\S^-$ coincide
with $(\ref{le6})$ and $(\ref{le10})$, up to the factor $\sqrt{C_{34}}$ that now
is removed by the normalization of $\S^-$. We can then act with the spectral flow
operator $\S^-(z_4)$ on one of the other fields $\F^{\pm}(z_i)$ taking the limit
$z_4 \rightarrow z_i$ and using the OPE to extract the three-point function.

Let us start with  $(\ref{le6})$ and consider the limit
$z_4 \rightarrow z_3$. As explained in section 3, a $\F^+_{p,\jh}$ representation
becomes a $\F^-_{1-p,\jh}$ representation after spectral flow by minus one unit.
However the states $R^+_{p,\jh;n,\bar{n}}$
that form the basis of the original representation
are not mapped under spectral flow to the states $R^-_{1-p,\jh;m,\bar{m}}$
but to the descendant states that form the right edge of the $\F^-_{1-p,\jh}$
representation. This is confirmed by the structure of the three-point function
obtained taking the limit $z_4 \rightarrow z_3$ in $(\ref{le6})$
\ba
& & \langle \F^+_{p_1,\jh_1}(z_1,x_1)\F^+_{p_2,\jh_2}(z_2,x_2)\w_{-1}(\F^-_{p_3,\jh_3})
(z_3,x_3) \rangle    \\
&=& \frac{1}{|L|!} \left ( \frac{\g(p_1+p_2)}{\g(p_1)\g(p_2)} \right )^{\frac{1}{2} + |L|}
\frac{|z_{12}z_{32}z^{-1}_{13}-x|^{2|L|}|x_3-x_1|^{2|L|}}
{|z_{12}|^{2(h_1+h_2-h_3)} |z_{13}|^{2(h_1+h_3-h_2)} |z_{23}|^{2(h_2+h_3-h_1)}} \nb \ .
\label{le12}
\ea
Here $h_3 = \jh_3(1-p_3) + \frac{p_3}{2}(1-p_3)$, the conformal dimension of
the ground states in $\F^-_{1-p_3,\jh_3}$ and the constant appearing in the second line
is ${\cal C}_{++-}(p_1,p_2,p_1+p_2)$ as expected.
Expanding the term $|z_{12}z_{32}z^{-1}_{13}-x|^{2|L|}$ in the previous expression
we obtain
\ba
& & \frac{1}{|L|!} \left ( \frac{\g(p_1+p_2)}{\g(p_1)\g(p_2)} \right )^{\frac{1}{2} + |L|}
\frac{1}{|z_{12}|^{2(h_1+h_2-h_3)} |z_{13}|^{2(h_1+h_3-h_2)} |z_{23}|^{2(h_2+h_3-h_1)}} \\
& & \sum_{n,\bar{n}=0}^{|L|}
\binom{|L|}{n}\binom{|L|}{\bar{n}}
(x_2-x_1)^{|L|-n}(x_3-x_1)^n
(\bar{x}_2-\bar{x}_1)^{|L|-\bar{n}}(\bar{x}_3-\bar{x}_1)^{\bar{n}}
\left ( \frac{z_{12}}{z_{13}z_{23}} \right )^n
\left ( \frac{\bar{z}_{12}}{ \bar{z}_{13}  \bar{z}_{23}} \right )^n \nb \ .
\label{le13}
\ea
The powers of $z_{ij}$ are precisely as required by a three-point coupling with a
descendant at level $n$ and therefore we can see that the action of the spectral flow
operator has mapped the ground states of $\F^+_{p_3,\jh_3}$ into the descendants
sitting on the right edge of the  $\F^-_{1-p_3,\jh_3}$ affine representation.

We can proceed in a similar way starting from $(\ref{le10})$. We take the limit
$z_4 \rightarrow z_2$ so that the operator $\S^-$ acts on $\F^-_{p_2,\jh_2}$
producing the spectral flowed representation $\w_{-1}(\F^-_{p_2,\jh_2})$.
The result is
\ba
& & \langle \F^+_{p_1,\jh_1}(z_1,x_1)\w_{-1}(\F^-_{p_2,\jh_2})(z_2,x_2) \F^+_{p_3,\jh_3}(z_3,x_3)
\rangle    \\
&=&  \frac{1}{L!} \left ( \frac{\g(p_1)}{\g(p_2)\g(p_1-p_2)} \right )^{\frac{1}{2} + L}
\frac{\left | x_2^L e^{-p_2x_1x_2 + p_2x_2(x_1-x_3)z_{12}z_{32}z^{-1}_{13}} \right |^2}
{|z_{12}|^{2(h_1+h_2-h_3)} |z_{13}|^{2(h_1+h_3-h_2)} |z_{23}|^{2(h_2+h_3-h_1)}}
\nb  \ ,
\label{le14}
\ea
where
\be
h_2 = \jh_2(1+p_2) +\frac{p_2}{2}(1-p_2) - L \ ,
\label{le15}
\ee
is the dimension of
a state in the representation $\w_{-1}(\F^-_{p_2,\jh_2})$
obtained by acting $L$ times with $P_{1}^+$ on the ground state. Note
that the three-point coupling is ${\cal C}_{+--}(p_1,1-p_3,p_2)$. This is as expected
since the previous three-point function can be written as
\be
\langle \F^+_{p_1,\jh_1}(z_1,x_1)\w_{-1}(\F^-_{p_2,\jh_2})(z_2,x_2)
\w_{1}(\F^-_{1-p_3,\jh_3})(z_3,x_3) \rangle   \ ,
\label{le16}
\ee
and then related to a three-point function of the form $<+-->$
between highest-weight representations. Moreover $L \ge 0$, as required by the fusion
rules in $(\ref{sf2})$. Finally expanding the exponential we can see that the components
in the $\F^+_{p_1,\jh_1}$ and $\F^+_{p_3,\jh_3}$ representations couple to descendant
fields of decreasing conformal dimension, a distinctive feature of spectral flowed
representations. In the next section we will find the same three-point couplings
studying the factorization of four-point functions between highest-weight representations
on spectral flowed states.

\section{Correlators between spectral-flowed states\label{sfll}}
\setcounter{equation}{0}

In the previous sections we derived all the four-point correlators between
states belonging to the highest-weight representations of the $H_4$ algebra.
In order to complete our discussion we have to consider also correlators between
spectral-flowed states.
Here we first show that spectral-flowed representations have to be added to the operator
content of the model in order to obtain a closed operator algebra, the reason being that
spectral-flowed states appear in the fusion of highest-weight representations
and therefore as intermediate states in the four-point amplitudes.
We then explain how to compute a correlator when the spectral-flowed states
appear as external states.

\subsection{Spectral-flowed states as intermediate states}
\setcounter{equation}{0}

The only correlators  between
highest-weight representations where spectral-flowed states can appear
in an intermediate channel are the $<+-+->$ correlators.
Indeed in a correlator of the form $<+-+->$ the states produced in the $t$-channel
carry $p=p_1+p_3$ which can be bigger than one, even when all the external states
have $p_i < 1$. For all the other types of correlator, having a spectral-flowed
state in an intermediate channel implies that at least one of the external
states carries $p \ge 1$.
In the previous section we explained that in the $s$ and in the $u$ channel a
$<+-+->$ correlator factorizes on highest-weight representations.
Here we want to study more closely the behavior of the correlator
when $z \sim \infty$. We will show that when $p_1+p_3 <1$ the correlator
factorizes on highest-weight representations and that when $p_1+p_3 \ge 1$ the states in the intermediate
channel belong to a spectral-flowed representation.

We recall that for $p_1+p_3 < 1$ we have the following fusion rules
\be
[\F^{\pm}_{p_1,\jh_1}] \otimes [\F^{\pm}_{p_3,\jh_3}] = \sum_{n=0}^\infty
[\F^{\pm}_{p_1+p_2,\jh_1+\jh_2 \pm n}] \ .
\label{lf1}
\ee
From this equation and from  (\ref{sf1}) and (\ref{sf2}) it follows
that when $p_1+p_3 > 1$
\be
[\F^+_{p_1,\jh_1}] \otimes [\F^+_{p_2,\jh_2}] = \sum_{n=0}^\infty [\w_1(\F^+_{p_1+p_2-1,\jh_1+\jh_2-n})] \ .
\label{sf3}
\ee

In the following analysis we use the same notation as in section 5.2 for $<+-+->$ correlators,
changing however variable in the KZ equation to $w = \frac{1}{z}$ so that it can be
written as
\ba
w(1-w)\p_w {\cal A} &=& w \left [ x\p^2_x +\left ( ax+1-k \right )\p_x
+\frac{x}{4}(a^2-b^2) + \ka_{12} \right ] {\cal A}  \nb \\
&+& \left [ -2ax \p_x +\frac{x}{4}(b^2-c^2)  - \ka_{12} - \ka_{14} \right ] {\cal A}  \ .
\label{lf2}
\ea
Let us study the behavior of the solutions to $(\ref{lf2})$
for $w \sim 0$. Using the OPE, the small $w$ behavior of the four-point function is
\be
\sum_{n=0}^\infty |w_{13}|^{2(h_1+h_3)} |w_{24}|^{2(h_2+h_4)} |w|^{2\d_n} \ ,
\label{lf3}
\ee
where the $\d_n$ are the conformal dimension of the intermediate states.
When $p_1+p_3 < 1$ and $L \le 0$ the intermediate states are in the representation
$V^+_{p_1+p_3,\jh_1+\jh_3+n-L}$ and $\d_n = h(p_1+p_3, j_1+j_3+n-L)$.
We decompose ${\cal A}$ in conformal blocks  ${\cal A} \sim \sum_{n=0}^\infty {\cal F}_n(w,x)
{\cal F}_n(\bar{w},\bar{x})$ and for $w \sim 0$ we set ${\cal F}_n \sim w^{\a_n}f_n(x)$ where
\be
\a_n = -n(p_1+p_3)-(1-L)p_3 \ , \hspace{1cm} n \ge 0 \ .
\label{lf4}
\ee
The KZ equation fixes the form of $f_n(x)$ and we get
\be
f_n(x) = x^n e^{\frac{x}{4}(p_1-2p_2-p_3) +\frac{xp_2p_3}{p_1+p_3}} \ .
\label{lf5}
\ee

Let us compare this result with the small $w$
behavior of the four-point correlator in $(\ref{pmk-corr})$.
In this limit
\ba
u &\sim& a_1 |x||w|^{-p_1-p_3} \ , \hspace{2cm}
\frac{\m_L}{S^{1-L}} \sim a_2 |w|^{-2(1-L)p_3}  \ , \nb \\
q(w) &-& (1-w) \p \ln{S} \ \sim \ \frac{p_1-2p_2-p_3}{4} +\frac{p_2p_3}{p_1+p_3} \ ,
\label{lf6}
\ea
where $a_1$ and $a_2$ are constants and therefore  ${\cal A} \sim
\sum_{n=0}^{\infty} |w|^{2 \a_n}|f_n(x)|^2$,
as expected.

We proceed in the same way when $p_1+p_3>1$, setting  ${\cal F}_n \sim w^{\a_n}e^{\frac{\b x}{w}}f_n(x)$
with
\be
\a_n = n(p_1+p_3)-(1-L)(1-p_1)-L+N \ ,  \hspace{1cm} n \ge 0 \ .
\label{lf7}
\ee
Let us call $|\sigma>$ the state obtained as the image under spectral flow of the ground state of the
$V^+_{p_1+p_3,\jh_1+\jh_3-n}$ representation.
In the previous equation we introduced an integer $N$ to take into account that
the conformal dimension of the intermediate state may differ  from
$h_{\s}$ by some integer due to the action of the current modes.
Note that $N \in \mathbb{Z}$ since we are dealing with spectral-flowed
representations.
The KZ equation requires $\b=p_1+p_3-1$ and
\be
f_n(x) = x^{n+\frac{N+2n-L}{p_1+p_3-2}}
e^{x \left( \frac{p_1-2p_2-p_3}{4} +\frac{(1-p_1)(1-p_2)}{2-p_1-p_3}
+(1-p_1) \right )} \ .
\label{lf8}
\ee
We recover an integer power in $x$ only for $N=-2n+L$. Since the intermediate state
now belongs to $\w_1(V^+_{p_1+p_3,j_1+j_3-n})$ we should interpret this result
as saying that the external states do not couple directly to $|\s>$
but to a state of the form $(P^-_1)^{-N}|\s>$ whose conformal dimension is
given by $h = h_\s + N$.

Let us compare again this result with
the behavior of the correlator when  $p_1+p_3 > 1$ using
\ba
& & u \sim b_1 |x||w|^{-2+p_1+p_3} \ , \hspace{1cm}
\frac{\n_L}{S^{1-L}} \sim b_2 |w|^{-2(1-L)(1-p_1)}  \ , \\
& & q(w) -(1-w) \p_w \ln{S}  \sim  \frac{p_1-2p_2-p_3}{4} +\frac{(1-p_1)(1-p_2)}{2-p_1-p_3}
+(1-p_1) + \frac{p_1+p_3-1}{w} \ , \nb
\label{lf9}
\ea
where $b_1$ and $b_2$ are constants.
Note first of all that from the last line in the previous set of equations
it follows that there
is a whole tower of states with arbitrary negative conformal dimension,
as expected for a spectral-flowed representation.
Moreover
\be
\d_n = h(p_1+p_3, j_1+j_3-n) -2n +L \ ,
\label{lf10}
\ee
that is the conformal dimension we expect according to (\ref{sf3}) up to
a negative integer $N = L-2n$.

That the intermediate state has to be a descendant is a simple consequence of
the conservation of the charge $J_0$. To see this explicitly consider
the following correlator between component fields
\be
<R^+_{p_1,\jh_1;r_1}(z_1)R^-_{p_2,\jh_2;r_2}(z_2)
R^+_{p_3,\jh_3;s_1}(z_3)R^-_{p_4,\jh_4;s_2}(z_4)> \ ,
\label{lf11}
\ee
with  $r_1+s_1=n-L$ and $r_2+s_2=n$. We know
that when $p_1+p_3 < 1$, in the intermediate channel we have
the ground state of the $\F^+_{p_1+p_3,j_1+j_3-n}$ representation.
Because of $J$-conservation, the state flowing in the intermediate channel
when $p_1+p_3>1$ has to be a state in the
conformal family of  $\w_1(V^+_{p_1+p_3,j_1+j_3-n})$
obtained acting on $|\s>$ with an operator carrying a $J$-charge $q$ such that
$q = 2n - L$. Since $P^-_1$ acts non trivially on $|\s>$
the natural candidate is
$(P^-_1)^{2n-L}|\s >$, which has both the correct
charge and dimension to match the behavior of the four-point function.

\subsection{Correlators with spectral-flowed states as external states}
\setcounter{equation}{0}

The vertex operators $\F^a_q(z,x,\bar{z},\bar{x})$ we used
so far collect in a single field an infinite number of
component fields generated starting from a highest (lowest) weight $R^a_{q;0}(z.\bar{z})$
and acting on it with the raising and lowering modes $P_0^{\pm}$.
Correlators between spectral-flowed states are however more easily discussed
using directly the component fields. As we explained in section 3,
if we bosonize the currents $J \sim \p v$ and $K \sim \p u$ we can write
\be
\F^{\pm}_{p,\jh}(z,x) = \sum_{n=0}^\infty e^{i(\jh \pm n)u} \tilde{R}^{\pm}_{p,\jh;n}(z)
\frac{(x \sqrt{p})^n}{\sqrt{n!}} \ ,
\label{lf13}
\ee
and similarly
\be
\F^{0}_{s,\jh}(z,x) = \sum_{n \in \mathbb{Z}} e^{i(\jh + n)u} \tilde{R}^{0}_{s,\jh;n}(z)
x^n \ .
\label{lf14}
\ee
Here the operators $\tilde{R}^a_{q;n}(z)$ carry zero $J$-charge and
are related to the component fields in $(\ref{lb9})$ simply by
$R^a_{q;n}(z)= e^{i n u}\tilde{R}^a_{q;n}(z)$.

Spectral flow by $w$ units is then represented as multiplication by $e^{iwv}$
and every correlator can be obtained by first contracting the exponentials
and then using our previous results for correlators between the highest-weights.
Therefore, up to resorting to component fields and to correlators between descendants of
highest-weight states (that can be obtained from the Ward identities)
we can compute correlators for arbitrary external states.
We note that one has to choose the amount of spectral flow in such a way that
the sum of fractional and the integer part of $p$ is conserved.

Let us consider as an explicit example
a correlator between four states carrying momenta
$p_i = \hat{p}_i+w_i$ with $0< \hat{p}_i <1$, $w_i \in \mathbb{N}$ and satisfying
\be
\hat{p}_1+\hat{p}_3=\hat{p}_2+\hat{p}_4 \ , \hspace{1cm}
w_1+w_3=w_2+w_4 \ .
\label{lf15}
\ee
The correlator is therefore of the form
\be
<\w_{w_1}(\F^+_{\hat{p}_1,\jh_1})\w_{-w_2}(\F^-_{\hat{p}_2,\jh_1})\w_{w_3}(\F^+_{\hat{p}_3,\jh_3})
\w_{-w_4}(\F^-_{\hat{p}_4,\jh_4})> \ .
\label{lf16}
\ee
Let us restrict our attention to the correlator between the
ground states of the spectral-flowed representations, assuming
for simplicity $L = \sum_i \jh_i = 0$.
Therefore, all we have to do is to compute the contraction of the exponentials in
\be
< e^{i(w_1v_1+\jh_1u_1)}e^{i(-w_2v_2+\jh_2u_2)}e^{i(w_3v_3+\jh_3u_3)}e^{i(-w_4v_4+\jh_4u_4)}
R^+_{\hat{p}_1,\jh_1;0}R^-_{\hat{p}_2,\jh_2;0}R^+_{\hat{p}_3,\jh_3;0}R^-_{\hat{p}_4,\jh_4;0}> \ ,
\label{lf17}
\ee
and then use $(\ref{pmk-corr})$.
The result can be written as
\be
\sqrt{C_{12}C_{34}} \prod_{i<j}^4 |z|_{ij}^{2 \left ( \frac{h}{3}-h_i-h_j \right) }
\frac{|z|^{2 \tilde{\ka}_{12}}
|1-z|^{2 \tilde{\ka}_{34}}}{S(z,\bar{z})} \ ,
\label{lf18}
\ee
where the function $S$ is as defined in $(\ref{ld25})$ and
\ba
\ka_{12} &=& h_1+h_2-\frac{h}{3}+\hat{p}_1 \hat{p}_2-\jh_2 (\hat{p}_1+w_1) + \jh_1 (\hat{p}_2+w_2)
-\hat{p}_2 \ ,  \nb \\
\ka_{14} &=&  h_1+h_4-\frac{h}{3}+\hat{p}_1\hat{p}_4-\jh_4(\hat{p}_1+w_1)+\jh_1(\hat{p}_4+w_4)-\hat{p}_4 \ .
\label{lf18b}
\ea
are similar to the analogous quantities defined in $(\ref{ka-pm})$. Note that here the
conformal dimensions $h_i$ are the conformal dimensions of the spectral-flowed states.

Another interesting example is provided by a correlator of the form $<+++->$ with $p_1+p_2+p_3=p_4$.
When all the $p_i$ are less than one,
this correlator can be decomposed in the sum of a finite number of conformal blocks.
However when $p_4 > 1$  it can be written as
\be
<\w_{1}(\F^-_{1-p_1}) \ \w_{1}(\F^-_{1-p_2}) \  \F^+_{p_3}  \ \w_{-2}(\F^+_{2-p_4})> \ ,
\label{lf20}
\ee
and therefore it is related to
a correlator of type $<--++>$, with an infinite number of blocks.
Similarly when $p_4 =1$ we can relate it to a $<++- \ 0>$ correlator
\be
<\F^+_{p_1} \ \F^+_{p_2} \ \w_{1}(\F^-_{1-p_3}) \ \w_{-1}(\F^0_{s_4})> \ .
\label{lf200}
\ee

\section{String amplitudes}
\setcounter{equation}{0}

We can combine the Nappi-Witten gravitational wave and  six flat coordinates
as well as the associated world-sheet fermions in order to
describe the superstring theory of the NS5 Penrose limit.

However for the purpose of studying the structure of the S-matrix
elements a non-supersymmetric version will suffice.
We will thus dress the NW theory with 22 flat coordinates
to obtain a critical string theory background
${\cal C} = {\cal C}_{H_4} \times {\cal C}_{int} \times {\cal C}_{gh}$
with
${\cal C}_{int} = \mathbb{R}^{22}$.
Consequently,
the internal part of a vertex operator is given by an exponential $e^{i \vec{p} \vec{X}}$ with
$h = \frac{\vec{p}^2}{2}$.
Representations of the $H_4$ current algebra contain negative-norm states
but once we impose the Virasoro constraint
\be
(L_n - \d_{n,0})|\psi> = 0 \ , \hspace{1cm} n \ge 0 \ ,
\label{lg1}
\ee
on the string states $|\psi>$, the norm in the physical Hilbert space
is positive definite, provided that the component in the $H_4$ CFT of the
state $|\psi>$ belongs to a highest-weight representation of $H_4$ with $|p|<1$
or to some spectral-flowed image of it.
Besides the mass shell condition $(L_0-1)|\psi>=0$ we also have to impose
the level matching condition $L_0 = \bar{L}_0$ on physical states.
For this purpose it is convenient to
introduce two helicity quantum numbers defined as follows
\be
\l = n_- - n_+ \ , \hspace{1cm} \bar{\l} = \bar{n}_- - \bar{n}_+ \ ,
\label{lg2}
\ee
where $n_{\pm}$ ($\bar{n}_{\pm}$) denote the number of modes of the $P^{\pm}$ ($\bar{P}^{\pm}$)
currents that are necessary to create the given state. The $J$-eigenvalue of
a generic state can then
be written as $j = \jh + \m \l$ and $\bar{j} = \jh + \m \bar{\l}$.
It is important to notice that for highest-weight
$\F^+$ representation $n_-$ contains a zero-mode
part, due to the action of $P_0^-$ on the ground state,
while for highest-weight $\F^-$ representation it is $n_+$ that contains a zero-mode
part, due to the action of $P_0^+$ on the ground state.
On the other hand for highest-weight
$\F^0$ representations both $n_+$ and $n_-$ contain a zero-mode contribution.

The dimension of a generic state belonging to a $\w_{w}(\F^{\pm}_{p,\jh})$
representation is
\be
h^w_{\pm}(p,\jh) =  \mp \jh \left ( p + \frac{w}{\m} \right )
+ \frac{\m p}{2}(1-\m p) +\frac{\vec{p}{}^2}{2} \mp w \l + N  \ ,
\label{lg3}
\ee
where $w \in \mathbb{N}$
is the amount of spectral flow, $\vec{p}$ the momentum in the additional
$22$ directions and $N$ the level before the spectral flow.
For $\w_w(\F^0_{s,\jh})$ representations we have
\be
h^w_{0}(s,\jh) = - \jh \frac{w}{\m} + \frac{s^2}{2} + \frac{\vec{p}{}^2}{2} -w \l+ N  \ ,
\label{lg4}
\ee
where now $w \in \mathbb{Z}$.
We can compare this spectrum with the spectrum of
a scalar field in the gravitational wave background (\ref{lab1}).
In radial coordinates the wave-functions have the form
\be
\psi_{p^-,p^+,n,m}=e^{i(p^-u+p^+v)}f_{n,m}(\r,\f) \ ,
\label{lg6}
\ee
where
\be
f_{n,m}(\r,\f) = \left ( \frac{n!}{2 \pi (n+|m|)!} \right )^{\frac{1}{2}}
e^{i m \f} e^{-\frac{\xi}{2}} \xi^{\frac{|m|}{2}} L_n^{|m|}(\xi) \ ,
\label{lg7b}
\ee
with $\xi = \frac{\m p \r^2}{2}$ and $n \in \mathbb{N}$, $m \in \mathbb{Z}$.
The functions $f_{n,m}$ solve the equation
\be
\left ( \p_\r^2 +\frac{1}{\r^2} \p^2_{\f}+\frac{1}{\r} \p_\r  +
2p^-p^+-\frac{(\m p^+)^2}{4}\r^2 \right ) f_{n,m}(\r,\f) = M^2 f_{n,m}(\r,\f)  \ .
\label{lg7a}
\ee
and the spectrum is
\be
-M^2 = -2p^-p^+ + \m \left | p^+ \right | (2n+|m|+1) \ .
\label{lg7c}
\ee
Comparing this result with $h + \bar{h}$ when $N=\bar{N}=0$
we can identify
\be
p^- =  2 \jh + \m(\l + \bar{\l}) \ , \hspace{1cm}
p^+ = p + \frac{w}{\m} \ , \hspace{1cm}  n_{\pm} + \bar{n}_{\pm} = 2n + |m| \ .
\ee
The main difference between the two spectra, (besides the term quadratic in $\m$ which is higher
order in $\alpha'$ and thus not visible in the field theory limit) ,
is that $\m p^+$ has to be separated in a fractional and an integer part,
respectively $\m p$ and $w$, the integer part corresponding to the amount of spectral flow.
As we explained, this is due to the fact that when the ``magnetic'' length
of the wave $(\m p)^{-1}$ becomes of the same order as the string length $\a^{'}$,
the stringy nature of the fundamental excitations becomes essential even at the
semiclassical level. This is at the origin of the quasi-periodic structure we observed
in $p^+$.

\subsection{Three-point amplitudes}
\setcounter{equation}{0}

In this section we use the three-point functions of the $H_4$ WZW model
to discuss the behavior of string amplitudes in the corresponding gravitational wave
background.
The string three-point couplings are directly related to the three-point couplings of the
CFT on the world-sheet, multiplied by a further
$\d$-function imposing momentum conservation in the transverse directions.
They can be written  as
\ba
& & <\F^a_{q_1,\vec{p}_1}(x_1,\bar{x}_1)\F^b_{q_2,\vec{p}_2}(x_2,\bar{x}_2)
\F^c_{q_3,\vec{p}_3}(x_3,\bar{x}_3)> \nb \\
&=&  (2 \pi)^{22} \d(\vec{p}_1+\vec{p}_2+\vec{p_3})
{\cal C}_{abc}(q_1,q_2,q_3)D_{abc}(x_1,x_2,x_3;\bar{x}_1,\bar{x}_2,
\bar{x}_3) \ ,
\ea
where we are using the same notation as in section 5 and therefore
$q$ denotes the collection of the $H_4$
quantum numbers while the additional label $\vec{p}$ stands for the momentum carried by the
vertex operator in the remaining $22$ directions.
As we change  the quantum numbers of the external states, we have
to be careful to use the correct three-point couplings.
Consider for instance the coupling (we discard the dependence on the
transverse momenta $\vec{p}_i$ and on the group variables)
\be
{\cal C}_{+--}(p_1,\jh_1;p_2,\jh_2;p_1-p_2,\jh_1+\jh_2-n) =
\frac{1}{n!} \left [ \frac{\g(p_1)}{\g(p_2)\g(p_1-p_2)}
\right ]^{\frac{1}{2}+n} \ .
\label{lg13}
\ee
This expression, valid for $p_1>p_2$, when naively continued to $p_1 < p_2$
becomes imaginary. However
we showed in the previous sections that the coupling for $p_1<p_2$ is
\be
{\cal C}_{+--}(p_1,\jh_1;p_2,\jh_2;p_2-p_1, \jh_1+\jh_2+n) =
\frac{1}{n!} \left [ \frac{\g(p_2)}{\g(p_1)\g(p_2-p_1)}
\right ]^{\frac{1}{2}+n} \ ,
\label{lg15}
\ee
which is real and can be obtained from the previous one substituting $p_i$ with $1-p_i$.
Moreover when $p_1=p_2$ the coupling is
\be
{\cal C}_{+-0}(p,s) = e^{\frac{s^2}{2}[\psi(p)+\psi(1-p)-2\psi(1)]} \ .
\label{lg14}
\ee

We may use these couplings to investigate the stability under
three-point decay of the states of the theory.
This is however a bit involved since there are several issues to
be clarified, namely the correspondence of the spectral-flow
images of the massless sector to the supergravity states, and the
associated gauge-invariance. We will leave this analysis for a
future publication.

\subsection{The superstring theory}
\setcounter{equation}{0}

The $H_4$ conformal current algebra, (\ref{c-ope})
\be
J^a(z)J^b (w)={g^{ab}\over (z-w)^2}+{f^{ab}}_c{J^c(w)\over
z-w}+{\rm RT}
\ee
where $a=1,2,3,4$ corresponding to the $P_1,P_2,J,K$ generators,
 must be supplemented,  with four free fermions $\psi^a$ with
\cite{kkl}
\be
\psi^a(z)\psi^b(w)={g^{ab}\over z-w}+{\rm RT}
\ee
where
\be
g=\left(\begin{matrix} 1&&0&&0&&0 \\ 0&&1&&0&&0\\ 0&&0&&0&&1\\0&&0&&1&&0
\end{matrix}\right)
\ee
is the invariant metric of the current algebra.

The super-current is
\be
G_4=E_{ab}\psi^a J^b-{1\over 6}f_{abc}\psi^a\psi^b\psi^c
\label{superc}\ee
with
\be
E=\left(\begin{matrix} 1&&0&&0&&0 \\ 0&&1&&0&&0\\ 0&&0&&0&&1\\0&&0&&1&&{1\over 2}
\end{matrix}\right)
\ee
and the only non-zero component of $f_{abc}$ is $f_{12K}=1$.
Indices are raised and lowered with $g^{ab}$.

The theory is supplemented by 6 extra coordinates $x^{\mu}$
and their associated free fermions $\psi^{\mu}$
so that the total super-current is
\be
G=G_4+\psi^{\mu}\p x^{\m}
\ee
The tachyon vertex operator is $V_T=R V_{k}$ where $R$ is an $H_4$
primary and $V_k=e^{ik\cdot x}$ is the standard free vertex
operator.

The (holomorphic) massless vertex operators in the minus-one
picture are (up to the standard ghost factors)
\be
\zeta_a {\cal V}^{-1}_a=\zeta_a\psi^a R~V_k\sp \e_{\mu}{\cal
V}^{-1}_{\mu}=\e_{\mu}\psi^{\mu}R~V_k
\label{minusp}\ee
satisfying the transversality conditions
\be
(\zeta_a ~{E^{a}}_{b}~T^b)R=0\sp \e_{\mu}k^{\mu}=0
\ee
where $T^a$ are the $H_4$ algebra generators in the appropriate
representation .

The operators of the zero picture are obtained to be
\be
{\cal V}^{0}_a=\left[{E^{a}}_{b}J^b-E_{bc}\psi^a\psi^b T^c-{1\over
2}{f^{a}}_{bc}\psi^b\psi^c-i\psi^a (k\cdot \psi)\right]RV_k
\ee
\be
{\cal V}^{0}_{\mu}=\left[\p x^{\mu}-i(k\cdot
\psi)\psi^{\mu}-\psi^{\m}E_{ab}\psi^a T^b\right]RV_k
\ee

We may now calculate the holomorphic part of the three-point
amplitudes.
\be
T_3(\e_1,\e_2,\e_3)=z_{13}z_{23}\e_1^{\mu}\e_2^{\nu}\e_3^{\rho}\langle
{\cal V}^{-1}_{\mu}{\cal V}^{-1}_{\nu}{\cal V}^{0}_{\rho}\rangle
\ee
where the prefactor is due to the bosonic and fermionic ghosts.
We obtain by direct computation (using the transversality
conditions, momentum conservation along $R^6$ and the global $H_4$
Ward identities
\be
T_3(\e_1,\e_2,\e_3)=-i\left[(\e_1\cdot \e_2)(\e_3\cdot k_{12})+(\e_2\cdot \e_3)(\e_1\cdot
k_{23})+(\e_1\cdot \e_3)(\e_2\cdot k_{31})\right]\times
\ee
$$\times (2\pi)^6\delta^{(6)}(k_1+k_2+k_3)
{\langle R_1R_2R_3\rangle\over
z_{12}^{-k_1\cdot
k_2}z_{13}^{-k_1\cdot
k_3}z_{23}^{-k_2\cdot
k_3}}
$$
Similarly, we obtain
\be
T_{3}(\e_1,\e_2,\zeta)=(2\pi)^6\delta^{(6)}(k_1+k_2+k_3)(\e_1\cdot\e_2)(\zeta_{a}{E^{a}}_{b}T_1^b)
{\langle R_1R_2R_3\rangle\over
z_{12}^{-k_1\cdot
k_2}z_{13}^{-k_1\cdot
k_3}z_{23}^{-k_2\cdot
k_3}}
\ee
\be
T_{3}(\zeta_1,\zeta_2,\e)=i(2\pi)^6\delta^{(6)}(k_1+k_2+k_3)(\zeta_{1,a}g^{ab}\zeta_{2,b})
(k_2\cdot \e){\langle R_1R_2R_3\rangle\over
z_{12}^{-k_1\cdot
k_2}z_{13}^{-k_1\cdot
k_3}z_{23}^{-k_2\cdot
k_3}}
\ee
\be
S_{3}(\zeta_1,\zeta_2,\zeta_3)=(2\pi)^6\delta^{(6)}(k_1+k_2+k_3)\left[(\zeta_{1,a} g^{ab} \zeta_{2,b})
(\zeta_{3,c} {E^{c}}_{d} T_{12}^d)+(\zeta_{2,a} g^{ab} \zeta_{3,b})
(\zeta_{2,c} {E^{c}}_{d} T_{23}^d)+
\right.\ee
$$
+\left.(\zeta_{1,a} g^{ab} \zeta_{3,b})
(\zeta_{2,c} {E^{c}}_{d} T_{31}^d)+\zeta_{1,a}\zeta_{2,b}\zeta_{3,c} f^{abc}\right]
{\langle R_1R_2R_3\rangle\over
z_{12}^{-k_1\cdot
k_2}z_{13}^{-k_1\cdot
k_3}z_{23}^{-k_2\cdot
k_3}}
$$
where $T_{ij}^a\equiv T_i^a-T_j^a$.

It is convenient to  define the following tensors
\be
V^{\m\n\r}=-i(\eta^{\m\n}k_{12}^{\r}+\eta^{\n\r}k_{23}^{\m}+\eta^{\m\r}k_{31}^{\n})
\ee
\be
V^{\m\n a}=\eta^{\m\n}{E^{a}}_{b}T^b_1\sp
V^{ab\m}=i~g^{ab}k_2^{\mu}
\ee
\be
V^{abc}=g^{ab}{E^{c}}_{d}T_{12}^{d}+g^{bc}{E^{a}}_{d}T_{23}^{d}+g^{ac}{E^{b}}_{d}T_{31}^{d}+f^{abc}
\ee

The ``massless" string states are of three-types:

(i) Those that have
both indices in the pp-wave part, with polarization tensors
$\zeta_{ab}$.

(ii) Those that have both indices in the $\Real^6$ part with
polarization tensors $\e_{\m\n}$.

(iii) Those that have one index in the pp-wave part and another in
the $\Real^6$ part with
polarization tensors $\xi_{\m a}$ and $\tilde\xi_{a\m}$.

The three-point S-matrix element can now be obtained,
\be
S_3(\e^1,\e^2,\e^3)=(2\pi)^6\delta^{(6)}(k_1+k_2+k_3)\e^1_{\m\tilde \m}\e^2_{\n\tilde \n}\e^3_{\r\tilde \r}
V^{\m\n\r}V^{\tilde\m \tilde \n \tilde \r}
\ee
$$\times
{\cal C}_{abc}(q_1,q_2,q_3)D_{abc}(x_1,x_2,x_3;\bar{x}_1,\bar{x}_2,
\bar{x}_3)
$$
\be
S_{3}(\e^1,\e^2,\zeta^3)=(2\pi)^6\delta^{(6)}(k_1+k_2+k_3)\e^1_{\m\tilde \m}\e^2_{\n\tilde \n}
\zeta^3_{a\tilde a}
V^{\m\n a}V^{\tilde \m \tilde \n \tilde a}~\times
\ee
$$\times
{\cal C}_{abc}(q_1,q_2,q_3)D_{abc}(x_1,x_2,x_3;\bar{x}_1,\bar{x}_2,
\bar{x}_3)
$$

\be
S_{3}(\e^1,\e^2,\xi^3)=(2\pi)^6\delta^{(6)}(k_1+k_2+k_3)\e^1_{\m\tilde \m}\e^2_{\n\tilde \n}\xi^3_{\r\tilde a}
V^{\m\n\r}V^{\tilde \m \tilde \n \tilde a}~\times
\ee
$$\times
{\cal C}_{abc}(q_1,q_2,q_3)D_{abc}(x_1,x_2,x_3;\bar{x}_1,\bar{x}_2,
\bar{x}_3)
$$

and so on.

So far, the above applies to non-spectral flowed states.
In the case of the superstring, the spectral flow on the bosonic
manifold, has to be accompanied with an associated spectral flow on
the free fermionic partners:

\be
\psi^{\pm}_r\to \psi^{\pm}_{r\mp w}\sp \psi^{J,K}_r\to \psi^{J,K}_r
\label{sff}\ee
so that the total supercharge (\ref{superc}) is invariant. This
should not confused with the fermionic spectral flows on all the
fermions (there are several since the model has N=4 superconformal
symmetry \cite{kkl}) and which implement the space-time
supersymmetry.

The spectral flow of the fermions (\ref{sff}) guarantees that the
-1 vertex operator of a spectral-flowed rep is similar to the usual case
(\ref{minusp}).

We shall now use these couplings to
investigate on-shell processes in the "massless" sector
that can happen at the 3-point level.
Let us start with the decay of a state belonging to $\w_{w}(\F^+_{\hat{p},\jh,\vec{p}})$ into two other
states belonging respectively to $\w_{w_1}(\F^+_{\hat{p}_1,\jh_1,\vec{p}_1})$ and
$\w_{w_2}(\F^+_{\hat{p}_2,\jh_2,\vec{p}_2})$. It is useful to discuss separately
processes where the amount of spectral flow is conserved and processes
where it is violated. In the first case
$\hat{p}=\hat{p}_1+\hat{p}_2$ and $w=w_1+w_2$. The three-point coupling vanishes
unless $\jh - \jh_1 - \jh_2 \ge 0$
and since $\jh + n = \jh_1+n_1 + \jh_2+n_2$,
where $n$, $n_1$ and $n_2$ are non-negative integers, we also have
$n_1+n_2-n \ge 0$.
Using the mass-shell conditions, we can solve for $\jh$,
$\jh_1$ and $\jh_2$ and we obtain the following constraint
\be
\frac{\hat{p}(1-\hat{p}+2n)}{p}-\frac{\hat{p}_1(1-\hat{p}_1+2n_1)}{p_1}
-\frac{\hat{p}_2(1-\hat{p}_2+2n_2)}{p_2}
=\frac{p_2}{p_1 p} \left (  \vec{p}_1 - \frac{p_1}{p_2} \vec{p}_2 \right )^2 \ ,
\label{lg155}
\ee
where $p_i = \hat{p}_i + w_i$.
When $w=w_1=w_2=0$ the lhs of the previous equation is always negative and therefore
the states are stable with respect to the decay in this channel.
On the other hand the decay is possible when the amount of spectral flow is non zero.
It is also possible
in the second case, when we have
$\hat{p} = \hat{p}_1+\hat{p}_2-1$ and $w = w_1+w_2+1$, since then
the three-point coupling is non zero for $\jh - \jh_1 - \jh_2 \le 0$
and therefore $n_1+n_2 \le n$.

We can discuss the decay
$\w_{w}(\F^+_{\hat{p},\jh,\vec{p}})
\rightarrow \w_{w_1}(\F^+_{\hat{p}_1,\jh_1,\vec{p}_1}) + \w_{w_2}(\F^-_{\hat{p}_2,\jh_2,\vec{p}_2})$
along similar lines. The constraint is
\be
\frac{\hat{p}_1(1-\hat{p}_1+2n_1)}{p_1}-\frac{\hat{p}(1-\hat{p}+2n)}{p}-
\frac{\hat{p}_2(1-\hat{p}_2+2n_2)}{p_2}
=\frac{p_2}{p_1 p} \left (  \vec{p} + \frac{p}{p_2} \vec{p}_2 \right )^2 \ ,
\label{lg156}
\ee
and the two possible cases $\hat{p}+\hat{p}_2=\hat{p}_1$, $w+w_2=w_1$ with $n_1-n_2-n \le 0$
and $\hat{p}+\hat{p}_2=1+\hat{p}_1$, $w+w_2=w_1-1$ with $n_1-n_2-n \ge 0$.
Only when $w=w_1=w_2=0$ the decay is forbidden, otherwise it is allowed.

Therefore also state with $w=0$ can decay through this channel.

Another possible process is  $\w_{w}(\F^0_{s,\jh,\vec{p}})
\rightarrow \w_{w_1}(\F^+_{\hat{p}_1,\jh_1,\vec{p}_1}) + \w_{w_2}(\F^{+}_{\hat{p}_2,\jh_2,\vec{p}_2})$,
with $\hat{p}_1+\hat{p}_2=1$ and $w_1+w_2+1=w$.
The constraint in this case reads
\be
s^2 = \frac{w\hat{p}_1(1-\hat{p}_1+2n_1)}{p_1}+\frac{w\hat{p}_2(1-\hat{p}_2+2n_2)}{p_2}
+\frac{p_2}{p_1}  \left (  \vec{p}_1 - \frac{p_1}{p_2} \vec{p}_2 \right )^2 \ ,
\ee
and we see that the decay is always possible. The decay
amplitude is proportional to
\be
{\cal A}_{0\to\pm}\sim \exp\left[-{s^2\over
2}(2\psi(1)-\psi(\hat{p})-\psi(1-\hat{p}))\right] \ ,
\label{lg43}
\ee
where the function $2\psi(1)-\psi(\hat{p})-\psi(1-\hat{p})$ is positive
for $0<\hat{p}<1$, with a minimum at $\hat{p}=1/2$ equal to 2.77259 and
diverges as $\hat{p}\to 0$ as $1/\hat{p}$.
Therefore a long string has
an appreciable decay rate only
into states carrying $\hat{p} \sim \frac{1}{2}$
and this gives a lower bound on $s^2$, namely
\be
\frac{s^2}{2} \ge \frac{w^2}{\m (2w_1+1)(2w_2+1)} \ .
\ee
To obtain the lifetime we need to sum over final states.
This is quite difficult to do directly. The proper method is to
compute the two-point function on the torus of a spectral-flowed
type-0 state and then compute the discontinuity of the cut
diagram, but we do not attempt this computation in the present paper.

The last process we have to discuss is
$\w_{w}(\F^+_{\hat{p},\jh,\vec{p}}) \rightarrow
\w_{w_1}(\F^{+}_{\hat{p},\jh_1,\vec{p}_1}) + \w_{w_2}(\F^0_{s,\jh_2,\vec{p}_2})$,
with $w=w_1+w_2$.
The constraint is
\be
\frac{\hat{p}(1-\hat{p})+2n}{p}-\frac{\hat{p}(1-\hat{p})+2n_1}{p_1}
-\frac{s^2}{w_2}
=\frac{w_2}{p p_1} \left (  \vec{p}_1 - \frac{p}{w_2} \vec{p}_2 \right )^2 \ ,
\label{lg157}
\ee
where now $w_2, n_2 \in \mathbb{Z}$. Again it seems that $w=0$ states can decay
through this channel.

Thus, unlike the flat space case the massless sector states are
unstable at the three-point level. So far we have imposed conservation
of the scalar particle quantum numbers.
There are extra constraints
coming from the kinematical factor of the three-point amplitudes.
For states however with polarization in the $\mathbb{R}^6$ part
there are no further constraints. An interesting outcome is that point-like $w=0$ states can decay to spectral
flowed states corresponding to long strings that are extended in space-time.
  A complete analysis of the full
decay amplitudes is left for a future publication.

\subsection{Four-point amplitudes}
\setcounter{equation}{0}

The four-point string amplitudes are given by the CFT four-point correlators
computed in section 5 and integrated on the world-sheet.
To simplify our discussion we consider four-point amplitudes for
tachyons in the bosonic case, since they capture the essential
ingredients of four-point scattering in pp-wave space-times.

As we explained, correlators between primary fields in the $H_4$ WZW model can be factorized, for
particular choices of their quantum numbers, either on the discrete or in the
continuous series. For each level we have, in the first case, a sum over a finite
or an infinite number of
states while in the second case an integral over a continuum of states.
Moreover, the amplitudes can also factorize on spectral-flowed states
and therefore long strings can propagate in the intermediate channels.
In order to study
which kind of singularity appears in the string
correlators, we have
first of all to include the contribution of the ghosts and
of the primary fields of the internal CFT.
Let $\vec{p}_i$ be the internal momentum of the vertex operator
inserted in $z_i$ and let
us introduce
\be
\s_{12} = \ka_{12} +\vec{p}_1  \vec{p}_2 \ , \hspace{1cm}
\s_{14} = \ka_{14} +\vec{p}_1  \vec{p}_4 \ ,
\label{lg16}
\ee
where the $\ka_{ij}$ are defined as in (\ref{ka-pp}), (\ref{ka-pm}), (\ref{ld41})and
(\ref{ld50})
but now using for the $h_i$ the complete conformal dimension of the states.
The string amplitude can then be written in general as
\be
{\cal A}_{string} = \int d^2 z |z|^{2 \s_{12} - \frac{4}{3}}|1-z|^{2 \s_{14} - \frac{4}{3}}
K(x_i,\bar{x}_i) {\cal A}(z,\bar{z};x,\bar{x}) \ ,
\label{lg17}
\ee
where $K(x_i,\bar{x}_i)$ is the part of the four-point correlator fixed by
the Ward identities for the global $H_4$ symmetry and
${\cal A}(z,\bar{z};x,\bar{x})$ is the non-trivial part of the $H_4$ four-point functions,
as in (\ref{gs-ci}).
The term $|z(1-z)|^{-4/3}$ results from the ghost contribution and
from the prefactor $\prod_{j>i=1}^4 z^{\frac{h}{3}-h_i-h_j}$ once
all external states are on shell, $h_i=1$.
In the following expressions we will omit the factor $K(x_i,\bar{x}_i)$.

Consider for comparison the four tachyon amplitude in flat space, given
by the well-known Shapiro-Virasoro formula
\ba
{\cal A}_{string} &=& (2\pi)^{26} \d \left ( \sum_{i=1}^4 p_i \right ) \int
d^2z~
|z|^{\frac{\a^{'}}{2}(p_1+p_2)^2-4}
|1-z|^{\frac{\a^{'}}{2}(p_2+p_3)^2-4}  \\
&=& (2\pi)^{26} \d \left ( \sum_{i=1}^4 p_i \right )
\frac{\G \left (\frac{\a^{'}s}{4}- 1\right)
\G \left (\frac{\a^{'}t}{4}- 1\right)\G \left (\frac{\a^{'}u}{4}- 1\right)}
{\G \left (\frac{\a^{'}(s+t)}{4}- 2\right)
\G \left (\frac{\a^{'}(s+u)}{4}- 2\right)
\G \left (\frac{\a^{'}(t+u)}{4}- 2\right)} \ , \nb
\label{lg18}
\ea
where $s = (p_1+p_2)^2$, $t = (p_1+p_3)^2$ and $u = (p_1+p_4)^2$.
In this case the $z$-integral can be done explicitly and the resulting
expression in terms of $\G$ functions makes
the location of the poles manifest. However we can also
expand the integrand around $z = 0$, $z=1$ and $z = \infty$ and check for
its convergence. For instance around $z = 0$ we have
\be
{\cal A}_{string} \sim (2\pi)^{26} \d \left ( \sum_{i=1}^4 p_i \right ) \int
d^2 z~
|z|^{\frac{\a^{'}}{2}(p_1+p_2)^2-4} \ ,
\label{lg19}
\ee
and we recognize that the amplitude has a pole when
\be
\a^{'}(p_1+p_2)^2 = 4 \ .
\label{lg20}
\ee
Keeping higher powers in $z$ in the expansion (\ref{lg19}) one recognizes that
the amplitude has a pole whenever
\be
\a^{'}(p_1+p_2)^2 = 4(1-N) \ , \hspace{1cm} N \in \mathbb{N} \ ,
\label{lg201}
\ee
and therefore the poles in the amplitude, due to the propagation of on-shell states in the
intermediate channel, precisely
match the spectrum of the bosonic string.
An identical set of poles is displayed by the amplitude in the
{\it t} and in the {\it u} channels.

In our case, we do not have closed form expressions for the integrated
correlators and
the study of their singularities is greatly simplified
by the use of the factorized form of the amplitudes
displayed in section 5.
Let us start with the simplest case, string amplitudes involving
$<+++->$ $H_4$ correlators. We can write
\be
\s_{12} = h_{12} - \frac{4}{3} \ , \hspace{1cm}
\s_{14} = h_{14} - \frac{4}{3} - L(p_4-p_1) \ ,
\label{lg21}
\ee
where $ h_{12} = h_+(\jh_1+\jh_2,p_1+p_2) + \frac{(\vec{p}_1+\vec{p}_2)^2}{2}$,
$ h_{14} = h_-(\jh_1+\jh_4,p_4-p_1) + \frac{(\vec{p}_1+\vec{p}_4)^2}{2}$
and therefore
\be
{\cal A}_{string} = \int d^2 z |z|^{2 (h_{12}- 2)}|1-z|^{2 (h_{14} -L(p_4-p_1) - 2)}
{\cal A}(z,\bar{z};x,\bar{x}) \ .
\label{lg22}
\ee
We expand the amplitude for $z \sim 0$ expressing the integrand
in terms of the corresponding conformal blocks
\ba
{\cal A}_{string} &\sim& \int d^2 z |z|^{2 (h_{12}- 2)}
\sum_{n=0}^{|L|} {\cal C}_{++}{}^+(q_1,q_2,n){\cal C}_{+-}{}^-(q_3,q_4,|l|-n) \nb \\
& &
|z|^{-2n(p_1+p_2)}|x|^{2n} \left | 1-\frac{xp_2}{p_1+p_2} \right |^{2(|L|-n)}
\left ( 1 + O(z,\bar{z}) \right ) \ ,
\label{lg23}
\ea
where the higher power of $z$ are due to the descendant fields. We then see
that the amplitude has a pole when the intermediate state is on shell
\be
h_{12}-n(p_1+p_2) = 1 - N \ ,  \hspace{1cm} n= 0, \dots |L| \ ,  \hspace{1cm} N \in \mathbb{N} \ ,
\label{lg25}
\ee
a condition that can be written, reintroducing $\a^{'}$ and the parameter $\m$, as
\be
-(p_1+p_2)(\jh_1+\jh_2+ \m n) + \frac{\m}{2}(p_1+p_2)(1-\m(p_1+p_2))
+\frac{(\vec{p}_1+\vec{p}_2)^2}{2} = \frac{1 - N}{\a^{'}} \ .
\ee

Consider now  a string amplitudes involving a
$<+-+->$ $H_4$ correlator. When $p_1>p_2$
\be
\s_{12} = h_{12} - \frac{4}{3} \ , \hspace{1cm}
\s_{14} = h_{14} - \frac{4}{3} \ ,
\label{lg27}
\ee
where $ h_{12} = h_+(\jh_1+\jh_2,p_1-p_2) +  \frac{(\vec{p}_1+\vec{p}_2)^2}{2}$,
$ h_{14} = h_+(\jh_1+\jh_2,p_1-p_4) + \frac{(\vec{p}_1+\vec{p}_4)^2}{2}$ and therefore
\be
{\cal A}_{string} = \int d^2 z |z|^{2 (h_{12}- 2)}|1-z|^{2 (h_{14}- 2)}
{\cal A}(z,\bar{z};x,\bar{x}) \ .
\label{lg28}
\ee
Again when $z \sim 0$ we can write it as
\ba
{\cal A}_{string} &\sim&  \int d^2 z |z|^{2 (h_{12}- 2)} \sum_{n=0}^{\infty}
\frac{ n!^2 }{(p_1-p_2)^{2n}} {\cal C}_{+-}{}^+(q_1,q_2,n){\cal C}_{+-}{}^-(q_3,q_4,n+|L|)
\nb \\
& & |z|^{2n(p_1-p_2)} \left | e^{-\frac{x}{4}(p_1-2p_2-p_3)}
L_n^{|L|}(x(p_2-p_1)) \right |^2 \left ( 1 + O(z,\bar{z}) \right ) \ ,
\label{lg29}
\ea
and we see clearly that we have a pole whenever
\be
h_{12}+n(p_1-p_2) = 1 - N \ ,  \hspace{1cm} n, N \in \mathbb{N} \ .
\label{lg30}
\ee

When $p_1=p_2=p$ and $p_3=p_4=l$, the amplitude factorize on the continuum and
can be written as
\be
{\cal A}_{string} \sim \int d^2 z |z|^{2 (h_{12}- 2)}
\int_0^\infty ds s \ {\cal C}_{+-0}(p,s){\cal C}_{+-0}(l,s)
|z|^{s^2} |x|^L \left | I_{|L|}(s \sqrt{2x}) \right |^2 \left ( 1 + O(z,\bar{z}) \right ) \ ,
\label{cut}
\ee
where now $ h_{12} = \frac{(\vec{p}_1+\vec{p}_2)^2}{2}$.
In this case, as a result
of the coalescence of the poles, the amplitude develops a branch cut.
In order to show its presence explicitly we fix $\s <1$ and
approximate the integral in(\ref{cut}) as follows
\be
{\cal A}_{string} \sim \int_{|z|<\s} d^2z |z|^{-4+2h_{12}+2L}
\Theta^{-1+L} \left | e^{x l z+\frac{x}{\Theta}} \right |^2 \sum_{n=0}^\infty \frac{1}{n!(n+|L|)!}
\left | \frac{x}{\Theta} \right |^{2n} \  ,
\label{lg31}
\ee
where
\be
\Theta = -\ln{|z|^2}-4\psi(1)-\psi(p)-\psi(1-p)-\psi(q)-\psi(1-q) \ .
\label{lg32}
\ee
The integrals in $(\ref{lg31})$ can be expressed in terms
of the Exponential Integral function. When $L=n=0$ for instance
we have
\be
{\cal A}_{string} \sim \int_0^\s dr \frac{1}{r^\d \ln{r}} \ ,
\label{lg33}
\ee
with $\d = 3 - 2h_{12}$. The integral
is convergent for $\d <1$ and can be written as
${\cal A}_{string} \sim Ei((1-\d) \ln{\s})$ so that
in the limit $\d \rightarrow 1^-$ the amplitude behaves as
\be
{\cal A}_{string} \sim \ln{(h-1)} \ ,
\label{lg34}
\ee
and develops a logarithmic branch cut starting from
$h_{12}=1$. Indeed $h_{12}$ is the dimension of an intermediate state
in a $\F^0$ representation with $s=0$, that is carrying zero radial
momentum and it is the presence of the continuum mass spectrum that gives rise to the cut.
Note that in this case only the tachyon can develop a branch
cut because the other string states are never on-shell. However
the same behavior
appears when the intermediate state belongs to a spectral-flowed continuous
representations and in this case there is a branch cut for each string level.
We can see this explicitly
factorizing a correlator $<+-+->$ with  $p_1+p_3=1$ around $z \sim \infty$,
so that the intermediate state belongs to $\w_{1}(\F^0_{s,\jh_1+\jh_3})$.
We proceed as before, setting $w = \frac{1}{z}$ and writing
\be
{\cal A}_{string} \sim \int_{|w|<\s} d^2w |w|^{-4+2h_{13}+2L}
\Theta^{-1+L} \left | e^{\frac{x}{\Theta \ w}} \right |^2 \sum_{n=0}^\infty \frac{1}{n!(n+|L|)!}
\left | \frac{x}{\Theta \ w} \right |^{2n} \  .
\label{lg35}
\ee
We have again an integral of the form
\be
{\cal A}_{string} \sim \int_0^\s dr \frac{1}{r^\d \ln{r}} \ ,
\label{lg36}
\ee
where now $\d = 3 - 2h_{13}$, $h_{13} = -\jh_1-\jh_3+\frac{(\vec{p}_1+\vec{p}_3)^2}{2}$.
Since $p=1$ for the intermediate state, we can have a branch cut
for each string level.

Logarithmic branch cuts appear in S-matrix elements of field
theory, signaling thresholds for massless states. As an example,
for the one-loop propagator in massive $\phi^3$ theory we obtain after
subtracting the logarithmic ultraviolet divergence
\be
G_2\sim {1\over p}\sqrt{p^2-4m^2}~\log{4m^2\over
(p+\sqrt{p^2-4m^2})^2}
\ee
The amplitude exhibits the order-two branch cut signaling the
presence of the
two-particle threshold.
In the limit $m\to 0$ the amplitude behaves as
\be
G_2\to \log{m^2\over p^2}+{\cal O}(m^2)
\ee
and develops  a logarithmic branch cut.
The difference here is that, unlike field theory, the logarithmic
branch cut appears at tree level.

String amplitudes having spectral-flowed states as external
states can be reduced to the amplitudes we
already discussed, up to the shift in the conformal dimension, as we showed in section 8.
So far we have analyzed amplitudes between primary vertex operators which correspond to
the scattering of tachyons in the bosonic string.
Other scattering processes, as those between gravitons,
require the computation of correlation functions between affine descendants
that can be obtained starting from the correlators
displayed in this paper and using the OPE in (\ref{c-ope}).

\section{The flat space limit}
\setcounter{equation}{0}

In this section we discuss how to recover
the flat space string spectrum starting from the Nappi-Witten
gravitational wave.
We first reintroduce the parameter $\m$  in the metric performing a boost
$u \rightarrow \m u$, $v \rightarrow \frac{v}{\m}$
\be
ds^2 = -2 du dv - \frac{\m^2 r^2}{4} du^2 + dx_T^2 \ .
\label{lii0}
\ee
There are two interesting limits to consider, $\m\rightarrow  0$ and
$\m \rightarrow \infty$. We will argue that in both cases one recovers string
theory in flat space, even though the states that survive in the two limits
are very different. Actually when $\m \rightarrow 0$ the spectral flowed states disappear
from the spectrum and flat space is reconstructed by states in the highest-weight
representations of the $H_4$ algebra.
On the other hand, when  $\m \rightarrow \infty$ the situation is reversed:
the spectral flowed states give rise to a continuum in $p$ while the
highest weights decouple. This behavior should be compared with
the one of a compactified boson at radius $R$ in the two limits $R \rightarrow 0$ and
$R \rightarrow \infty$.

Since in most of the previous sections we set $\m = 1$, let us
show explicitly how the current algebra and the conformal dimensions depend on
$\m$. We have
\ba
P^+(z)P^-(w) &\sim& \frac{2}{(z-w)^2} - \frac{2i \m K(w)}{z-w}+{\rm RT} \ , \\
J(z)P^{\pm}(w) &\sim& \mp i \m \frac{P^\pm(w)}{z-w}+{\rm RT} \ , \\
J(z)K(w) &\sim& \frac{1}{(z-w)^2}+{\rm RT} \ ,
\label{lii1}
\ea
and the stress-energy tensor
\be
T={1\over2}\left[{1\over 2}\left(P^+P^-+P^-P^+\right)+2JK+ \m^2 K^2\right] \ .
\label{lii2}
\ee
From the previous expressions it follows that in order
to reintroduce the parameter $\m$ in our formulas
we have to make the following rescalings
\be
p \rightarrow \m p \ , \hspace{1cm}
\jh  \rightarrow \frac{\jh}{\m} \ .
\label{lii3}
\ee
In particular the conformal dimension of a state in a $\w_{\pm w}(\F^{\pm}_{p,\jh})$ representation
becomes
\be
h = \mp \left ( p+\frac{w}{\m} \right ) \jh + \frac{\m p}{2}
\left ( 1 - \m p \right ) \ ,
\label{lii4}
\ee
with  $p \in (0,\frac{1}{\m})$  and $w \in \mathbb{N}$ .
For a state in a $\w_w(\F^0_{s,\jh})$ representation we obtain
\be
h =  -\frac{w \jh}{\m} + \frac{s^2}{2}  \ ,
\label{lii5}
\ee
with $\jh \in [-\m/2,\m/2)$ and $w \in \mathbb{Z}$.

Let us start from the limit  $\m \rightarrow 0$. The metric in $(\ref{lii0})$
reduces to the flat space metric and the current algebra in $(\ref{lii1})$
becomes the current algebra of four free bosons.
This flat-space limit can be considered as a further contraction of
the $H_4$ algebra and it can be studied following the same
approach as in section 2 and in appendix A, considering the
behavior of the three and four-point functions. We do not perform
such a detailed analysis here but we limit ourselves to explaining
how one can recover the usual flat-space vertex operators.
At the intuitive level we expect that the $V^{\pm}$ states will asymptote to plane waves,
since the harmonic oscillator potential flattens in this limit.
Since $0 < p < \frac{1}{\m}$, in the limit
we obtain arbitrary values for $p$ for the highest-weight representations.
Moreover, spectral-flowed states are scaled out of the spectrum.
We can also consider the limit
of the semi-classical wave-functions
for the various components of
a $V^{\pm}_{p,\jh}$ representation. They can be extracted from $(\ref{laa2})$ and read
\be
\F^{\pm}_{p,\jh;n, \bar{n}} =  e^{ \mp ipv + iju - \frac{\m p}{2} \z \tilde{\z}}
\sqrt{\frac{\bar{n}!}{n!}}
e^{\frac{i \m u}{2}(n+\bar{n})} ~ (\sqrt{\m p})^{n-\bar{n}}
~\z^{n-\bar{n}}~
L_{\bar{n}}^{n-\bar{n}}(\m p \r^2) \ ,
\label{lii51}
\ee
where $\z = \r e^{i \f}$, $\tilde{\z} = \r e^{- i \f}$.
We now scale the quantum numbers as follows
\be
\jh = p^- \mp \frac{s^2}{2 p} \ , \hspace{1cm} n = m +
\frac{s^2}{2 p \m} \ ,
\ee
respectively for $V^{\pm}_{p,\jh}$ representations.
Note that in the limit  $\m \rightarrow 0$ we have $m,\bar{m} \in \mathbb{Z}$ and
the conformal dimension is $h = -pp^- + \frac{s^2}{2}$.
The wave-function in $(\ref{lii51})$ becomes
\be
\F^{\pm}_{p,\jh;n, \bar{n}} \rightarrow e^{\mp i p v + i p^- u }
e^{i(m-\bar{m})\f}J_{m-\bar{m}}(\sqrt{2} s \r ) \ ,
\label{lii52}
\ee
and it can be recognized as one of the components of the usual
flat-space vertex operator expanded in a series of Bessel
functions,
\be
e^{i\frac{s}{\sqrt{2}}(\z e^{-i \theta} + \tilde{\z} e^{i \theta})}
= \sum_{k \in \mathbb{Z}} i^k e^{i k (\f-\theta)} J_k( \sqrt{2} s \r) \ .
\ee
Let us finally consider the
limit of the two-point function between a $\F^+$ and a $\F^-$
vertex operator $(\ref{lc5})$. Scaling the quantum numbers as explained
before we obtain
\be
\lim_{\m \rightarrow 0}
\langle
R^+_{p_1,\jh_1;n_1,\bar{n}_1}(z_1,\bar{z}_1)R^-_{p_2,\jh_2;n_2,\bar{n}_2}(z_2,\bar{z}_2)\rangle
\label{lii53}\ee
$$=
p \m \d (p_1-p_2) \d (p^-_1 + p^-_2) \frac{\d (s_1 - s_2)}{s_1} \d_{n,m}  \d_{\bar{n},\bar{m}}
\frac{(-1)^{n+\bar{n}}}{|z_{12}|^{4h}} \ ,
$$
as expected
since the original wave functions were confined in a volume $v \sim \frac{1}{\m p}$
in the transverse plane.

Consider now the case $\m \rightarrow \infty$. From $(\ref{lii4},\ref{lii5})$
it follows that states in spectral flowed continuous representations
have $p = w/\m$, which becomes a continuous variable in the limit, and $\jh \in \mathbb{R}$.
On the other hand,
all operators with $p \notin \mathbb{Z}$ behave as if $\m p \rightarrow 1$
and thus become null vectors and decouple.

Therefore we have two different pictures for the flat space limit.
In the first ($\m \rightarrow 0$) the potential flattens and flat space arises
as suggested by the classical intuition, with the confined
states describing larger and larger orbits until they become free.
In the second less intuitive case ($\m \rightarrow \infty$), flat space is recovered
trough a stringy mechanism, namely the spectral flow. Indeed states with $p \notin \mathbb{Z}$
are so strongly trapped by the wave that they disappear from the spectrum. On the other
hand the spectral flowed states with $p = w/\m$ do not feel the potential, remain free
and form a continuum in the limit.

The presence of the spectral flowed states is due in our case to the coupling between
the string world-sheet and the $NS$ antisymmetric tensor field that supports the wave
and it is precisely for this reason that we can see the long sting states
already in the perturbative string spectrum.

In the case of the RR pp-wave obtained from $AdS_5\times S^5$,
the analogue of the long strings are the giant gravitons and they correspond to
$D$-branes, not to fundamental string states.
We would expect that in the limit $\mu\to\infty$ giant graviton
scattering might reproduce the flat space result.

\section{Comparison with light-cone quantization}
\setcounter{equation}{0}

The light-cone quantization of the NW theory was performed in
\cite{forgacs}.
We briefly review it here in order to compare with the covariant
quantization we presented.

We start from the $\s$-model action in a Minkowskian world-sheet
\be
S_{NW}={1\over 2\pi}\int d^2z\left[2\p_+ v\p_- u+2\p_+ u\p_-
v-(x^ix^i)\p_+ u\p_- u+\p_+x^i \p_- x^i+\right.
\ee
$$\left.+
\epsilon^{ij}x^i(\p_+ u\p_-x^j
-\p_+x^j\p_- u)\right] \ ,
$$
where
\be
\p_{\pm}=\p_{\tau}\pm\p_{\s} \ .
\ee

The equations of motion are
\be
\p_+\p_- u=0\sp \p_+\p_-
x^i+x^i\p_+u\p_-u-\epsilon^{ij}(\p_+u\p_-x^j-\p_-u\p_+x^j)=0 \ ,
\ee
\be
4\p_+\p_-v=\p_+(x^ix^i\p_-u)+\p_-(x^ix^i\p_+u)-\epsilon^{ij}(\p_+(x^i\p_-x^j)-\p_-(x^i\p_+x^j)) \ .
\ee
We may now pick the light cone gauge $u=p_+~\tau$ to obtain
\be
\p_+\p_-
x^i+p_+^2  x^i+2p_+\epsilon^{ij}\p_{\s}x^j=0 \ ,
\label{eq}\ee
\be
2\p_+\p_-v=p_+\left[\p_{\tau}(x^ix^i)-\epsilon^{ij}\p_+x^i\p_-x^j\right] \ .
\ee
As usual the Virasoro constraints can be used to determine $v$
in terms of $p_+$ and the transverse variables $x^i$ as well as
those of the rest of the space $x^A$, that we take to be flat.
The reparametrization constraint reads
\be
\int_{0}^{2\pi}d\s \left[\p_{\tau}x^i\p_{\s}x^i
+\p_{\tau}x^A\p_{\s}x^A\right]=0
\ee

The light-cone Hamiltonian can be computed to be
\be
H={1\over p_+}\int_0^{2\pi}{d\s\over
2\pi}\left[(\p_{\tau}x^A)^2+(\p_{\s}x^A)^2+(\p_{\tau}x^i)^2+(\p_{\s}x^i)^2+p_+^2x^ix^i
+2p_+\epsilon^{ij}x^i\p_{\s}x^j\right] \ .
\ee
The equations of motion for the transverse plane coordinates $x^i$
(\ref{eq})
can be written as
\be
\p_+\p_-X+p_+^2~ X-2ip_+\p_{\s}X=0 \ ,
\ee
for the complex filed $X=x^1+ix^2$.
defining $X=e^{-ip_+\s}\Phi$ we obtain
\be
\p_+\p_-\Phi=0 \ .
\ee
Thus, $\Phi$ is a free field albeit with twisted boundary
conditions.
Indeed, in order for $X$ to be periodic, $\Phi$ must have
non-trivial monodromy
\be
\Phi(\s+2\pi)=e^{2\pi i p_+}\Phi(\s) \ .
\ee
This matches well with the free-field resolution of the covariant theory.

We can write
\be
\Phi=\phi(\tau+\s)+\chi(\tau-\s) \ ,
\ee
with
\be
\phi=\sum_{n\in Z}{a_n\over n+p_+}e^{i(n+p_+)(\tau+\s)}\sp \chi=\sum_{n\in Z}{b_n\over
n-p_+}e^{i(n-p_+)(\tau-\s)} \ ,
\ee
and finally
\be
X=\sum_{n\in Z}\left[{a_n~e^{ip_+\tau}\over n+p_+}e^{in(\tau+\s)}+{b_n~e^{-ip_+\tau}\over
n-p_+}e^{in(\tau-\s)}\right] \ ,
\ee
\be \bar X=\sum_{n\in Z}\left[{a^{\dagger}_n~e^{-ip_+\tau}\over n+p_+}e^{-in(\tau+\s)}+
{b^{\dagger}_n~e^{ip_+\tau}\over
n-p_+}e^{-in(\tau-\s)}\right] \ .
\ee
For the free part we have the usual expansions
\be
X^A=x^A+p^A\tau +\sum_{n\in Z-\{0\}}\left[{a^A_n\over
n}e^{in(\tau+\s)}+{b^A_n\over n}e^{in(\tau-\s)}\right] \ .
\ee
Canonical quantization imposes the commutation relations
\be
[a_n,a^{\dagger}_m]=(n+p_+)\delta_{m,n}\sp [b_n,b^{\dagger}_m]=(n-p_+)\delta_{m,n} \ .
\ee
In the case $0<p_+<1$    ~~$a_{n~\geq 0},a^{\dagger}_{n< 0}$,
$b_{n>
0},b^{\dagger}_{n\leq 0}$ are annihilation operators and the ground
state  satisfies
\be
a_{n\geq 0}|p_+\rangle=a^{\dagger}_{n< 0}|p_+\rangle=b_{n>
0}|p_+\rangle=b^{\dagger}_{n\leq 0}|p_+\rangle=0 \ .
\ee
The light-cone hamiltonian becomes
\be
p_+H=(p^A)^2+\sum_{n=1}^{\infty}(a^A_{-n}a^A_{n}+b^A_{-n}b^A_{n}+a_{-n}a^{\dagger}_{-n}+b^{\dagger}_{n}b_n)
+\sum_{n=0}^{\infty}(a^{\dagger}_na_n+b_{-n}b^{\dagger}_{-n})
+{1-p_+\over 2}\ee
Note that the transverse plane part is the standard Virasoro
operator for a complex twisted boson as expected.

Spectral flow is visible in the light-cone quantization scheme.
When the integer part of $p_+$ is $N>0$ a rearrangement of the
commutation relations is in order.
Now
$a_{n~\geq -N},a^{\dagger}_{n< -N}$,
$b_{n>
-N},b^{\dagger}_{n\leq -N}$ are annihilation operators and the new (spectral-flowed) ground
state satisfies
\be
a_{n\geq -N}|p_+,N\rangle=a^{\dagger}_{n< -N}|p_+,N\rangle=b_{n>
-N}|p_+,N\rangle=b^{\dagger}_{n\leq -N}|p_+,N\rangle=0 \ .
\ee

\section{Generalizations}
\setcounter{equation}{0}

The NW WZW model is the first of a class of WZW models with
generalized Heisenberg symmetry \cite{ks2}.
All of them can be solved using the techniques developed here.
The associated  pp-wave space-times have two light-cone directions, and n transverse two-planes.
The case $n=1$ is the NW model studied here. $n=2$ corresponds to
the Penrose limit of the near-horizon region of an NS5-F1
bound-state \cite{kp}.
The other cases $n=3,4$ correspond to the Penrose limit of
intersecting NS5's and F1's.

They have metrics of the form
\be
ds^2=-2dudv-{1\over 4}\left[\sum_{i=1}^n~y^i\bar
y^i\right]du^2+\sum_{i=1}^n~dy^i
d\bar y^i+\sum_{I=1}^{8-2n}~dx^Idx^I \ ,
\ee
and NS-antisymmetric tensor field strengths
\be
H_{uy^i\bar y^i}=1 \ .
\ee
The associated $H_{2+2n}$ Heisenberg current algebra has
generators $K,J,P_i^{\pm}$ with OPEs

\ba
P^+_i(z)P^-_j(w) &\sim& \frac{2\delta_{ij}}{(z-w)^2} - \frac{2i\delta_{ij}~K(w)}{z-w}+{\rm RT} \ , \nb \\
J(z)P^{\pm}_{i}(w) &\sim& \mp i\frac{P^\pm_{i}(w)}{z-w}+{\rm RT} \ , \nb \\
J(z)K(w) &\sim& \frac{1}{(z-w)^2}+{\rm RT} \ .
\label{cg-ope}
\ea

The zero mode algebra is the same as the one generated by $n$
harmonic oscillators.
We can identifying $P^{\pm}_i$ with the creation and annihilation operators of the
i-th harmonic oscillator, J with the total hamiltonian and $K$
with $\hbar$.
The Cartan subalgebra is as before, while now we have several
raising and lowering operators.
There are again two classes of unitary representations: $V^{\pm}$
with $p\not=0$ and $V^0$ representations with $p=0$.

There is a free-field resolution generalizing the one given in
(\ref{lb16}),
namely
\ba
J &=& \p v \ , \hspace{2cm} K =  \p u \ , \nb \\
P^+_i &=& i e^{-iu} \p y^i \ , \hspace{1cm} P^- = i e^{iu} \p \bar y^i \ ,
\label{lb166}
\ea
where $u,v,y^i,\bar y^i$ are free bosons.

The ground state of a $V^{\pm}$ representation is given by
\be
R^{\pm}_{p,\jh;0}(z) = e^{i (\jh~u(z) \pm p v(z))} \prod_{i=1}^n~H^{i,\mp}_{p}(z) \ ,
\label{t111}
\ee
where $H^{i,\mp}_{p}$ is the associated twist field of the $i$-th
plane.
The other states of the $V^{\pm}$ representations are obtained through the action of $P_i^{\mp}$.
Here also $0<p<1$.
The conformal dimension of the level-zero states is
\be
h_{p,\jh}=-p\jh +{n\over 2}|p|(1-|p|) \ .
\ee

Similarly, the operators of the $p=0$ representations $V^0$ can be
written in terms of free vertex operators,
\be
R_{p^-,p_i}\sim e^{ip^- u+i\sum_{i=1}^n(p_i\bar y^i+\bar p_i y^i)} \ .
\ee

Arbitrary values for $p$ are again recovered
through the action of the spectral flow that shifts $p$ by integers.

The full level-zero operators can be assembled into a single field at the expense of introducing $n$
auxiliary
charge variables $x^i$
\be
\f_{\pm}(\vec x) = \sum_{n_i=0}^{\infty} \f_{n_i} \prod_{i=1}^n~\frac{(x^i\sqrt{|p|})^{n_i}}{\sqrt{n_i!}} \ .
\label{lb111}
\ee
When $p>0$, the zero-mode operators realizing the $H_{2+2n}$ algebra are
\ba
P^+_i &=&  \sqrt{2} p ~x^i \ , \hspace{1cm}
P^-_i = \sqrt{2}~ \p_{x^i} \ , \nb \\
J &=& i(\jh + \sum_{i=1}^n~x^i \p_{x^i} ) \ , \hspace{1cm}
K = i p \ ,
\label{h111}
\ea
and similar expressions for the other representations, generalizing eqs. $(\ref{h1})$,
$(\ref{h2})$ and $(\ref{h3})$.

Finally, the calculation of the correlators proceeds along similar
lines and they are simply given by the product of the basic
correlators computed in this paper.

\section{Holography}
\setcounter{equation}{0}

In \cite{kp} a holographic correspondence was proposed to link the
on-shell data of string theory in a pp-wave background to
off-shell data of the limiting boundary theory. Alternative
holographic proposals were also advanced in \cite{hol}.
The weak link so far in holographic correspondence is the lack of
knowledge of well defined off-shell data  in the corresponding
boundary theories.

In \cite{kp} it was argued that the proper on-shell data of the
associated pp-string theory are S-matrix elements.
They were defined semiclassically by going to a conformally flat
coordinate system that covers a strip in $x^+$ of the full pp-wave
space-time. The semiclassical scattering problem was defined with
boundary conditions in the past wedge $x^+\to -\infty$, $x^-\to
-\infty$. It was shown that the boundary condition in the $x^-\to
-\infty$ of the wedge must be trivial in order for the Heisenberg
symmetry to be unbroken.
The basis used in this paper is the oscillator basis described in
appendix E of \cite{kp}.

The classical supergravity action was used to derive the
appropriate scattering amplitude. Once the amplitudes are derived
in this coordinate system, they could be extended to the full
pp-wave space-time by imposing $p_-$ conservation. This works
because the semiclassical amplitudes are phase-shifts.

What we have shown here is that the semiclassical scattering
amplitudes can be extended to bona-fide scattering amplitudes in the
string theory on NW pp-wave backgrounds. It is also clear from the
setup, that this should also be true in more general pp-wave backgrounds
supported by NS-antisymmetric tensor flux.

Moreover, the auxiliary current algebra coordinates $x,\bar x$
should correspond to coordinates of the boundary theory.
This expectation is substantiated by the fact that this is the
case in the $AdS_3/CFT$ correspondence \cite{ads3,oo}.
Indeed consider the $AdS_3\times S_3$ dual of the NS5-F1 CFT.
Introduce auxiliary variables $y,\bar y$ to keep track of the
SL(2,R) quantum numbers and $x,\bar x$ for the SU(2) R-symmetry
quantum numbers.
The $y,\bar y$ turn into the world-sheet coordinates of the
boundary CFT, while $x,\bar x$ are charge coordinates in the
R-symmetry  space of the CFT.

The Penrose limit of this theory gives the $H_6\times R^4$ pp-wave theory
which differs from the one presented here by the presence of an
extra transverse plane.
The free-field resolution is identical, involving an extra set of
twist fields, and the scattering amplitudes similarly defined.
They are functions of $x,\bar x,y,\bar y$ as well as $p_+$ and the
momenta of the flat four-dimensional part.
The holographic proposal identifies $x,\bar x,y,\bar y$ as well as
the conjugate (via Fourier transform) coordinates $x^-,x^i$ of $p_+,p^i$
with the coordinates of the off-shell boundary theory.
Moreover, the free field realization of $H_6$ indicates a symmetry
in the interchange $x\leftrightarrow y$, which implies that in
the associated pp-wave background, world-sheet and charge variables
appear on an equal footing.

In the NW case described in this paper, the S-matrix elements
depend on $x,\bar x$ the charge coordinates of $SU(2)$ turned
$H_4$, $p_+$, and the 22 flat momenta $p^i$ (in the bosonic case),
or 6 flat momenta in the supersymmetric case.

The Heisenberg symmetry generators of the holographic dual theory were
shown \cite{kp} to involve inverse powers of  $p_+$ and it is thus
expected that there is a non-locality in $x^-$ in the dual
data.

This approach
indicates that an auxiliary charge-space must be
introduced in the boundary theory in order to properly define the
holographic data to be compared with string theory. This is also
the spirit of the approach in \cite{sok}.
An alternative formalism is to dispense with the charge space
coordinates at the cost of introducing infinite matrices.
It is not clear yet which of the two parameterizations is suitable
for the description of the correspondence, although our feeling is
that charge-space coordinates give an easier formalism since the group
acts by differential operators rather than infinite dimensional
matrices.

\section{Conclusions}

In this paper, we have analyzed string theory in the pp-wave background corresponding to the
Penrose limit of LST.
The associated CFT is a product of the Nappi-Witten WZW model and six free flat non-compact coordinates.
We have
calculated the  tree-level
three- and four-point scattering amplitudes, analyzed their structure and
studied some consequences for the physics .

Our results are as follows.

There are two basic set of operators of the CFT corresponding to
standard Heisenberg current algebra representations:

$\bullet$ The
``discrete-like" $V^{\pm}$ sector involving operators with
$0<|p^+|<1$. The superscript $\pm$ indicates positive or negative $p^+$.
$V^+$ is conjugate to $V^-$.
$V^{\pm}$ are organized into semi-infinite lowest-weight
or highest-weight representations of the Heisenberg algebra.
The transverse plane spectrum of such representations is discrete
in agreement with the fact that in this sector there is an
effective harmonic oscillator potential confining them around the
center of the plane.

The cutoff $p^+=1$ is stringy in origin and descents from the
$l\leq {k\over 2}$ cutoff of the parent $SU(2)_k$ theory.
It is also also a unitarity cutoff: standard current algebra
representations with $p^+>1$ will contain physical negative-norm states in
the associated string theory.

$\bullet$ The ``continuous-like $V^0$ sector involving operators
with $p^+=0$. The transverse plane spectrum of such operators is continuous in agreement
with the fact that there is no potential for such operators.
The interactions among states in this sector are
similar to the flat space theory.

There is a spectral flow that shifts $p^+$ by integers.
The spectral-flowed current algebra representations have
$L_0$ eigenvalues not bounded from below.
Such representations correspond to states deep-down the $SU(2)_k$
current algebra representations, related to the top states by the
action of the affine Weyl group of $\widehat{SU(2)}$.
They are crucial for the closure of the full
operator algebra and the consistency of interactions, as they
are in the parent $SU(2)_k$ theory.
At $p_+=1$, the harmonic oscillator
potential fails to squeeze the string since it is compensated by
the NS-antisymmetric tensor background. Such long strings generate
a sequence of new ground-states that are related to the current
algebra ground states by spectral flow.

Both the $V^0$ and the $V^{\pm}$ sectors contain spectral-flowed
operators.
There are  three types of three-point functions: $\langle
V^{+}V^{+}V^{-}\rangle$, $\langle
V^{+}V^{-}V^{0}\rangle$ and $\langle
V^{0}V^{0}V^{0}\rangle$.
The first two have quantum structure constants that are
non-trivial functions of the $p^+_i$ of $V^{\pm}$ and $\vec p$ of
$V^{0}$.

The third is the same as in flat space (with $p^+=0$) since the
$V^0$ operators are standard free vertex operators.

There are several types of four-point functions among the $V^{\pm,0}$
operators:

\bigskip
$\bullet$ $\langle
V^+(p^+_1)V^+(p^+_2)V^+(p^+_3)V^-(-p^+_1-p^+_2-p^+_3)\rangle$ and their conjugates.
Such correlators factorize on $V^{\pm}$ representations and they
have a finite number of conformal blocks. They are given by powers of
$_2F_1$-hypergeometric functions.

\bigskip
$\bullet$ $\langle
V^+(p^+_1)V^+(p^+_2)V^-(-p^+_3)V^-(p^+_3-p^+_1-p^+_2)\rangle$.
These
correlators factorize on $V^{\pm}$ representations and their spectral-flowed images
for generic values of $p^+_i$.
They have an infinite number of conformal blocks and they are given by
exponentials of hypergeometric functions.
For special values of the momenta, for example $p^+_1=p^+_3$, they
factorize onto the $V^0$ representations and their spectral-flowed
images.
In this case, they develop logarithmic behavior in the cross-ratio, which signals the presence
of the continuum of intermediate operators.
It should be noted that the logarithmic behavior here does not
indicate that we are dealing with a logarithmic CFT (argued to
describe cosets of pp-wave nature \cite{lcft}) but rather
the presence of a continuum of intermediate states.

\bigskip
$\bullet$ $\langle
V^+(p^+_1)V^+(p^+_2)V^-(-p^+_1-p^+_2)V^0(\vec p)\rangle$.
Such correlators have an infinite number of conformal blocks. They factorize
on $V^{\pm}$ representations, and they are given in terms of the
$_3F_2$-hypergeometric functions.

\bigskip
$\bullet$ $\langle
V^+(p^+)V^-(-p^+)V^0(\vec p_1)V^0(\vec p_2)\rangle$. These
correlators have an infinite number of conformal blocks. They
factorize on $V^{\pm}$ or $V^0$ operators depending on the
channel.

\bigskip
$\bullet$ $\langle
V^0(\vec p_1)V^0(\vec p_2)V^0(\vec p_3)V^0(-\vec p_1-\vec p_2-\vec p_3)\rangle$.
These correlators have a single conformal block and are the same
as in flat space.

Starting from the solution of the NW CFT we have calculated the
three-point and the four-point S-matrix elements of the associated
(super)string theory on this background.
We found them to have the following salient features.

\bigskip
$\bullet$ The S-matrix elements exist, i.e. they are free of
spurious singularities.

\bigskip
$\bullet$ They are {\tt non-analytic} as functions of the external $p^+$
momenta.
In particular the amplitudes vanish when one external $p_+\to 0$
and do not asymptote to the associated amplitude with the
insertion of a $p_+=0$ state.

\bigskip
$\bullet$ They are ``dual", i.e. crossing symmetric as in string
theory in flat space. Crossing symmetry was explicitly verified in most cases.

\bigskip
$\bullet$ The four-point S-matrix elements have two types of
singularities: Poles associated to the propagation of intermediate
physical states with $p^+\not =0$, as well as logarithmic branch
cuts signaling the propagation of a continuum of intermediate
physical states which are spectral-flowed $p^+=0$ states. Such branch cuts are very much
similar to those appearing in field theory S-matrix elements with
massless on-shell intermediate states.
However, because of momentum conservation such branch cuts can
appear only in loops in field theory, while here they already
appear at tree-level.

The states corresponding to $p_+=0$, do not give rise to physical
states except when there is a tachyon in the spectrum.
This would be the case in a bosonic NW$\times \mathbb{R}^{22}$ string
theory, but not in the superstring NW$\times \mathbb{R}^{6}$ corresponding
to the Penrose limit of LST.
However, spectral-flowed
$V^0$ states with $p^+=$ integer, can be physical already in the
massless (i.e. level-zero) sector.

In the flat space limit, the spectrum of the theory remains as
before, the limit is smooth and the spectral-flowed states decouple.
This should be compared with the decompactification limit of a
torus. There, the momentum states combine to make the non-compact
propagating degrees of freedom, while the winding states decouple.

The S-matrix elements we have computed, are naturally functions of the
auxiliary charges variables $x,\bar x$ used to keep track of the
Heisenberg symmetry of the pp-wave as well as of $p^+$.
In analogy with $AdS_3$ we can identify these variables as
auxiliary coordinates of the holographic dual. Introducing
also $x^-$, the Fourier conjugate of $p^+$, we obtain ``boundary" coordinates
that match the holographic setup of \cite{kp}.

The corresponding light-cone quantization \cite{forgacs} of this theory can
be compared with the covariant quantization employed in this
paper.

In the light-cone gauge one obtains the same spectrum for $p^+\not
=0$. The $p^+=0$ states are not accessible in the light-cone
gauge.

The possibility and structure of the spectral flow is also visible
in the light-cone gauge, but again {\tt not} the spectral-flow
images of the $p^+=0$ sector.

Taking all the above into account we may extrapolate and speculate
on the structure of the associated tree S-matrix elements of
RR-supported pp-waves relevant for super-YM.
So far, such amplitudes are accessible only in the light-cone GS
formalism. There is no signal for spectral flow in this case. This is expected
since the RR-flux does not couple directly to fundamental strings.

We should expect again that in analogy with our case, the S-matrix
elements will also be non-analytic in $p^+$.

Four-point S-matrix elements also are expected to have branch cuts
associated to on-shell intermediate (spectral-flowed) $p^+=0$ states. It is an
interesting question, whether the structure of such branch cuts,
factorization and consistency can determine S-matrix elements of
external $p_+=0$ states.

One of the lessons we have learned here is that, as suspected
\cite{kp,jabari}, there can exist S-matrix amplitudes in
some time-dependent space-times. Asymptotic flatness does not seem
to be necessary for this.

There are several avenues for exploration along the lines followed
in this paper.
The techniques used here can be applied to a wider class of
pp-wave space-times, WZW models or cosets thereof.
In particular, it would be interesting to study one of the two
simplest cosets \cite{kk} of the NW WZW model, due to the claim
\cite{lcft} that it corresponds to a bona-fide logarithmic CFT.

As we have shown, there are two inequivalent marginal
deformations of the NW theory. One corresponds to a space-time
which is generically not of the pp-wave type. The other
describes a class of pp-waves with time-dependent masses in the
light-cone gauge.
It is an interesting problem whether the space-time stringy
dynamics of such backgrounds is amenable to an exact treatment
using the solution of the unperturbed theory.

The structure and dynamics of D-branes in the NW background
has not given us yet all its secrets, although there has been work in this direction
 \cite{dbr}. It is certainly an interesting laboratory for the study
of D-branes, in non-trivial curved/time-dependent backgrounds.
In particular, analogs of long strings and ``transparency"
phenomena are expected also for D-branes. For example, $D_1$
strings on $S^3$ with a RR 2-form flux are expected to lead to
spectral flow phenomena, and long-string ground states.
This effect also exists in other cases like $D_3$ branes in
$AdS_5$ with a 4-form flux \cite{sw}. It would be interesting to investigate
such phenomena since they have important implications for D-brane
dynamics in non-trivial background fields.

An important open problem that remains is the elucidation of the
proper holographic observables on the YM/LST side.
It is plausible,that the use of  auxiliary charge coordinates on
the field theory side will make the choice of such off-shell
observables more transparent.

\enlargethispage{3mm}

\acknowledgments The authors are grateful to C. Bachas, E. Cremmer, V. Dotsenko, J. Iliopoulos,
C. Kounnas, and G. Papadopoulos for discussions.
   The work of G. D. is supported by an European Commission Marie Curie Postdoctoral
Fellowship, contract HMPF CT 2002-01908.
The work of E. K. was partially supported by  RTN contracts HPRN--CT--2000--00122 and --00131 and INTAS contract
INTAS contract N 99 1 590.

\vskip 1.5cm

\appendix

\vskip 10mm

\centerline{\it\bf Appendices}

\section{The SU(2)$_k$ CFT and its Penrose limit\label{b}}
\renewcommand{\theequation}{\Alph{section}.\arabic{subsection}.\arabic{equation}}

In this Appendix we will review basic results concerning the  SU(2)$_k$ operator algebra
following the original reference \footnote{We correct on the way some misprints
in that paper.}
\cite{zf}.We will further take the Penrose
limit and compare the result with our direct calculation in the Nappi-Witten model.

\subsection{SU(2)$_k$ CFT\label{b1}}
\def\xb{\bar x}
\def\mb{\bar m}
\def\zb{\bar z}

The SU(2)$_k$ current algebra was  displayed in (\ref{la4},\ref{la5}).
It is convenient to introduce auxiliary coordinates $x,\xb$ in
order to keep track of the group structure.
The classical SU(2) generators act on a spin $l$ representation as the following differential
operators
\be
T^{+}=\partial _x \sp T^{-}=-x^2\partial_x+2l~x\sp
T^3=x\partial_x-l \ ,
\label{a2}\ee
and similarly for the anti-holomorphic algebra.
The affine primary fields $\Phi^{l}_{m,\mb}(z,\zb)$ whose conformal dimension is
$h = \frac{l(l+1)}{k+2}$ can be
organized as
\be
\Phi^l(z,\zb;x,\xb)\equiv\sum_{m,\mb=-l}^l~~\sqrt{\left({2l\atop
l+m}\right)\left({2l\atop
l+\mb}\right)}~x^{l+m}~\xb^{l+\mb}~\Phi^l_{m,\mb}(z,\zb) \ .
\label{a1}\ee
The two-point function is
\be
<\Phi^{l_1}(z_1,\zb_1;x_1,\xb_1)\Phi^{l_2}(z_2,\zb_2;x_2,\xb_2)> = \d_{l_1,l_2}
\frac{(x_{12}\bar{x}_{12})^{2l_1}}{(z_{12}\bar{z}_{12})^{2h_1}} \ .
\ee
We also define the following $x$ dependent current
\be
J(x,z)= J^-(z)+2xJ^3(z)-x^2 J^+(z) \ ,
\label{a3}\ee
and then using the OPE of the $SU(2)$ currents with the primary fields
one can write the Ward identity in a compact form
\be
\langle ~J(x,z)\prod_{i=1}^M\Phi^{l_i}(z_i,x_i)~\rangle
=-\sum_{j=1}^M\left[{2l_j(x-x_j)\over (z-z_j)}+{(x-x_j)^2\over
(z-z_j)}{\partial\over \partial x_j}\right]\langle~ \prod_{i=1}^M\Phi^{l_i}(z_i,x_i)~\rangle \ .
\label{a4}\ee
Solving the global SU(2)$_L\times $SU(2)$_R$ and the conformal
Ward identities we can write the three-point functions as
\be
\langle~
\prod_{i=1}^3\Phi^{l_i}(z_i,\zb_i;x_i,\xb_i)~\rangle
=C(l_1,l_2,l_3)\prod_{i<j}^3{(x_{ij}~\xb_{ij})^{l_{ij}}\over
(z_{ij}~\zb_{ij})^{h_{ij}}} \ ,
\label{a5}\ee
where $x_{ij}=x_i-x_j$, $z_{ij}=z_i-z_j$, $l_{12}=l_1+l_2-l_3$,
$h_{12}=h_1+h_2-h_3$ and cyclic permutations of the indexes.
Similarly the four-point function is given by
\be
\langle~
\prod_{i=1}^4\Phi^{l_i}(z_i,\zb_i;x_i,\xb_i)~\rangle
=(x_{14}~\xb_{14})^{2l_1}(x_{24}~\xb_{24})^{l_2+l_4-l_1-l_3}
(x_{34}~\xb_{34})^{l_3+l_4-l_1-l_2}\times
\label{a6}\ee
$$
\times
(x_{23}~\xb_{23})^{l_1+l_2+l_3-l_4}(z_{14}~\zb_{14})^{-2h_1}
(z_{24}~\zb_{24})^{\nu_1}(z_{34}~\zb_{34})^{\nu_2}(z_{23}~\zb_{23})^{\nu_3}U(z,\zb,x,\xb) \ ,
$$
with
\be
\nu_1=+h_1+h_3-h_2-h_4\sp
\nu_2=+h_1+h_2-h_3-h_4\sp
\nu_3=-h_1-h_2-h_3+h_4 \ ,
\label{a7}\ee
and
\be
z={z_{12}z_{34}\over z_{14}z_{32}} \ , \hspace{2cm}  x = {x_{12}x_{34}\over x_{14}x_{32}} \ .
\label{a8}
\ee
Note that the above cross-ratio differs from the one used in the paper and
defined in Eq. (\ref{ld2}) by the exchange $z_1 \leftrightarrow z_2$.
We will take, without loss of generality, the $l_i$ to be ordered as $l_1\leq l_2\leq l_3\leq
l_4$.

The correlation functions satisfy two types of differential
equations. The first, the Zamolodchikov-Fateev  equation, comes from the
pure affine null vectors and reads
\be
\left[\sum_{n=2}^M{1\over
z_1-z_n}\left\{(x_1-x_n)^2\partial_{x_n}+2l_n(x_1-x_n)\right\}\right]^{k-2l_1+1}\langle~
\prod_{i=1}^M\Phi^{l_i}(z_i,\zb_i;x_i,\xb_i)~\rangle \ ,
\label{a9}\ee
as well as (M-1) similar ones.
The others, the Knizhnik-Zamolodchikov equations, come from the
fact that the stress tensor is of the affine-Sugawara form and
read
\be
\left[(k+2)\partial_{z_i}- \sum_{i<j}^M{2T^3_iT^3_j+T^+_iT^-_j+T^-_iT^+_j\over
z_i-z_j}\right]~~\langle~\prod_{r=1}^M\Phi^{l_r}(z_r,\zb_r;x_r,\xb_r)~\rangle = 0 \ .
\label{a10}\ee

The KZ equations on the three-point functions relate the spin to the
conformal dimension as $h={l(l+1)\over k+2}$
while the ZF equations give
\be
\prod_{n=1}^{k+1}(l_1+l_2-l_3+n-1)~~ C(l_1,l_2,l_3) \ ,
\label{a11}\ee
which in turn imply the truncation of the operator algebra at $l={k\over
2}$
\be
[\Phi^{l_1}]~\otimes~[\Phi^{l_2}]=\sum_{l=|l_1-l_2|}^{{\rm
min}~(l_1+l_2,k-l_1-l_2)}~[\Phi^{l}] \ ,
\label{a12}\ee

For the four-point function the ZF equation imply
\be
\prod_{n=0}^{k-2l_4}\left[(x-z)\partial_x
+l_4-l_1-l_2-l_3+n\right]~~U(z,\zb;x,\xb)=0 \ .
\label{a13}\ee
This equation can be solved as
\be
U(z,\zb;x,\xb)=\sum_{p,\bar p=p_0}^{p_1}~(x-z)^p~(\xb-\zb)^{\bar
p}~~U_{p,\bar p}(z,\zb) \ ,
\label{a14}\ee
where
\be
p_0 = {\rm max}(0,l_1+l_2+l_3+l_4-k) \ , \hspace{1cm}
p_1 = {\rm min}(2l_1,l_1+l_2+l_3-l_4) \ .
\label{a14bis}
\ee
On the other hand the KZ equation implies
\be
(k+2)z(z-1)\partial_z U(z,x)=x(x-1)(x-z)\partial_x^2
U(z,x)-
\label{a15}\ee
$$
\left[(l_1+l_2+l_3-l_4-1)(x^2-2zx+z)+2(l_1x(x-1)+l_2x(z-1)+l_3(x-1)z)\right]\partial_xU(z,x)
$$
$$
-\left[
2(l_1+l_2+l_3-l_4)l_1(z-x)-2l_1l_2(z-1)-2l_1l_3z\right]U(z,x) \ ,
$$
and upon substituting (\ref{a14}) we obtain the system of ordinary
differential equations
\be
(k+2)z(z-1)\partial_z U_{p,\bar p}=(p+1)(p+1+k-j)z(z-1)U_{p+1,\bar
p}
\label{a16}\ee
$$
-(a_pz+b_p(z-1))U_{p,\bar p}+(2l_1-p+1)(j+1-2l_4-p)U_{p-1,\bar p} \ ,
$$
and a similar one for $\zb$ where
\be
j=l_1+l_2+l_3+l_4\sp a_p=p(p+1)-2(p-l_1)(p-l_3)\sp
b_p=p(p+1)-2(p-l_1)(p-l_2) \ .
\label{a17}\ee

Fateev and Zamolodchikov noticed that (\ref{a15}) can be connected to the
differential equation for a five-point correlator of a conformal minimal
model and therefore they were able to present the solution as a
Dotsenko-Fateev integral
\be
U(z,\zb,x,\xb)={\cal N}(l_1,l_2,l_3,l_4)~|z|^{4l_1l_2\over k+2}~|1-z|^{4l_1l_3\over
k+2}\times
\label{a18}\ee
$$
\times \int \prod_{i=1}^{2l_1}~{dt_id\bar t_i\over (2\pi
i)}~|t_i-z|^{-{2\beta_1\over k+2}}~|t_i|^{-{2\beta_2\over k+2}}~
 |t_i-1|^{-{2\beta_3\over k+2}}~|x-t_i|^2~|D(t)|^{4\over
k+2} \ ,
$$
where
\be
D(t)=\prod_{i<j} (t_i-t_j) \ ,
\label{a19}\ee
\be
\beta_1=l_1+l_2+l_3+l_4+1\sp \beta_2=k+l_1+l_2-l_3-l_4+1\sp \beta_3=k+l_1+l_3-l_2-l_4+1 \ .
\label{a20}\ee
The normalization is (we define from now on $N\equiv k+2$)
\be
{\cal N}^2(l_1,l_2,l_3,l_4)=\left[{\Gamma\left({1\over
N}\right)\over \Gamma\left(1-{1\over
N}\right)}\right]^{4l_1+2}{\Gamma\left(1-{2l_1+1\over
N}\right)~P^2(l_1+l_2+l_3+l_4+1)\over\Gamma\left({2l_1+1\over
N}\right)P^2(2l_1)}\times
\label{a21}\ee
$$
\times \prod_{i=2}^4~{\Gamma\left(1-{2l_i+1\over
N}\right)~P^2(l_2+l_3+l_4-l_1-2l_i)\over\Gamma\left({2l_i+1\over
N}\right)P^2(2l_i)} \ ,
$$
with
\be
P(n)=\prod_{m=1}^n~{\Gamma\left({m\over
N}\right)\over \Gamma\left(1-{m\over
N}\right)}\sp P(0)=1 \ .
\label{a22}\ee
The quantum structure constants of the operator algebra are
\be
C^2(l_1,l_2,l_3)={\Gamma\left({1\over
N}\right)\over \Gamma\left(1-{1\over
N}\right)}~P^2(l_1+l_2+l_3+1)\prod_{i=1}^3~{\Gamma\left(1-{2l_i+1\over
N}\right)~P^2(l_1+l_2+l_3-2l_i)\over\Gamma\left({2l_i+1\over
N}\right)~P^2(2l_i)} \ .
\label{a23}\ee
In \cite{dot}, the four-point functions of the $SU(2)$ WZW model were computed
using the Wakimoto free-field representation.

\subsection{The asymptotics of the P-function\label{b2}}
\setcounter{equation}{0}

As explained in section 2, the Penrose limit involves taking $N=k+2\to\infty$ while scaling
\be
2l=k|p|\mp 2\jh=N|p|-2(|p|\pm \jh)\equiv N|p|+a \ ,
\label{a24}\ee
 for the $V^{\pm}$
representations.
For the $V^0$ representations we have instead
\be
2l=\sqrt{2N}~s \ .
\label{a25}\ee

As a first step we need the asymptotics of the P-function in order
to obtain the structure constants of the Penrose limit operator
algebra.

 The P-function satisfies
\be
P(0)=1\sp P(N-1)=1\sp P(n)=0~~~~{\rm for}~~~~n>N-1\sp P(N-n)=P(n-1) \ .
\label{a26}\ee

Using the  asymptotic formula

\be
{\Gamma\left(p+{a\over N}\right)\over \Gamma\left(1-p-{a\over N}\right)}
={\Gamma(p)\over \Gamma(1-p)}\left[1+{a\over N}\partial_p\log {\Gamma(p)\over
\Gamma(1-p)}+{\cal O}\left({1\over N^2}\right)\right] \ ,
\label{a27}\ee
we obtain that for  $n\sim {\cal O}(1)<<N$
\be
P(n)={N^n\over n!}\left[1+{\cal O}\left({1\over N}\right)\right]\label{a28} \ .
\ee
To proceed further we will need the following asymptotic
expansions
\be
\log \Gamma(1-x)={1\over 2}\log {\pi x\over \sin \pi x}+Cx+
\sum_{n=1}^{\infty}{x^{2n+1}\over {2n+1}}\zeta(2n+1) \ ,
\label{a29}\ee
\be
\log {\Gamma(x)\over \Gamma(1-x)
}=-\log x-2\left[Cx+\sum_{n=1}^{\infty}{x^{2n+1}\over {2n+1}}\zeta(2n+1)\right] \ ,
\label{a30}\ee
valid for $x\in [0,1]$, as well as the asymptotic formula for the $\Gamma$-function
\be
\log[\Gamma(x)]=\left(x-{1\over 2}\right)\log x-x+\log\sqrt{2\pi}
+{\cal O}\left({1\over x}\right) \ ,
\label{a31}\ee
valid for large $x$.

The strategy to obtain the asymptotics is to use (\ref{a30}) to expand the
logarithm of $P$ as a series of sums of integers
and then do the leading parts of the sums using
\be
\sum_{i=1}^M~i^{2n+1}={1\over 2(n+1)}M^{2n+2}+{1\over 2}M^{2n+1}+{2n+1\over 12}M^{2n}
+{\cal O}\left(M^{2n-1}\right) \ .
\label{a32}\ee

In this way we obtain

\be
\log P(p\sqrt{N})=p\sqrt{N}\left({1\over 2}\log N+1-\log p\right)-\log\sqrt{2\pi p\sqrt{N}}-Cp^2+
{\cal O}\left({1\over N}\right) \ ,
\label{a33}\ee
\be
\log{P(p\sqrt{N}+n)\over P(p\sqrt{N})}=n\log{\sqrt{N}\over p}+{\cal O}\left({1\over N}\right) \ ,
\label{a34}\ee
where $C$ is the Euler constant.

Define further
\be
F(p)=\int_0^p dx\log\left[{\Gamma(x)\over \Gamma(1-x)}\right]\sp
F(p)=F(1-p) \ .
\label{a35}\ee
$F$ is a monotonically increasing function in [0,1/2].
Then,
\be
\log P(Np)=N~F(p)-{1\over 2}\log N+{1\over 2}\log{\Gamma(p)
\over \Gamma(1-p)}-\log\sqrt{2\pi}+{\cal O}\left({1\over N}\right) \ ,
\label{a36}\ee
\be
\log{P(Np_1+\sqrt{N}p_2)\over P(Np_1)}=\sqrt{N}p_2\log{\Gamma(p_1)\over
\Gamma(1-p_1)}+{p_2^2\over 2}\partial_{p_1}
\log{\Gamma(p_1)\over \Gamma(1-p_1)}+{\cal O}\left({1\over N}\right) \ ,
\label{a37}\ee

\be
\log P(Np_1+\sqrt{N}p_2+n)=\log P(Np_1+\sqrt{N}p_2)+n\log{\Gamma(p_1)\over
\Gamma(1-p_1)}+{\cal O}\left({1\over N}\right) \ ,
\label{a38}\ee
where in that last equation $n\sim{\cal O}(1)$

\subsection{The Penrose limit of the classical Clebsch-Gordan coefficients\label{b3}}
\setcounter{equation}{0}

We can now study the limiting behavior of the three-point couplings of the
$SU(2)_k$ model. As for the $z$ dependence, we already explained that
the conformal dimensions when dressed by the U(1) part of the primaries trivially asymptotes to
the appropriate $H_4$ limit.
Let us then consider the part proportional to the Clebsch-Gordan (CG) coefficients
that should reproduce
the corresponding quantities for the $H_4$ algebra.
The classical CG part is
\be
CG_3(l_1,l_2,l_3)=|x_{12}|^{2l_{12}}~|x_{13}|^{2l_{13}}~|x_{23}|^{2l_{23}} \ .
\label{a43}\ee
To obtain the Penrose limit we must define
\be
\hat \Phi^{-}_{p,\jh}(y,z)=\lim_{N\to\infty}~\Phi^l\left({y\over
\sqrt{2l}},z\right)\sp 2l=N|p|-2(|p|- \jh) \ ,
\label{a44}\ee
\be
\hat \Phi^{+}_{p,\jh}(y,z)=
\lim_{N\to\infty}~\left({\sqrt{2l}\over y}\right)^{-2l}~
\Phi^l\left({\sqrt{2l}\over y},z\right)\sp 2l=N|p|-2(|p|+ \jh) \ ,
\label{a45}\ee
\be
\Phi^0_s(y,z)=\lim_{N\to\infty}~y^{-l-\jh}~\Phi^l(y,z)\sp 2l=\sqrt{2N}~s \ .
\label{a46}\ee
If we further define
\be
\Phi^{\pm}_{p,\jh}(x,z)=\hat \Phi^{\pm}_{p,\jh}(\sqrt{|p|}x,z) \ ,
\label{a47}\ee
then $\Phi^{\pm,0}(x,z)$ have the expansion used in the main part
of the paper (\ref{lb9}).

In the $<++->$ case we have
\be
CG_3\Longrightarrow
N^{\jh_1+\jh_2+\jh_3}\left|~\exp\left[-x_3(p_1x_1+p_2x_2)\right]
~~\left({x_1-x_2}\right)^{-\jh_1-\jh_2-\jh_3}\right|^2+{\rm subleading} \ ,
\label{a49}\ee
that agrees up to the divergent normalisation with the appropriate
$H_4$ CG coefficient (\ref{lc10}).

For a $<+-0>$ coupling we have
\be
CG_3\Longrightarrow
\left|~x_{3}^{-L}~e^{-px_1x_2-{s\over \sqrt{2}}
\left({x_2x_3}+{x_1\over x_3}\right)}\right|^2+{\rm subleading} \ ,
\label{a50}
\ee
matching (\ref{lc16}).
Finally in the $<000>$ case one can use
the relation between Jacobi polynomials and Bessel functions
\cite{cg} to take the Penrose limit of the CG coefficients of $SU(2)$
and the result is $(\ref{lc20})$.

\subsection{Penrose limit of the quantum structure constants\label{b4}}
\setcounter{equation}{0}

We consider now the Penrose limit of the quantum structure constants $C(l_1,l_2,l_3)$
of the operator algebra of SU(2)$_k$ and we will identify the result with
the $H_4$ three-point couplings, since the structure constants of the U(1) part are trivial.
We will also see how the selection rules for $l$, $m$ and $q$ translates into selection rules
for $p$ and $\jh$.

Let us start with the coupling between three $V^{\pm}$ representations
and without loss of generality consider a $++-$ configuration.
The standard inequalities of the Clebsch-Gordan decomposition of SU(2)
imply that $L \equiv \jh_1+\jh_2+\jh_3\leq 0$ and charge conservation requires
$p_3=p_1+p_2$. Using the formulae of the previous section we can compute
\be
C_{++-}={N^{{1\over 2}-L}\over
\Gamma(1-L)}\left[{\Gamma(p_1+p_2)\over
\Gamma(1-(p_1+p_2))}{\Gamma(1-p_1)\over
\Gamma(p_1)}{\Gamma(1-p_2)\over \Gamma(p_2)}\right]^{{1\over 2}-L} \ .
\label{a39}
\ee
Note that when combined with the $CG$ coefficients in $(\ref{a49})$,
the three-point coupling diverges with $N$ as $\sqrt{N}$.
The power divergences reflects the presence of the continuum
$p$-conserving $\delta$-function evaluated at zero argument.
It is for this reason that although the direct limit gives the
correct $p$-dependence of the amplitudes, the normalisation
is ambiguous. We observe that (\ref{a39}) matches perfectly (\ref{cppp}).

Proceeding in the same way for two $V^{\pm}$  representations
with $2l_1=Np_1+a_1$, $2l_2=Np_2+a_2$, and one $V^0$ representation
with $2l_3=s\sqrt{2N}$ we can see that
for generic $p_{1,2}$ the amplitude blows up exponentially
\be
C_{+-0}^2\sim e^{4N\left[F\left({p_1+p_2\over 2}\right)+
F\left({p_1-p_2\over 2}\right)-{1\over 2}\left(F(p_1)+F(p_2)\right)\right]} \ ,
\label{a40}
\ee
but when we impose conservation of $p$ the leading exponential cancels
and we obtain
\be
C_{+-0}={2^{1+s\sqrt{2N}}\over \sqrt{2\pi}}
\exp\left[s^2\left(C+{1\over 2}\partial_p\log{\Gamma(p)\over
\Gamma(1-p)}\right)\right]=
\label{a41}\ee
$$={2^{1+s\sqrt{2N}}\over
\sqrt{2\pi}}\exp\left[{s^2\over 2}(\psi(p)+\psi(1-p)-2\psi(1))\right] \ .
$$
The amplitude diverges as expected but it is
proportional to the correct structure constant (\ref{lc23}).

The last case that remains to be discussed is the coupling between
three $V^0$ representations with $2l_i=\sqrt{2N}s_i$. The result is
\be
C_{000}=4\sqrt{2\over \pi}{s_1~s_2~s_3\over
\sqrt{s_{12}~s_{13}~s_{23}~(\sum_i s_i)^{3}}}\times
\label{a42}\ee
$$
\times\exp\left[-\sqrt{N\over 2}\left\{\left(\sum_i s_i\right)
\log{\sum_i s_i}+\sum_{i<j}s_{ij}\log{s_{ij}}-
2\sum_i s_i\log s_i-2\log2\sum_i s_i\right\}\right] \ .
$$
When $s_{ij}\geq 0$ as implied by the classical Clebsch-Gordan,
we find that the amplitude diverges exponentially
as expected.
However this exponentially divergent part does not reflect the
true structure constant that turns out to come entirely from the
limit of the GC coefficient of SU(2).

\subsection{Penrose limits of the four-point correlators\label{b5}}
\setcounter{equation}{0}

We have seen so far that, scaling the $SU(2)_k \times U(1)$ charges in the way required
by the Penrose limit, we can derive the three-point couplings of the $H_4$ model.
We can also proceed a bit further and take the limit of the four-point correlators.
This is extremely simple when we scale the quantum numbers in such a way to end up
with a correlator of type $<+++->$ since in this case there is only a finite number
of conformal blocks.

Let us start from (\ref{a18}). Remembering that the cross-ratios used in this paper
and the one in \cite{zf} differ by the exchange
$z_1 \leftrightarrow z_2$ let us set
\ba
2l_1 &=& kp_2 - 2 \jh_2 \ , \hspace{2cm} 2l_2 = kp_1 - 2 \jh_1 \ ,  \nb \\
2l_3 &=& kp_3 - 2 \jh_3 \ , \hspace{2cm} 2l_4 = kp_4 + 2 \jh_4 \ ,
\label{a53}
\ea
with $p_1+p_2+p_3=p_4$.
Let us discard for the moment the prefactor ${\cal N}(l_1,l_2,l_3,l_4)$. According
to $(\ref{a14bis})$ we have, when  $k \rightarrow \infty$,
$p_0=0$ and $p_1 = -\jh_1-\jh_2-\jh_3-\jh_4 \equiv L$ and therefore
the number of blocks is $|L|+1$, as expected.
The index $i$ in the product in $(\ref{a18})$ ranges from $i=1$ to $i=|L|$ and since
in taking the limit the term $|D(t)|^{\frac{4}{k+2}} \rightarrow 1$, the various
integrals become independent and we obtain
\ba
U(z,\bar{z},x,\bar{x}) &\sim&
|z|^{-2(p_1p_2+p_1 \jh_2+p_2\jh_1)}|1-z|^{-2(p_2p_3+p_2 \jh_3+p_3\jh_2)} \nb \\
& & \left ( \int \frac{dt d \bar{t}}{2 \pi i} |t-z|^{-2p_4}|t|^{-2(1-p_3)}|t-1|^{-2(1-p_1)}
|x-t|^{2} \right )^{|L|} \ .
\label{a54}
\ea
The integral can be performed using
\ba
& &\int  \frac{dt d \bar{t}}{2 \pi i} |t-z|^{2(c-b-1)}|t|^{2(b-1)}|t-1|^{-2a} =
\frac{\g(b)\g(c-b)}{\g(c)}|z|^{2(c-1)}|F(a,b,c;z)|^2  \nb \\
&-& \frac{\g(c)\g(1+a-c)}{(1-c)^2\g(a)}|F(1+a-c,1+b-c,2-c;z)|^2 \ .
\label{a55}
\ea
and the result is
\be
\frac{1}{(C_{12}C_{34})^{|L|}} \left ( C_{12}|f(x,z)|^2+C_{34}|g(x,z)|^2 \right )^{|L|} \ ,
\label{a56}
\ee
where we used the definitions in $(\ref{ld9})$, $(\ref{ld10})$ and  $(\ref{ld13})$.

On the other hand, the limit leading to a correlator of the form $<+-+->$ is more subtle,
since there is an infinite number of conformal blocks. It is however possible to verify
that the limit of the KZ equation for $SU(2)$ leads to the correct KZ equation
for $H_4$.

\subsection{Spectral flow\label{b6}}
\setcounter{equation}{0}
\def\T{\Theta}
\def\vt{\vartheta}
\def\te{\tilde\e}

We would like to explain how the spectral-flowed representations
arise in the contraction of the $SU(2)_k \times U(1)$ CFT.

There, we have two independent spectral flows acting on the
$SU(2)_k$ and $U(1)$.
\be
J^{\pm}_n\to J^{\pm}_{n\mp \e}\sp J^3_n\to J^3_n-\e{k\over 2}\delta_{n,0}
\sp L^{SU(2)}_{n}\to L^{SU(2)}_{n}-\e J^3_{n}+{k\over
4}\e^2\delta_{n,0}
\ee
and
\be
J^0_n\to J^0_n+\te{k\over 2}\delta_{n,0}\sp L^{U(1)}_{n}\to L^{U(1)}_{n}+\te J^0_{n}+{k\over
4}\te^2\delta_{n,0}
\ee
The SU(2) spectral flow when $\e$ is an integer is essentially
generating the symmetry under affine Weyl translations.
In terms of the $l,m$ quantum numbers, affine-Weyl translations act on characters as $m\to m+nk$.
We also have the non-trivial external automorphism $l\to k/2-l$ accompanied by a affine translation
(see \cite{dual} for more details).

For the total stress tensor $L^{SU(2)\times
U(1)}_m=L^{SU(2)}_{n}-L^{U(1)}_{n}$ we obtain
\be
L^{SU(2)\times
U(1)}_m\to L^{SU(2)\times
U(1)}_m-\e J^3_{n}-\te J^0_n+{k\over
4}(\e^2-\te^2)\delta_{n,0}
\ee
On the other hand, the most general spectral flow of the $H_4$
algebra is described also by two parameters
\be
P^{\pm}_n \to  P^{\pm}_{n\mp a}\sp
K_n\to K_n -i a \d_{n,0} \sp
J_n \to J_n -i b \d_{n,0}
\label{gspec}\ee
$$
 L_n \to L_n -i a J_n-ibK_n-ab\delta_{n,0} \ .
$$
Taking into account the relation of the $SU(2)\times U(1)$ and
$H_4$ Cartan currents
\be
J^3_n=iJ_n-i{k\over 2}K_n\sp J^0_n=-i{k\over 2}K_n
\ee
we see that the we can relate the two spectral flows by
choosing $a=\e$ and $2b=-k(\e+\te)$ with $\e\in Z$ and $b\sim
{\cal O}(1)$.

As we show in the next subsection, the relevant spectral flow in
the $H_4$ theory has $b=0$.

The spectral-flowed states are generated as follows in the $SU(2)_k$ theory:

Consider the state $|\chi\rangle =(J^-_{-1})^{k-2l}|l;m=-l\rangle$
This state is a lowest-weight state of a different SU(2) algebra:
$J^{\pm}_{\pm 1}, \hat J_3=J^3_{0}+{k\over 2}$ since
$J^-_{-1}|\chi\rangle=0$ (non-trivial affine null vector).
It has $\hat J|\chi\rangle =\left(l-{k\over 2}\right)|\chi\rangle$
and generates an SU(2) representation with $L={k\over 2}-l$.
Unitarity of such representations implies the affine cutoff $l\leq {k\over 2}$
This state is obtained from the highest-weight states of affine
representation by an  affine Weyl translation.
In the scaling limit and with appropriate dressing of the U(1) charge, this state is a lws of a
spectral-flowed
$H_4$ representation.

In the scaling limit it still has $l={k\over 2}p-\jh$ but has $m=-{k\over 2}(2-p)-\jh$.
This class of states are the  spectral-flowed once, lowest-weight representations.

The analogous highest-weight representations are generated by the state
\be
|\hat \chi\rangle =
(J^+_{-1})^{k-2l}|l;m=l\rangle \ ,
\ee
highest-weight with respect to the algebra
$J^{\pm}_{\mp 1}, \hat J_3=J^3_{0}-{k\over 2}$ since $J^+_{-1}|\hat\chi\rangle=0$
It has $m=k-l$ which in the scaling limit becomes $m={k\over 2}(2-p)+\jh$
For $0\leq p \leq 1$, $1\leq 2- p \leq 2$.

Thus, the spectral
flow by $\pm 1$ steps generates the spectral-flowed $V^{\pm}$ representations
with $1\leq p \leq 2$.

\subsection{Penrose contraction of characters\label{b7}}
\setcounter{equation}{0}

We would like here to indicate the precise decomposition of the
$SU(2)_k\times U(1)$ affine representations into $H_4$ ones.
The starting point are the affine $SU(2)_k$ characters.
Define the $\t$-functions
\be
\T_{a,b}(\tau,z)=\sum_{n\in Z}~~ e^{2\pi i a\left[\left(n+{b\over
2a}\right)^2\tau-\left(n+{b\over
2a}\right)z\right]}
\ee
Then, the $SU(2)_k$ character for the representation with spin $l$ is
\be
\chi_{l}(\tau,z)\equiv Tr_l[e^{2\pi i\tau ~L_0}~e^{2\pi i~z
J^3_0}]={\T_{2l+1,k+2}-\T_{-2l-1,k+2}\over \T_{1,2}-\T_{-1,2}}
\ee We also have
\be
\T_{1,2}-\T_{-1,2}=-i~\vt_1(z|\tau)
\ee
We will decompose the affine character as
\be
\chi_{l}(\tau,z)=\sum_{n\in Z}(\chi_{l,n}^++\chi_{l,n}^-)
\ee
where
\be
\chi_{l,n}^+=i{e^{2\pi i\tau(k+2)\left(n+{2l+1\over
2(k+2)}\right)^2}~e^{-2\pi iz\left((k+2)n+{2l+1\over
2}\right)}\over ~\vt_1(z|\tau)}
\ee
\be
\chi_{l,n}^-=-i{e^{2\pi i\tau(k+2)\left(n+{2l+1\over
2(k+2)}\right)^2}~e^{2\pi iz\left((k+2)n+{2l+1\over
2}\right)}\over ~\vt_1(z|\tau)}
\ee
In this representation, if the affine SU(2) representation had no
null vectors, then the character would be given by
$\chi_{l,0}^++\chi_{l,0}^-$
The other terms in the series come from subtracting the embedded
Verma modules (due to the presence of the null vectors) in order
to obtain the irreducible character.
The summation over the integer n imposes the affine Weyl symmetry
on the character \cite{dual}.

The time-like U(1) character is
\be
\psi_q(\tau,w)={e^{-2\pi i{q^2\over k}\tau+2\pi i w~q}\over
\eta(\tau)}
\ee

The $H_4$ $V^{\pm}$ characters are
\be
\zeta^{\pm}_{p,\jh}(\tau,z,w)\equiv Tr[e^{2\pi i\tau ~L_0}~e^{-2\pi ~(z
J_0+wK_0)}]=\pm i{e^{2\pi i\tau(-p\jh+p(1-p)/2)}~e^{\mp 2\pi
i~wp\pm 2\pi i z\left(\jh - {1\over 2}\right)}\over
\eta(\tau)\vt_1(z|\tau)}
\ee
where $0<p<1$.
For the spectral-flowed representations we obtain from
(\ref{gspec})
\be
\zeta^{\pm}_{p,\jh;\mp a,\pm b}(\tau,z,w)=
e^{2\pi i\tau ~ab\pm 2\pi i(bz-aw)}~\zeta^{\pm}_{p,\jh}(\tau,z\pm a\tau,w\mp b\tau)
\ee
$$
=e^{\mp i\pi a}~e^{2\pi i \tau\left[{a^2\over 2}+ab+bp+a\left(\jh
-{1\over 2}\right)\right]}~e^{\pm 2\pi i z(a+b)\mp2\pi i w ~a}\zeta^{\pm}_{p,\jh}(\tau,z,w)
$$

Using the fact that
\be
w~J^0+zJ^3=(w+z){k\over 2}iK+z(iJ)
\ee
we will redefine the characters
\be\hat\chi_{q,l}(w,z,\tau)=\psi_q({2w\over
k}-z,\tau)\chi_l(z,\tau)=Tr[e^{2\pi i\tau ~L_0}~e^{-2\pi w
~K_0-2\pi z~J_0}]
\ee
This redefinition guarantees that in the limit we will obtain the
$H_4$ characters.

For $l={k\over 2}p-\jh$, we obtain
\be
\lim_{k\to\infty}~~ \psi_{\mp k\left({p\over 2}+n\right)}({2w\over
k}-z,\tau)\chi_{l,n}^{\pm}(z,\tau)=\zeta^{\pm}_{p,\jh;n,0}(\tau,z,w)
\ee

This describes the precise decomposition of the characters of the
parent theory into those of the pp-wave theory. In particular,
the SU(2) representation with $2l\sim kp$ provides the even
integer spectral flow of the representations with $p^+=p$ and the
odd integer spectral flow of the representations with $p^+=1-p$.
On the other hand, the representation with $2l\sim k(1-p)$ provides
the even
integer spectral flow of the representations with $p^+=1-p$ and the
odd integer spectral flow of the representations with $p^+=p$.

\section{Calculation of general twist-field correlators\label{c}}
\renewcommand{\theequation}{\Alph{section}.\arabic{equation}}
\setcounter{equation}{0}

In terms of the free-field realization of
the $H_4$ algebra \cite{kk} reviewed in section 3.2,
the primary vertex operators for $V^{\pm}$ representations
are given by the product of twist fields $H_p^{\mp}$ and
exponentials $e^{i(p v+\jh u)}$
while those for $V^0$ representations are essentially
of the form $e^{i\jh u + i \vec{p} \vec{y}}$.
As a consequence, the computation
of correlation functions for the $H_4$ model
reduces to the computation of correlators between twist fields \cite{dfms},
the main difference being that we do not restrict the twist parameter to be a
rational number and that we consider
these amplitudes as describing scattering processes in space-time and not in some internal
compactification manifold. In this appendix we will
briefly review the method used in \cite{dfms} to compute these correlators and then
compare them to the ones resulting from the solution of the KZ equation.

Consider a complex boson $y = y_1+iy_2$ and its conjugate  $\tilde{y} = y_1-iy_2$,
with the propagator $<y(z) \tilde{y}(w)> = -2 \log{(z-w)}$.

The method exploits the fact that the Green's function for the fields $\p y$ and
$\p \tilde{y}$ in the presence of a collection of twist fields $H^{\s_i}_{p_i}$ with $\s_i = \pm$
\be
g(z,w;z_i,\bar{z}_i) = \frac{< - \frac{1}{2} \p y(z) \p \tilde{y}(w)
\prod_{i=1}^4 H^{\s_i}_{p_i}(z_i,\bar{z}_i)>}{<\prod_{i=1}^4 H^{\s_i}_{p_i}(z_i,\bar{z}_i)>} \ ,
\label{green1}
\ee
is almost completely fixed by the monodromies around the points where the twist fields
are inserted and by the pole structure required by the OPE.
The interest of this function lies in the fact that if we take
the limit $z \rightarrow w$ and remove the double pole, we obtain
the expectation value of the stress-energy tensor in the presence of four twist fields
\be
<T(z)> = \frac{<T(z) \prod_{i=1}^4 H^{\s_i}_{p_i}(z_i,\bar{z}_i)>}{<\prod_{i=1}^4
H^{\s_i}_{p_i}(z_i,\bar{z}_i)>} \ ,
\label{bb1}
\ee
and therefore, using the standard OPE between
stress-energy tensor and primary fields and
taking the limit $z \rightarrow z_i$, we can extract a first-order
differential equation for the correlator
$ <\prod_{i=1}^4 H^{\s_i}_{p_i}(z_i,\bar{z}_i)>$ we want to compute.
Let us show how to determine the function $g(z,w;z_i,\bar{z}_i)$.
We first introduce two holomorphic functions which have the same
monodromies around the twist fields as  $\p y(z)$ and $\p \tilde{y}(z)$.
They are
\ba
\w(z) &=& \prod_{i=1}^4 (z-z_i)^{\nu_i} \ , \nb \\
\tilde{\w}(z) &=& \prod_{i=1}^4 (z-z_i)^{-1 - \nu_i} \ ,
\ea
where $\nu_i = -p_i$ if the field in $z_i$ is $H^-_{p_i}$ and $\nu_i=-1+p_i$ if it is $H^+_{p_i}$,
We then
split the Green function $(\ref{green1})$ in a singular and in a regular part writing
\be
g(z,w;z_i,\bar{z}_i) = \w(z) \tilde{\w}(w) \left ( \frac{P(z,w;z_i,\bar{z}_i)}{(z-w)^2} +
A(z_i,\bar{z}_i) \right ) \ ,
\ee
where the function $P$, which parameterizes the singular part, is fixed by the pole structure
up to shifts in the regular part $A$.
In order to fix the function $A$ we introduce another Green function
\be
h(\bar{z},w;z_i,\bar{z}_i) = \frac{< - \frac{1}{2} \bar{\p} y(\bar{z}) \p \tilde{y}(w)
\prod_{i=1}^4 H_i(z_i,\bar{z}_i)>}{<\prod_{i=1}^4 H_i(z_i,\bar{z}_i)>} \ ,
\label{green}
\ee
which can be parameterized as
\be
h(z,w;z_i,\bar{z}_i) = \w(\bar{z}) \tilde{\w}(w) B(z_i,\bar{z}_i) \ ,
\ee
since there are no singular terms in the OPE between $\bar{\p} y $ and $\p \tilde{y}$.

The functions $A$ and $B$ are fixed by
the requirement
that the fields $y$ and $\tilde{y}$ do not change when
carried around a collection of twist fields with net twist zero. This constraint
can be written as
\be
\oint_{\gamma_a} \left ( dz \p y + d \bar{z} \bar{\p} y \right ) = 0 \ , \hspace{1cm} a = 1, 2 \ ,
\label{monodromy}
\ee
where the $\gamma_a$ form a basis for the closed loops on the cut
complex plane. In terms of the functions $g(z,w)$ and $h(\bar{z},w)$ the
condition (\ref{monodromy}) amounts to
\be
\oint_{\gamma_a} \left ( dz g(z,w) + d \bar{z} h(\bar{z},w) \right ) = 0 \ ,
\ee
and leads to a system of two linear equations that can be solved for $A$ and $B$.
Further details on this method can be found in \cite{dfms}.

We compute using this method the following correlator
\be
{\cal K}(p_1,p_2,p_3,p_4) = <R^+_{p_1,\jh_1;0}(z_1)R^-_{p_2,\jh_2;0}(z_2)
R^+_{p_3,\jh_3;0}(z_3)R^-_{p_4,\jh_4;0}(z_4)> \ ,
\ee
with $p_1+p_3=p_2+p_4$ and $\sum_{i=0}^4 \jh_i = 0$.
The result is
\be
{\cal K}(p_1,p_2,p_3,p_4)
\sim \frac{\kappa}{|S(z,\bar{z})|} \ ,
\label{t22}
\ee
where $z=\frac{z_{12}z_{34}}{z_{13}z_{24}}$ and
\ba
S(z,\bar{z}) &=& \kappa [ B(p_2,1-p_3)B(p_1,1-p_2)|1-z|^{p_2-p_3}F_2(1-z)\bar{G}_1(\bar{z}) \nb \\
&+& B(p_3,1-p_2)B(p_2,1-p_1)|1-z|^{p_3-p_2}F_1(z)\bar{G}_2(1-\bar{z}) ] \ ,
\ea
where $B(a,b)$ is the Euler beta function and $F_i$, $G_i$ are
hypergeometric functions
\ba
F_1(z) &=& F(p_3,1-p_1,1+p_2-p_1,z) \ , \hspace{0.4cm}
G_1(z) = F(1-p_3,p_1,1-p_2+p_1,z) \ , \nonumber \\
F_2(z) &=& F(1-p_4,p_2,1+p_2-p_3,z) \ , \hspace{0.4cm}
G_2(z)=F(p_4,1-p_2,1-p_2+p_3,z) \ .
\ea
Finally the constant $\kappa$ is given by
\be
\kappa = \left [ \frac{\G(p_4)\G(1-p_4)}{\prod_{i=1}^3 \G(p_i)\G(1-p_i)} \right ]^{1/2} \ .
\ee
This correlator coincides with the corresponding one that can be extracted
from $(\ref{pmk-corr})$ when $L=0$.

When $p_1=p_4=p$ and $p_2=p_3=l$ the correlation function
in (\ref{t22}) reduces to
\be
{\cal K}(p,l,l,p) \sim \frac{1}{|S(z,\bar{z})|} \ ,
\ee
with
\be
S(z,\bar{z})=B(l,1-p)F_1(z)\bar{G}_2(1-\bar{z})+B(p,1-l)F_2(1-z)\bar{G}_1(\bar{z}) \ ,
\ee
and
\ba
F_1(x) &=& F(l,1-p,1+l-p,x) \ , \hspace{1cm}
G_1(x) = F(1-l,p,1+p-l,x)  \ , \nonumber \\
F_2(x) &=& F(1-p,l,1,x) \ , \hspace{2.4cm}
G_2(x)=F(p,1-l,1,x) \ .
\ea
Finally when $p=l$ the correlator reduces to
\be
{\cal K}(p,p,p,p) \sim
\frac{1}{\G(p)\G(1-p)[F(z)\bar{F}(1-\bar{z})+F(1-z)\bar{F}(\bar{z})]} \ ,
\ee
where $F(z)=F(p,1-p;1,z)$.

\section{Four-point amplitudes\label{d}}
\renewcommand{\theequation}{\Alph{section}.\arabic{subsection}.\arabic{equation}}
\setcounter{equation}{0}

In this appendix we provide more details concerning the solutions of the
KZ equations relevant for the different types of four-point correlators.

\subsection{$<+++->$ correlators\label{d1}}
\setcounter{equation}{0}

Here we use the same notation as in section $5.1$. Consider a correlator
of the form
\be
<\F^+_{p_1,\jh_1}(z_1,\bar{z}_1,x_1,\bar{x}_1)
\F^+_{p_2,\jh_2}(z_2,\bar{z}_2,x_2,\bar{x}_2)\F^+_{p_3,\jh_3}(z_3,\bar{z}_3,x_3,\bar{x}_3)
\F^-_{p_4,\jh_4}(z_4,\bar{z}_4,x_4,\bar{x}_4)> \ ,
\label{ca1}
\ee
with
\be
p_1+p_2+p_3 = p_4 \ .
\label{ca2}
\ee
The relevant KZ equation for the conformal blocks is
\ba
\p_z F_n(x,z) &=& \frac{1}{z} \left [ -(p_1x+p_2x(1-x))\p_x +Lp_2x \right ]
F_n(x,z) \nb \\
&-& \frac{1}{1-z} \left [ (1-x)(p_2x+p_3)\p_x+Lp_2(1-x) \right ] F_n(x,z) \ ,
\label{ca3}
\ea
where $L=\jh_1+\jh_2+\jh_3+\jh_4$ and
\be
x = \frac{x_2-x_1}{x_3-x_1} \ , \hspace{1cm}
\bar{x} = \frac{\bar{x}_2-\bar{x}_1}{\bar{x}_3-\bar{x}_1} \ .
\label{ca4}
\ee
The correlator is non vanishing only for $L \le 0$ and in this case
the number of conformal blocks is $N = |L| + 1$.
We look for solutions that behave for small $z$ as
$F_n(z,x) \sim z^{-n(p_1+p_2)}f_n(x)$, $n = 0, ..., |L|$
so that we can identify them with
the conformal blocks corresponding to intermediate
$\F^+_{p_1+p_2,\jh_1+\jh_2+n}$ representations. The previous equation
requires that
\be
f_n(x) = x^n \left ( 1-\frac{p_2}{p_1+p_2}x \right )^{|L|-n}\,\,.
\ee
The behavior for small $z$, together with the structure of the conformal
blocks for $SU(2)_k$ suggest the following ansatz
\be
F_n(z,x) = f^n(z,x) (g(z,x))^{|L|-n} \ ,
\label{ca5}
\ee
where  $f(z,x) = f_0(z)+xf_1(z)$, $g(z,x)=g_0(z)+xg_1(z)$.
The KZ equation leads to an hyper-geometric equation for the four functions
$f_i$ and $g_i$ and we obtain
\be
f(z,x) = \frac{z^{1-p_1-p_2} p_3}{1-p_1-p_2}\f_0 - x z^{-p_1-p_2}\f_1 \ , \hspace{1cm}
g(z,x) = \g_0 -\frac{xp_2}{p_1+p_2} \g_1 \ ,
\label{ca6}
\ee
where
\ba
\f_0 &=& F(1-p_1,1+p_3,2-p_1-p_2,z) \ , \hspace{1cm}
\g_0 = F(p_2,p_4,p_1+p_2,z) \ , \nb \\
\f_1 &=& F(1-p_1,p_3,1-p_1-p_2,z) \ , \hspace{1cm}
\g_1 = F(1+p_2,p_4,1+p_1+p_2,z) \ .
\label{ca7}
\ea
From the Penrose limit of the $SU(2)_k$ couplings we know that the
quantum structure constants of the $H_4$ WZW model have a very
simple dependence on the quantum number $p$ and $\jh$.
Taking this dependence into account, it is natural to
consider the following combination of conformal blocks
\be
{\cal A} \sim \sum_{n=0}^{|L|} \frac{|L|!}{n!(|L|-n)!} R^n |F_n(z,x)|^2 \ ,
\label{ca8}
\ee
where the constant $R$ is fixed by
monodromy invariance.
Once the resulting expression is properly normalized we obtain
the correlator displayed in $(\ref{ld12})$
\be
{\cal A}(z,\bar{z},x,\bar{x})
= |z|^{2\ka_{12}}|1-z|^{2\ka_{14}}
\frac{ \sqrt{C_{12}C_{34}} }{|L|!} \left ( C_{12}|f(z,x)|^2+C_{34}|g(z,x)|^2 \right )^{|L|} \ ,
\label{ca9}
\ee
where
\be
C_{12} =  \frac{\g(p_1+p_2)}{\g(p_1)\g(p_2)} \ , \hspace{1cm}
C_{34} = \frac{\g(p_4)}{\g(p_3)\g(p_4-p_3)} \ .
\label{ca10}
\ee

\subsection{$<+-+->$ correlators\label{d2}}
\setcounter{equation}{0}

Here we use the same notation as is section $5.2$.
We consider a correlator of the form
\be
<\F^+_{p_1,\jh_1}(z_1,\bar{z}_1,x_1,\bar{x}_1)
\F^-_{p_2,\jh_2}(z_2,\bar{z}_2,x_2,\bar{x}_2)
\F^+_{p_3,\jh_3}(z_3,\bar{z}_3,x_3,\bar{x}_3)
\F^-_{p_4,\jh_4}(z_4,\bar{z}_4,x_4,\bar{x}_4)> \ ,
\label{cb1}
\ee
with $p_1+p_3=p_2+p_4$.
The KZ equation is
\ba
z(1-z) \p_z F_n(x,z) &=& \left [ x\p^2_x +\left ( ax+1-L \right )\p_x
+\frac{x}{4}(a^2-b^2) + \r_{12} \right ] F_n(x,z) \nb \\
&+& z \left [ -2ax \p_x +\frac{x}{4}(b^2-c^2)  - \r_{12} - \r_{14} \right ]   F_n(x,z) \ ,
\label{cb2}
\ea
where $L=\jh_1+\jh_2+\jh_3+\jh_4$ and $x = (x_1-x_3)(x_2-x_4)$. Moreover
\be
2a = p_1+p_3 \ , \hspace{0.5cm} b = p_1-p_2 \ , \hspace{0.5cm}
c= p_2-p_3 \ ,
\label{cb3}
\ee
and
\be
\r_{12} = \frac{(1-L)}{2}(a-b) \ ,  \hspace{1cm}
\r_{14} = \frac{(1-L)}{2}(a-c)\ .
\label{cb4a}
\ee
According to (\ref{tensor}) when $p_1 > p_2$ and $L \le 0$
the states flowing in the $s$-channel belong to the representations
$V^+_{p_1-p_2, \jh_1+\jh_2+n}$ with $n \in \mathbb{N}$. In this case
we require
\be
F_n(z,x) \sim z^{n(p_1-p_2)}\f_n(x)  ~~~~{\rm for}~~~~ z \sim 0\ee
The  KZ equation then
implies
\be
\f_n(x) = e^{-\frac{a-b}{2}x} L_{n}^{|L|}(t) \ , \hspace{1cm} t = -(p_1-p_2)x \ ,
\label{cb4}
\ee
where $L_n^{|L|}(t)$ is the $n$-th generalized Laguerre polynomials.

If we insert in the KZ equation a function of the form
\be
F(x,z) = \frac{e^{xg(z)}}{(f(z))^{1-L}} \ ,
\label{cb5}
\ee
we derive from (\ref{cb2}) the following equations for
$f$ and $g$
\ba
g &=& -z(1-z) \p \ln{[z^{ \frac{a-b}{2}}(1-z)^{ \frac{a-c}{2}} f(z)]}  \ , \nb \\
z(1-z)\p g &=& g (g+a(1-z)-az)+\frac{a^2-(1-z)b^2-zc^2}{4} \ ,
\label{cb6}
\ea
and a hyper-geometric equation
for $f$ whose two solutions are
\ba
f_1(z) &=& F(p_3,1-p_1,1-p_1+p_2,z) \ , \nb \\
f_2(z) &=& z^{p_1-p_2}F(p_4,1-p_2,1-p_2+p_1,z) \ .
\label{cb7}
\ea
We have thus found the conformal block for $n=0$
\be
F_0(z,x) = \frac{e^{xg_1(z)}}{(f_1(z))^{1-L}} \ .
\label{cb8}
\ee
The other blocks can be expressed in terms of
$F_0(z,x)$, Laguerre polynomials and the two solutions $f_1$, $f_2$
of the hyper-geometric equation
\be
F_n(z,x) = \n_n F_0(z,x) L^{|L|}_n(x  \g_{\psi}(z)) \psi(z)^n \ ,
\label{cb9}
\ee
where
\be
\psi(z) = \frac{f_2(z)}{f_1(z)}  \ , \hspace{1cm}  \g_{\psi}(z) =
-z(1-z)\p \ln{\psi} \ ,  \hspace{1cm} \n_n = \frac{n!}{(p_1-p_2)^n} \ .
\label{cb10}
\ee
All of them are solutions of the KZ equation.
Also in this case, the dependence on the $p$ and $\jh$ quantum number
of the three-point couplings of the $H_4$ WZW model obtained from
the Penrose limit of the $SU(2)_k$ structure constants, suggests
to look for a combination of the conformal blocks having the following
form
\be
{\cal A} \sim \sum_{n=0}^{\infty} \frac{1}{n!(n-L)!} R^n |F_{n}(z,x)|^2 \ ,
\label{cb11}
\ee
where the constant $R$ is fixed by monodromy invariance.
The sum can be performed using the identity
\be
\sum_{n=0}^\infty \frac{n!}{\G(n+\a+1)}
L^{\a}_n(x)L^{\a}_n(y)z^n = \frac{e^{-\frac{z(x+y)}{1-z}}}{1-z}
(xyz)^{-\frac{\a}{2}} I_{\a} \left (\frac{2\sqrt{xyz}}{1-z} \right ) \ ,
\label{cb12}
\ee
where  $I_{\a}$ is a Bessel function of imaginary argument
\be
I_{\a}(w) = \left ( \frac{w}{2} \right )^{\a} \sum_{n=0}^{\infty}
\left ( \frac{w}{2} \right )^{2n} \frac{1}{n!\G(\a+n+1)} \ ,
\label{cb13}
\ee
and we obtain the four-point correlator displayed in
$(\ref{pmk-corr})$
\be
{\cal A}(z,\bar{z}, x, \bar{x})
= \frac{\mu_L  |z|^{2\ka_{12}}|1-z|^{2\ka_{14}}}{S^{1+|L|}}
\left |e^{xq(z)-xz(1-z)\p \ln{S}} \right |^2
\left ( \frac{u}{2} \right )^{-|L|}
I_{|L|}(u) \ ,
\label{cb14}
\ee
where
\be
r = \frac{C_{12}C_{34}}{(p_1-p_2)^2} \ , \hspace{1cm}
C_{12} = \frac{\g(p_1)}{\g(p_2)\g(p_1-p_2)} \ , \hspace{1cm}
C_{34} = \frac{\g(p_4)}{\g(p_3)\g(p_4-p_3)} \ ,
\label{cb15}
\ee
and we defined
\be
S = |f_1|^2 - r |f_2|^2 \ , \hspace{1cm}
\mu_{L} = C_{12}^{\frac{1}{2}} C_{34}^{\frac{1}{2}+|L|} \ ,
\label{cb16}
\ee
as well as
\be
u = \frac{2\sqrt{r}|xz(1-z)\p \psi|}{1-r|\psi|^2}  = \frac{2\sqrt{r}|xz(1-z)W(f_1,f_2)|}{S} \ ,
\label{cb17}
\ee
with $W(f_1,f_2)$ the Wronskian of the two solutions of the hyper-geometric equation,
\be
W(f_1,f_2) = (p_1-p_2)z^{p_1-p_2-1}(1-z)^{p_2-p_3-1} \ .
\label{cb18}
\ee

The case  $p_2>p_1$, when
the representations $V^-_{p_2-p_1,\jh_1+\jh_2-n}$ with $n \in \mathbb{N}$,
 flow in the $s$-channel,
can be treated in exactly the same way.

When $p_1=p_2=p$ and $p_3=p_4=l$, the intermediate states in the $s$-channel
belong to $V^0_{s,\jh}$ representations and the amplitudes can be
represented as an integral over the Casimir $s^2$.
One needs the following integral
\be
\int_0^\infty dt \ e^{-\a t} J_{\n}(2\b\sqrt{t}) J_{\n}(2\g\sqrt{t}) =
\frac{e^{\frac{\b^2+\g^2}{\a}}}{\a}I_{\n} \left ( \frac{2\b\g}{\a} \right ) \ .
\label{cb19}
\ee

We can verify that, as we vary the external momenta,
the correlator $(\ref{cb14})$, which in the $s$ channel factorizes
on discrete representations, changes continuously
to the correlator $(\ref{cont-k-corr})$, which factorizes on continuous
representations.
The two solutions of the hypergeometric equation for the correlator with $p_1=p_4=p$ and
$p_2=p_3=l$
\ba
c_1(z) &=& F(l,1-p,1,z) \ , \nb \\
c_2(z) &=& [ \ln{z}+2\psi(1)-\psi(l)-\psi(1-p)]  c_1(z) \nb \\
&+& \sum_{n=0}^{\infty} \frac{(l)_n(1-p)_n}{n!^2}
[\psi(l+n)+\psi(1-p+n) - 2 \psi(n+1)] z^n \ ,
\label{cb20}
\ea
can indeed be obtained as limits of the two
solutions $f_1$ and $f_2$ in (\ref{cb7}). Setting $p_1=p$,
$p_2=p-\e$, $p_3=l$ and $p_4=l+\e$, we obtain
\be
f_1(z) = c_1(z) + \e b_1(z) \ , \hspace{1cm}
f_2(z) = c_1(z) + \e ( c_2(z) + b_1(z))   \ ,
\label{cb21}
\ee
where
\be
b_1(z) = \sum_{n=0}^{\infty} (l)_n (1-p)_n \left [ \psi(n+1)-\psi(1) \right ]
\frac{z^n}{(n!)^2} \ ,
\label{cb22}
\ee
and we used that
\ba
& & F(a+\e,b+\e,c+\e,z) = F(a,b,c,z)  \\
&+& \e \sum_{n=0}^{\infty} \frac{(a)_n(b)_n}{(c)_n}
\left [ \psi(a+n)-\psi(a)+\psi(b+n)-\psi(b)-\psi(c+n)+\psi(c)) \right ]
\frac{z^n}{n!} +O(\e^2) \nb \ .
\label{cb23}
\ea
Taking into account the behavior of the three-point couplings as $\e \rightarrow 0$
it is easy to verify that
\be
\lim_{\e \rightarrow 0} \e^{-1}S(p,-p+\e,l,-l-\e) = S(p,-p,l,-l) \ , \hspace{1cm}
\lim_{\e \rightarrow 0}  \e^{-1} W(f_1,f_2) = W(c_1,c_2) \ .
\label{cb24}
\ee
The necessary powers of $\e$ are provided by $\m_L$ that behaves as $\m_L \sim \e^{1-L}$
in the limit.

\subsection{$<++- \ 0>$ correlators}
\setcounter{equation}{0}

Here we use the same notation as in section 5.3. Consider a correlator of the form
\be
<\F^+_{p_1,\jh_1}(z_1,\bar{z}_1,x_1,\bar{x}_1)
\F^+_{p_2,\jh_2}(z_2,\bar{z}_2,x_2,\bar{x}_2)\F^-_{p_3,\jh_3}(z_3,\bar{z}_3,x_3,\bar{x}_3)
\F^0_{s,\jh_4}(z_4,\bar{z}_4,x_4,\bar{x}_4)> \ ,
\label{cc1}
\ee
with
\be
p_1+p_2 = p_3 \ .
\label{cc2}
\ee
The KZ equation reads
\be
z(1-z) \p_z F_n = -\left[ p_3x\p_x+\frac{s}{2\sqrt{2}}(p_1-p_2)x \right ]
F_n + z \left [ \left ( p_2x-\frac{s}{\sqrt{2}} \right ) \p_x
- \frac{sp_2}{2\sqrt{2}}x \right ] F_n \ ,
\label{cc3}
\ee
where  $L=\jh_1+\jh_2+\jh_3+\jh_4$ and $x= (x_1-x_2)x_4$.
The ansatz for the conformal blocks is
\be
F_n(z,x) = (s \f(z)+x\g(z))^ne^{s^2 \h(z) +sx \psi(z)} \ ,
\label{cc4}
\ee
with $n \ge 0$.
From $(\ref{cc3})$ it follows that the functions $f$, $g$ $\psi$ and
$\h$ have to satisfy the following equations
\ba
z(1-z) \p \psi &=& (zp_2-p_3) \psi - \frac{(p_1-p_2)+p_2z}{2\sqrt{2}}  \ , \nb \\
z(1-z) \p \g &=& (zp_2-p_3)\g \ , \nb \\
\g &=& -\sqrt{2}(1-z)\p \f \ , \nb \\
\psi &=& - \sqrt{2}(1-z) \p \h \ .,
\label{cc5}
\ea
and are given by
\ba
\f(z) &=& \frac{z^{1-p_3}}{\sqrt{2}(1-p_3)}F(1-p_1,1-p_3,2-p_3,z) \ , \nb \\
\g(z) &=& -z^{-p_3}(1-z)^{p_1} \ , \nb \\
\psi(z) &=& -\frac{1}{2\sqrt{2}} +\frac{p_2}{\sqrt{2}p_3}(1-z)F(1+p_2,1,1+p_3,z) \ , \nb \\
\h(z) &=& -\frac{z p_2}{2p_3} \ {}_3F_2(1+p_2,1,1;1+p_3,2;z)  - \frac{1}{4} \ln{(1-z)} \ .
\label{cc6}
\ea

\subsection{$<+- \ 0 \ 0>$ correlators}
\setcounter{equation}{0}

Here we consider a correlator of the form
\be
<\F^+_{p,\jh_1}(z_1,\bar{z}_1,x_1,\bar{x}_1)
\F^-_{p,\jh_2}(z_2,\bar{z}_2,x_2,\bar{x}_2)\F^0_{s_3,\jh_3}(z_3,\bar{z}_3,x_3,\bar{x}_3)
\F^0_{s_4,\jh_4}(z_4,\bar{z}_4,x_4,\bar{x}_4)> \ .
\label{cd1}
\ee
As we showed in section 5.4 this correlator is given by
\be
{\cal A}(u,\bar{u},x,\bar{x})
=  {\cal C}_{+-0}(p,s_3)  {\cal C}_{+-0}(p,s_4) |u|^{2\ka_{12}}|1-u|^{2\ka_{14}}
\left | e^{x \omega(u) + \frac{\psi(u)}{x}} \right |^2 \sum_{n \in \mathbb{Z}}
\left | x u^{-p} \right |^{2n} \ ,
\label{cd2}
\ee
where $u=1-z$, $L = \sum_{i=1}^4 \jh_i$, $x = \frac{x_4}{x_3}$ and
\be
\ka_{12} = h_1+h_4-\frac{h}{3} -p \jh_4 - \frac{s_4^2}{2} \ ,
\hspace{1cm}
\ka_{14} =   \frac{s_3^2+s_4^2}{2} - \frac{h}{3} \ .
\label{cd3}
\ee
Moreover
\be
\omega(u) = - \frac{s_3s_4}{2p} F_p(u) \ ,  \hspace{0.4cm}
\psi(u) =  - \frac{s_3s_4}{2(1-p)} u F_{1-p}(u) \ ,
\label{cd4}
\ee
where $F_p(u) = F(p,1,1+p,u)$ and $F_{1-p}(u) = F(1-p,1,2-p,u)$.
We can also rewrite $(\ref{cd2})$ as
\be
\left | e^{x \omega(u) + \frac{\psi(u)}{x}} \right |^2 \sum_{n \in \mathbb{Z}} \left | x u^{-p} \right |^{2n}
= \sum_{l,m \in \mathbb{Z}} x^l \bar{x}^m |u|^{-p(l+m)} \chi^{\frac{m-l}{2}} I_{|m-l|}(R) \ ,
\label{cd5}
\ee
where
\ba
\chi &=& \frac{u p F_{1-p} + (1-p) |u|^{2p} \bar{F}_p}{\bar{u} p \bar{F}_{1-p} + (1-p) |u|^{2p} F_p} \ , \nb \\
R^2 &=& s^2_3 s^2_4 |u|^{-2p} \left ( u \frac{F_{1-p}}{1-p} + |u|^{2p} \frac{\bar{F}_p}{p} \right )
\left ( \bar{u} \frac{\bar{F}_{1-p}}{1-p} + |u|^{2p} \frac{F_p}{p} \right ) \ .
\label{cd6}
\ea
Here we want to
reproduce this expression using the free-field representation for the vertex
operators. For simplicity let us concentrate on the following subset of components
\be
< R^+_{p,\jh_1;0,0}(z_1,\bar{z}_1)R^-_{p,\jh_2;0,0}(z_2,\bar{z}_2)
\F^0_{s_3,\jh_3}(z_3,\bar{z}_3,x_3,\bar{x}_3)
\F^0_{s_4,\jh_3}(z_4,\bar{z}_4,x_4,\bar{x}_4) > \ ,
\label{cd20}
\ee
that is, we keep only the ground state for the $\F^\pm$ representations.
In terms of free-fields, as explained in section 3.2, we can write
\ba
R^+_{p,\jh_1;0;0} &=& e^{i(\jh u + p v)} H^-_p \ ,  \hspace{1cm}
R^-_{p,\jh_2;0;0} = e^{i(\jh u - p v)} H^+_p \ , \nb \\
\F^0_{s_i,\jh_i} &=& e^{\frac{i s_i}{2} \left ( y
\sqrt{\frac{x_i}{\bar{x_i}}} + \tilde{y}
\sqrt{\frac{\bar{x_i}}{x_i}} \right ) + i \jh_i u}
\sum_{n_i \in \mathbb{Z}} (x_i \bar{x}_i)^{n_i} e^{i n_i u}  \ .
\label{cd21}
\ea
Remember that the momentum in the two-plane carried by the $\F^0$ representations
is $s_i e^{i \theta_i}$ where $ e^{2i \theta_i}=\frac{\bar{x}_i}{x_i}$.
We can expand the $\F^0_{s_i,\jh_i}$ vertex operators in powers of $x$ and $\bar{x}$,
writing $y = \r e^{i\f}$ and
\be
\F^0_{s_i,\jh_i} = \sum_{n, \bar{n} \in \mathbb{Z}}
(-i x)^n (-i \bar{x})^{\bar{n}}e^{-i\frac{u}{2}(n+\bar{n})}e^{i \f (\bar{n}-n)}
J_{\bar{n}-n}(s \r) \ .
\ee

We now substitute these expressions in $(\ref{cd20})$ and
compute the contractions between the free fields using
$<u(z)v(w)> = \ln{(z-w)}$ and $<y(z)\tilde{y}(w)> = -2 \ln{(z-w)}$.
The only non-trivial correlators are of the form

\be
< H^-_p(z_1)  H^+_p(z_2) e^{i  (m-\bar{m})\f(z_3)}
J_{m-\bar{m}}(s \r(z_3)) e^{i  (m-\bar{m})\f(z_4)}
J_{\bar{m}-m}(s \r(z_4)) > \ ,
\ee
which can be rewritten as
\be
\frac{1}{(2 \pi)^2} \int_0^{2 \pi} d\theta_3 d\theta_4
e^{-i(m-\bar{m})(\theta_3-\theta_4)}
< H^-_p  H^+_p e^{\frac{i s_3}{2} \left [ y
e^{-i\theta_3}+ \tilde{y}e^{i \theta_3}
\right ] } e^{\frac{i s_4}{2} \left [ y
e^{-i\theta_4}+ \tilde{y} e^{i \theta_4} \right ] }> \ ,
\label{cd22}
\ee
and then computed using the propagator for the complex boson
$y$ in the presence of the two twist fields
\be
< H^-_p  H^+_p e^{\frac{i s_3}{2} \left [ y
e^{-i\theta_3}+ \tilde{y}e^{i \theta_3}
\right ] } e^{\frac{i s_4}{2} \left [ y
e^{-i\theta_4}+ \tilde{y} e^{i \theta_4} \right ] }> \sim
e^{-\frac{s_3s_4}{4} \left ( e^{i \theta} C +  e^{- i \theta} \tilde{C} \right )} \ ,
\label{cd23}
\ee
where $\theta = \theta_3 - \theta_4$ and
\be
C = 2 \left [ |u|^p \frac{F_p}{p}
+ \bar{u} |u|^{-p} \frac{\bar{F}_{1-p}}{1-p} \right ] \ , \hspace{1cm}
\tilde{C} =  2 \left [ |u|^{p}\frac{\bar{F}_p}{p}
+ u |u|^{-p} \frac{F_{1-p}}{1-p} \right ]\ .
\label{cd24}
\ee
We then substitute $(\ref{cd24})$ in $(\ref{cd22})$, compute the integral
and combine the result with the other terms coming from the contraction of the
exponential in $u$ and $v$.
The result, up to overall powers of the $z_{ij}$, is
\be
\sum_{m,\bar{m} \in \mathbb{Z}} x^{m} \bar{x}^{\bar{m}}
|u|^{-p(m+\bar{m})} \left ( \frac{\tilde{C}}{C} \right )^{\frac{\bar{m}-m}{2}}
I_{m-\bar{m}} \left ( \frac{s_3s_4}{2} \sqrt{C \tilde{C}} \right )   \ ,
\label{cd27}
\ee
and so we reproduce precisely $(\ref{cd5})$.

\section{Deformations of the Nappi-Witten model\label{e}}
\renewcommand{\theequation}{\Alph{section}.\arabic{equation}}
\def\bp{\bar\partial}
\def\p{\partial}
\def\t{\theta}
\def\r{\rho}
\def\ph{\phi}

We will study here marginal deformations of the NW model.
Since the Cartan of $H_4$ is two-dimensional the moduli space of deformations is
four dimensional.
There are however two inequivalent choices for the Cartan subalgebra, not
related by the action of the group.
The first choice consists of the generators $K,J$. The other of
$K, P_1$.
To describe the former it is convenient to start from the NW in
coordinates exhibiting the polar angle in the transverse plane:
\be
S_{NW}={1\over 2\pi}\int d^2z\left[2\p u\pb v+2\p v\pb
u-r^2\p u\pb u+\p \r\pb \r+\right.
\ee
$$\left.+
\r^2\p \ph\pb \ph+\r^2(\p u\pb
\ph-\p\ph\pb u)\right]
$$
We first wish to perform the finite deformation associated with
$J\bar J$ where
\be
J=2\p v-\r^2(\p u+\p \ph)\sp \bar J=2\pb v-\r^2(\pb u-\pb\ph)
\ee
The currents are associated with the transformations
\be
u\to u+\epsilon_1+\epsilon_2\sp
\ph\to\ph+\epsilon_1-\epsilon_2\sp \delta S={1\over 2\pi}\int d^2z\left[
2\p \epsilon_1~J+2\pb \epsilon_2~\bar J\right]\ee

The deformation  can be obtained by the standard trick of multiplying with an
extra U(1), and gauging the combined symmetry. The result is
\be
S_{NW'}(R)=S_{NW}+{R^2\over 2\pi}\int d^2z\left[{J\bar J\over
R^2\r^2+1}\right]
\ee
along with the dilaton
\be
\Phi=-{1\over 2}\log[R^2 \r^2+1]
\ee
The deformed theory still carries part of the original symmetry with conserved
currents
\be
J(R)={J\over R^2\r^2+1}\sp \bar J(R)={\bar J\over R^2\r^2+1}
\ee

The deformed $\sigma$ model satisfies
\be
S_{NW'}(0)=S_{NW}\sp S_{NW'}(R+\delta R)=S_{NW'}(R)+{\delta
R^2\over 2\pi}\int d^2z~~J(R)\bar J(R)
\ee
as it should.
Notice that the new geometry described by the metric
\be
ds^2={4R^2(dv)^2+4dudv+\r^2[-du^2+d\ph^2]\over R^2
\r^2+1}+d\r^2
\label{orig}\ee
and antisymmetric tensor
\be
B_{+\ph}={2R^2\r^2\over R^2\r^2+1}\sp B_{-\ph}={\r^2\over R^2\r^2+1}
\ee
is regular for $R^2$ positive and are not of the pp-wave type.
For $R\not=0$ we can change coordinates as
\be
v={x-t\over 2R}\sp u=t~R
\label{cha}\ee
the metric becoming
\be
ds^2=-dt^2+d\r^2+{dx^2+\r^2d\t^2\over R^2
\r^2+1}
\label{def}\ee
and
\be
B_{x\ph}={R\r^2\over R^2\r^2+1}\ee
The manifold is asymptotically flat as $r\to \infty$.

The change of variables (\ref{cha}) is singular at $R=0$ and this is the
reason that the $R\to 0$ limit of the metric (\ref{def}) is flat.
In fact the original metric (\ref{orig}) can be considered as
containing the Penrose parameter (R), so that as $R\to 0$ the
Penrose limit is performed and the Nappi-Witten pp-wave is obtained.

If the deformation is done in the opposite direction, $R^2\to -R^2$,
the metric develops a naked singularity
(coming in from infinity)
as attested by the scalar curvature
\be
{\cal R}=-2R^2{2R^2\r^2+5\over (R^2\r^2-1)^2}\ee

There are two non-trivial T-dualities acting on the deformed
geometry.
Dualizing with respect to the coordinate $x$ we obtain flat space
\be
d\tilde s^2_x=-dt^2+(Rdx+d\ph)^2+dx^2+d\r^2
\ee
with trivial antisymmetric tensor and dilaton.
Thus, the fact that the NW pp-wave is T-dual to flat space
\cite{kk} persists for the whole line of deformations.

Dualizing with respect to $\ph$ we obtain
\be
d\tilde s^2_{\t}=-dt^2+(dx+Rd\ph)^2+d\r^2+{d\ph^2\over \r^2}\sp
\Phi=-\log \r
\ee
which is the dual of the two-dimensional flat plane.

This deformation can be generalized to a four dimensional family
by also introducing the central currents
\be
K=\p u\sp \bar K =\pb u
\ee
\be
S_{NW''}(R_{ij})=S_{NW}-{1\over 2\pi}\int d^2z~\left(\bar
J , \bar K\right)\cdot M^{-1}\cdot \left(\begin{matrix} J\\
K\end{matrix}\right)
\ee
where
\be
M=\left(\begin{matrix} -\r^2+R_{11}^2+R_{21}^2 &
2+R_{11}R_{12}+R_{21}R_{22}\\
2+R_{11}R_{12}+R_{21}R_{22}& R_{12}^2+R_{22}^2\end{matrix}\right)
\ee
and a dilaton
\be
\Phi=-{1\over 2}\log\det M
\ee
It reduces to the previous $\sigma$-model when
$R_{12}=R_{21}=R_{22}=0$ and $R_{11}^2\to -1/R^2$.
This class of string backgrounds are still not of the pp-wave
type.

The other inequivalent set of deformations are generated by the
$K$
and $P_1$ currents. It is convenient here to change coordinates
and write the NW $\sigma$-model as \cite{kk}:

\be
S={1\over 2\pi}\int d^2z\left[\p y^i\bp y^i+2\cos u\pb y^1\p
y^2+\p u\pb v+\p v\bp u\right]
\ee
which bears a strong similarity to the $SU(2)$ WZW model.
This can be used to generate a line of marginal deformations,
parameterized by a positive real variable R,
given
by \cite{gk}
\be
S(R)={1\over 2\pi}\int d^2z\left[{\sin^2(u/2)~\p w^1\pb w^1+\cos^2(u/2)(R^2\p w^2\pb
w^2+\p w^1\pb w^2-\pb w^1\p w^2)\over
\cos^2(u/2)+R^2\sin^2(u/2)}+\right.\ee
$$
\left.
+\p u\pb v+\p v\bp u\right]$$
where
$w^1=y^2-y^1$, $w^2=y^1+y^2$
along with a dilaton
\be
\Phi=-{1\over 2}\log\left[{1\over R}\cos^2(u/2)+R~\sin^2(u/2)\right]+{\rm constant}\ee

Changing coordinates as
\be
u=2x^{-}\sp v=x^{+}+{1\over 4}{\cot (x^-) ~(x^{1})^2-R^2\tan
(x^-)~(x^{2})^2\over \cos^2(x^-)+R^2\sin^2(x^-)}
\ee
\be
w^1={\sqrt{\cos^2(x^-)+R^2\sin^2(x^-)}\over \sin (x^-)}~x^1\sp
w^2={\sqrt{\cos^2(x^-)+R^2\sin^2(x^-)}\over R~\cos (x^-)}~x^2\ee
we obtain
\be
S(R)={1\over 2\pi}\int d^2z\left[2\p x^+\pb x^-+2\p x^-\pb x^+ +
\p x^1\pb x^1+\p x^2\pb x^2+\right.
\ee
$$
+{[1-2R^2+(1-R^2)\cos(2x^-)](x^1)^2+R^2[R^2-2+(1-R^2)\cos(2x^-)](x^2)^2\over
(\cos^2(x^-)+R^2\sin^2(x^-))^2}\p x^-\pb x^--
$$
$$
\left. +B_{12}(\p x^1\pb x^2-\pb x^1\p x^2)+B_{-1}(\p x^-\pb
x^1-\p x^1\pb x^-)+B_{-2}(\p x^-\pb
x^2-\p x^2\pb x^-)\right]
$$
with
\be
B_{12}={\cot(x^-)\over R}
\ee
$$
B_{-1}=-{R~x^2\over \cos^2(x^-)+R^2\sin^2(x^-)}\sp
B_{-2}=-{\cot^2(x^-)~x^1\over R(\cos^2(x^-)+R^2\sin^2(x^-))}
$$
and the dilaton becomes
\be
\Phi=-{1\over 2}\log\left[{1\over R}\cos^2(x^-)+R~\sin^2(x^-)\right]+{\rm constant}\ee

Dropping total derivatives the antisymmetric tensor can be
simplified:
\be
S(R)={1\over 2\pi}\int d^2z\left[2\p x^+\pb x^-+2\p x^-\pb x^+ +
\p x^1\pb x^1+\p x^2\pb x^2+\right.
\ee
$$
+{[1-2R^2+(1-R^2)\cos(2x^-)](x^1)^2+R^2[R^2-2+(1-R^2)\cos(2x^-)](x^2)^2\over
(\cos^2(x^-)+R^2\sin^2(x^-))^2}\p x^-\pb x^--
$$
$$\left.
-2\arctan[R\tan(x^-)](\p x^1\pb x^2-\pb x^1\p x^2)\right]
$$
This conformal $\sigma$-model describes the finite deformation
generated by the $\int P_1\bar P_1$ deformation.

Note that there is an $R\to 1/R$ duality of the background fields
accompanied by the reparametrization $x^1\leftrightarrow x^2$ and
$x^-\to x^-+\pi/2$.
At the ends of the line the theory becomes a direct product of a
real line and the coset of the NW theory obtained by gauging $P_1$
\cite{kk}.
\be
ds^2(R=0)=4dx^+dx^-+(dx^1)^2+(dx^2)^2+2{(x^1)^2\over
\cos^2(x^-)}(dx^-)^2
\ee

This deformation  can be also be generalized to a four-parameter family
\be
S(R,R_{ij})=\int {d^2z\over 2\pi}\left[{\sin^2(u/2)~\p w^1\pb w^1+\cos^2(u/2)(R^2\p w^2\pb
w^2+\p w^1\pb w^2-\pb w^1\p w^2)\over
\cos^2(u/2)+R^2\sin^2(u/2)}+\right.\ee
$$
{R_{12}(\pb u(-\sin^2(u/2)\p w^1+\cos^2(u/2)\p w^2)+R_{21}(\p u(\sin^2(u/2)\pb w^1+\cos^2(u/2)\pb w^2)
\over \cos^2(u/2)+R^2\sin^2(u/2)}+
$$
$$
\left.+R_2^2\p u\pb u
+\p u\pb v+\p v\bp u\right]$$
and the same dilaton as above.
Changing again coordinates as
\be
u=2x^{-}
\ee
$$ v=x^{+}+{1\over 4}{\cot (x^-) ~(x^{1})^2-R^2\tan
(x^-)~(x^{2})^2\over
\cos^2(x^-)+R^2\sin^2(x^-)}+
$$
$$+
{(R_{12}-R_{21})R\sin(x^-)x^1-(R_{12}+R_{21})\cos(x^-)x^2\over
2R\sqrt{\cos^2(x^-)+R^2\sin^2(x^-)}}
$$
\be
w^1={\sqrt{\cos^2(x^-)+R^2\sin^2(x^-)}\over \sin (x^-)}~x^1\sp
w^2={\sqrt{\cos^2(x^-)+R^2\sin^2(x^-)}\over R~\cos (x^-)}~x^2\ee
we obtain
\be
S(R,R_2,R_{12},R_{21})={1\over 2\pi}\int d^2z\left[2\p x^+\pb x^-+2\p x^-\pb x^+ +
\p x^1\pb x^1+\p x^2\pb x^2+\right.
\ee
$$
+\left\{{[1-2R^2+(1-R^2)\cos(2x^-)](x^1)^2+R^2[R^2-2+(1-R^2)\cos(2x^-)](x^2)^2\over
(\cos^2(x^-)+R^2\sin^2(x^-))^2}  +\right.
$$
$$\left.
+{4(R_{12}-R_{21})\cos(x^-)~x^1-4(R_{12}+R_{21})\sin(x^-)~x^2\over
[\cos^2(x^-)+R^2\sin^2(x^-)]^{3/2}}+4R_2^2\right\}
\p x^-\pb x^--
$$
$$
\left. +B_{12}(\p x^1\pb x^2-\pb x^1\p x^2)+B_{-1}(\p x^-\pb
x^1-\p x^1\pb x^-)+B_{-2}(\p x^-\pb
x^2-\p x^2\pb x^-)\right]
$$
with
\be
B_{12}={\cot(x^-)\over R}
\ee
$$
B_{-1}=-{R~x^2\over
\cos^2(x^-)+R^2\sin^2(x^-)}+{2(R_{12}+R_{21})~\sin(x^-)\over
\sqrt{\cos^2(x^-)+R^2\sin^2(x^-)}}
$$
$$
B_{-2}=-{\cot^2(x^-)~x^1\over R(\cos^2(x^-)+R^2\sin^2(x^-))}-{2(R_{12}-R_{21})~\cos(x^-)\over
R~\sqrt{\cos^2(x^-)+R^2\sin^2(x^-)}}
$$
By a gauge transformation on the antisymmetric tensor it can be
replaced by
\be
B_{12}=-2\arctan[R\tan(x^-)]
\ee
This class of space-times are of the pp-wave type.


\begin{thebibliography}{99}


\bibitem{orig}R.~G\"uven,
Phys.\ Lett.\ B {\bf 191} (1987) 275;\\
D.~Amati and C.~Klimcik,
Phys.\ Lett.\ B {\bf 210} (1988) 92;
Phys.\ Lett.\ B {\bf 219} (1989) 443;\\
H.~J.~de Vega and N.~Sanchez,
Nucl.\ Phys.\ B {\bf 317} (1989) 731;\\
G.~T.~Horowitz and A.~R.~Steif,
Phys.\ Rev.\ Lett.\  {\bf 64} (1990) 260;
Phys.\ Rev.\ D {\bf 42} (1990) 1950;\\
A.~R.~Steif,
Phys.\ Rev.\ D {\bf 42} (1990) 2150;\\
A.~A.~Tseytlin,
Phys.\ Lett.\ B {\bf 288} (1992) 279
[arXiv:hep-th/9205058];
Nucl.\ Phys.\ B {\bf 390} (1993)
153 [arXiv:hep-th/9209023];
Phys.\ Rev.\ D {\bf 47} (1993) 3421 [arXiv:hep-th/9211061]

\bibitem{nw}
C.~R.~Nappi and E.~Witten,
Phys.\ Rev.\ Lett.\  {\bf 71} (1993) 3751.

\bibitem{kk}
E.~Kiritsis and C.~Kounnas,
Phys.\ Lett.\ B {\bf 320} (1994) 264
[Addendum-ibid.\ B {\bf 325} (1994) 536].

\bibitem{kkl}
E.~Kiritsis, C.~Kounnas and D.~L\"ust,
Phys.\ Lett.\ B {\bf 331} (1994) 321.

\bibitem{wzw}
C.~Klimcik and A.~A.~Tseytlin,
Phys.\ Lett.\ B {\bf 323} (1994) 305
[arXiv:hep-th/9311012];
, Nucl.\ Phys.\ B {\bf 424} (1994) 71 [arXiv:hep-th/9402120];\\
K.~Sfetsos,
{\bf 324} (1994) 335 [arXiv:hep-th/9311010];
Int.\ J.\ Mod.\ Phys.\ A {\bf 9} (1994) 4759
[arXiv:hep-th/9311093];\\
N.~Mohammedi,
Phys.\ Lett.\ B {\bf 325} (1994) 371 [arXiv:hep-th/9312182];\\
A.~Kumar and S.~Mahapatra,
Mod.\ Phys.\ Lett.\ A {\bf 9} (1994) 925 [arXiv:hep-th/9401098];\\
J.~M.~Figueroa-O'Farrill and S.~Stanciu,
Phys.\ Lett.\ B {\bf 327} (1994) 40 [arXiv:hep-th/9402035];\\
I.~Antoniadis and N.~A.~Obers,
Nucl.\ Phys.\ B {\bf 423} (1994) 639 [arXiv:hep-th/9403191];\\
K.~Sfetsos and A.~A.~Tseytlin,
Nucl.\ Phys.\ B
{\bf 427} (1994) 245 [arXiv:hep-th/9404063];\\
O.~Jofre and C.~Nunez,
Phys.\ Rev.\ D {\bf 50} (1994) 5232 [arXiv:hep-th/9311187].

\bibitem{ors}
D.~I.~Olive, E.~Rabinovici and A.~Schwimmer,
Phys.\ Lett.\ B {\bf 321} (1994) 361.

\bibitem{sm}
A.~Hatzinikitas and I.~Smyrnakis,
[arXiv:hep-th/0303043].

\bibitem{rt}
J.~G.~Russo and A.~A.~Tseytlin,
Nucl.\ Phys.\ B {\bf 448}, 293 (1995)
[arXiv:hep-th/9411099].


\bibitem{forgacs}
P.~Forgacs, P.~A.~Horvathy, Z.~Horvath and L.~Palla,
Heavy Ion Phys.\  {\bf 1} (1995) 65.

\bibitem{pen}
R.~Gueven,
Phys.\ Lett.\ B {\bf 482} (2000) 255
[arXiv:hep-th/0005061];\\
M.~Blau, J.~Figueroa-O'Farrill, C.~Hull and G.~Papadopoulos,
arXiv:hep-th/0201081;\\
JHEP {\bf 0201}, 047 (2002)
[arXiv:hep-th/0110242].

\bibitem{metsaev}
R.~R.~Metsaev,
Nucl.\ Phys.\ B {\bf 625}, 70 (2002)
[arXiv:hep-th/0112044];\\
R.~R.~Metsaev and A.~A.~Tseytlin,
[arXiv:hep-th/0202109];\\
J.~Maldacena and L.~Maoz,
JHEP {\bf 0212} (2002) 046
[arXiv:hep-th/0207284];\\
I.~Bakas and J.~Sonnenschein,
JHEP {\bf 0212} (2002) 049
[arXiv:hep-th/0211257];\\
G.~Papadopoulos, J.~G.~Russo and A.~A.~Tseytlin,
Class.\ Quant.\ Grav.\  {\bf 20} (2003) 969
[arXiv:hep-th/0211289].

\bibitem{bmn}
D.~Berenstein, J.~Maldacena and H.~Nastase,
[arXiv:hep-th/0202021].

\bibitem{pp}C.~Kristjansen, J.~Plefka, G.~W.~Semenoff and M.~Staudacher,
Nucl.\ Phys.\ B {\bf 643} (2002) 3
[arXiv:hep-th/0205033];\\
D.~J.~Gross, A.~Mikhailov and R.~Roiban,
Annals Phys.\  {\bf 301} (2002) 31
[arXiv:hep-th/0205066];
[arXiv:hep-th/0208231];\\
N.~R.~Constable, D.~Z.~Freedman, M.~Headrick, S.~Minwalla, L.~Motl, A.~Postnikov and W.~Skiba,
JHEP {\bf 0207} (2002) 017
[arXiv:hep-th/0205089];\\
A.~Santambrogio and D.~Zanon,
Phys.\ Lett.\ B {\bf 545} (2002) 425
[arXiv:hep-th/0206079];\\
N.~Beisert, C.~Kristjansen, J.~Plefka, G.~W.~Semenoff and M.~Staudacher,
Nucl.\ Phys.\ B {\bf 650} (2003) 125
[arXiv:hep-th/0208178];\\
D.~Vaman and H.~Verlinde,
[arXiv:hep-th/0209215];\\
J.~Pearson, M.~Spradlin, D.~Vaman, H.~Verlinde and A.~Volovich,
[arXiv:hep-th/0210102];\\
N.~R.~Constable, D.~Z.~Freedman, M.~Headrick and S.~Minwalla,
JHEP {\bf 0210} (2002) 068
[arXiv:hep-th/0209002];\\
N.~Beisert, C.~Kristjansen, J.~Plefka and M.~Staudacher,
Phys.\ Lett.\ B {\bf 558} (2003) 229
[arXiv:hep-th/0212269].

















\bibitem{lc}
M.~Spradlin and A.~Volovich,
Phys.\ Rev.\ D {\bf 66} (2002) 086004
[arXiv:hep-th/0204146];JHEP {\bf 0301} (2003) 036
[arXiv:hep-th/0206073];\\
J.~H.~Schwarz,
JHEP {\bf 0209} (2002) 058
[arXiv:hep-th/0208179];\\
A.~Pankiewicz,
JHEP {\bf 0209} (2002) 056
[arXiv:hep-th/0208209];\\
A.~Pankiewicz and B.~.~Stefanski,
Nucl.\ Phys.\ B {\bf 657} (2003) 79
[arXiv:hep-th/0210246];\\
Y.~H.~He, J.~H.~Schwarz, M.~Spradlin and A.~Volovich,
[arXiv:hep-th/0211198];\\
P.~Di Vecchia, J.~L.~Petersen, M.~Petrini, R.~Russo and A.~Tanzini,
arXiv:hep-th/0304025;\\
A.~Pankiewicz,
arXiv:hep-th/0304232.



\bibitem{seiberg}
O.~Aharony, M.~Berkooz, D.~Kutasov and N.~Seiberg,
JHEP {\bf 9810} (1998) 004
[arXiv:hep-th/9808149].

\bibitem{gomis}
J.~Gomis and H.~Ooguri,
Nucl.\ Phys.\ B {\bf 635} (2002) 106
[arXiv:hep-th/0202157].

\bibitem{kp}
E.~Kiritsis and B.~Pioline,
JHEP {\bf 0208} (2002) 048
[arXiv:hep-th/0204004].

\bibitem{sj}
D.~Sadri and M.~M.~Sheikh-Jabbari,
arXiv:hep-th/0304169.


\bibitem{hol}
S.~R.~Das, C.~Gomez and S.~J.~Rey,
Phys.\ Rev.\ D {\bf 66} (2002) 046002
[arXiv:hep-th/0203164];\\
R.~G.~Leigh, K.~Okuyama and M.~Rozali,
Phys.\ Rev.\ D {\bf 66} (2002) 046004
[arXiv:hep-th/0204026];\\
D.~Berenstein and H.~Nastase,
[arXiv:hep-th/0205048].




\bibitem{ads3}
A.~Giveon, D.~Kutasov and N.~Seiberg,
Adv.\ Theor.\ Math.\ Phys.\  {\bf 2} (1998) 733;
[arXiv:hep-th/9806194].
D.~Kutasov and N.~Seiberg,
JHEP {\bf 9904} (1999) 008;
[arXiv:hep-th/9903219]

\bibitem{oo}
J.~de Boer, H.~Ooguri, H.~Robins and J.~Tannenhauser,
JHEP {\bf 9812} (1998) 026
[arXiv:hep-th/9812046].


\bibitem{bk}
I.~Bakas and E.~Kiritsis,
Nucl.\ Phys.\ B {\bf 343} (1990) 185
[Erratum-ibid.\ B {\bf 350} (1991) 512];
Mod.\ Phys.\ Lett.\ A {\bf 5} (1990) 2039.

\bibitem{dfms}
L.~J.~Dixon, D.~Friedan, E.~J.~Martinec and S.~H.~Shenker,
Nucl.\ Phys.\ B {\bf 282} (1987) 13.

\bibitem{zf}
A.~B.~Zamolodchikov and V.~A.~Fateev,
Sov.\ J.\ Nucl.\ Phys.\  {\bf 43} (1986) 657
[Yad.\ Fiz.\  {\bf 43} (1986) 1031].

\bibitem{maloog1}
J.~M.~Maldacena and H.~Ooguri,
J.\ Math.\ Phys.\  {\bf 42} (2001) 2929;
[arXiv:hep-th/0001053].

\bibitem{maloog2}
J.~M.~Maldacena, H.~Ooguri and J.~Son,
J.\ Math.\ Phys.\  {\bf 42} (2001) 2961;
[arXiv:hep-th/0005183].

\bibitem{maloog3}
J.~M.~Maldacena and H.~Ooguri,
Phys.\ Rev.\ D {\bf 65} (2002) 106006;
[arXiv:hep-th/0111180].

\bibitem{lcft}
I.~Bakas and K.~Sfetsos,
Nucl.\ Phys.\ B {\bf 639} (2002) 223;
[arXiv:hep-th/0205006];\\
K.~Sfetsos,
Phys.\ Lett.\ B {\bf 543} (2002) 73;
[arXiv:hep-th/0206091].

\bibitem{ns5}
C.~Kounnas, M.~Porrati and B.~Rostand,
Phys.\ Lett. {\bf B258} (1991) 61;\\
C.G.~Callan, J.A.~Harvey and A.~Strominger,
 \np{367}{1991}{60};
  [arXiv:hep-th/9112030].

\bibitem{dvv}
R.~Dijkgraaf, H.~Verlinde and E.~Verlinde,
Nucl.\ Phys.\ B {\bf 371} (1992) 269.


\bibitem{cg} W. Miller Jr.,  ``Lie Theory and Special functions", Academic Press, NY 1968;\\
N. Ja. Vilenkin, ``Fonctions Sp\'eciales et Th\'eorie de la Repr\'esentation des Groupes", Dunod, Paris 1969.


\bibitem{df}
V.~S.~Dotsenko and V.~A.~Fateev,
Nucl.\ Phys.\ B {\bf 240} (1984) 312;
Nucl.\ Phys.\ B {\bf 251} (1985) 691.

\bibitem{dot}
V.~S.~Dotsenko,
Nucl.\ Phys.\ B {\bf 358} (1991) 547.

\bibitem{kz}
V.~G.~Knizhnik and A.~B.~Zamolodchikov,
Nucl.\ Phys.\ B {\bf 247} (1984) 83.

\bibitem{gaberdiel}
M.~R.~Gaberdiel,
Nucl.\ Phys.\ B {\bf 618} (2001) 407;
[arXiv:hep-th/0105046].


\bibitem{feb}
J.~Teschner,
{\it Crossing symmetry in the H(3)+ WZNW model},
Phys.\ Lett.\ B {\bf 521} (2001) 127; \\
B.~Ponsot,
{\it Monodromy of solutions of the Knizhnik-Zamolodchikov equation:  SL(2,C)/SU(2) WZNW model},
Nucl.\ Phys.\ B {\bf 642} (2002) 114.



\bibitem{sok}
G.~Arutyunov and E.~Sokatchev,
JHEP {\bf 0208} (2002) 014
[arXiv:hep-th/0205270].


\bibitem{ks1}
K. Sfetsos,
Phys.\ Rev.\ D {\bf 50} (1994) 2784 [arXiv:hep-th/9402031];

\bibitem{ks2}
A.~A.~Kehagias and P.~A.~Meessen,
Phys.\ Lett.\ B {\bf 331}, 77 (1994)
[arXiv:hep-th/9403041].


\bibitem{jabari}
D.~s.~Bak and M.~M.~Sheikh-Jabbari,
JHEP {\bf 0302} (2003) 019
[arXiv:hep-th/0211073].

\bibitem{dbr}
S.~Stanciu and A.~A.~Tseytlin,
JHEP {\bf 9806} (1998) 010
[arXiv:hep-th/9805006];\\
J.~M.~Figueroa-O'Farrill and S.~Stanciu,
JHEP {\bf 0001} (2000) 024
[arXiv:hep-th/9909164];\\
O.~Bergman, M.~R.~Gaberdiel and M.~B.~Green,
[arXiv:hep-th/0205183];\\
S.~Stanciu and J.~Figueroa-O'Farrill,
[arXiv:hep-th/0303212];\\
Y.~Hikida, H.~Takayanagi and T.~Takayanagi,
[arXiv:hep-th/0303214].


\bibitem{dual}
E.~Kiritsis,
Nucl.\ Phys.\ B {\bf 405} (1993) 109
[arXiv:hep-th/9302033].

\bibitem{gk}
A.~Giveon and E.~Kiritsis,
Nucl.\ Phys.\ B {\bf 411} (1994) 487
[arXiv:hep-th/9303016].

\bibitem{sw}
J. Maldacena, J. Michelson and A. Strominger,
JHEP {\bf 9902} (1999) 011
[arXiv:hep-th/9812073],
\\
N. Seiberg and E. Witten,
JHEP {\bf 9904} (1999) 017
[arXiv:hep-th/9903224].

\end{thebibliography}
\end{document}